\documentclass[oneside]{amsart}
\usepackage[skip=3pt]{parskip}
\usepackage{enumitem}
\setlist[enumerate]{label=\arabic*., ref=\arabic*}
\usepackage[utf8]{inputenc} 
\usepackage[T1]{fontenc}    
\usepackage{url}            
\usepackage{booktabs}       
\usepackage{amsfonts}       
\usepackage{nicefrac}       
\usepackage{microtype}      
\usepackage{xcolor}         

\usepackage[textsize=tiny]{todonotes}
\usepackage{subfigure}
\usepackage{algorithm}
\usepackage{algorithmic}
\usepackage {graphicx}
\usepackage{amsmath}
\usepackage{amsthm}
\usepackage{amsfonts}
\usepackage{amssymb}
\usepackage{hyperref}
\hypersetup{
colorlinks=true, linkcolor=blue, citecolor=blue, urlcolor=blue}
\usepackage[capitalize,noabbrev]{cleveref}
\makeatletter
\def\subsection{\@startsection{subsection}{2}%
  \z@{.5\linespacing\@plus.7\linespacing}
{.5\baselineskip}%
  {\normalfont\centering\scshape}%
}
\makeatother
\usepackage{lineno}
\usepackage{mathptmx}

\usepackage{xcolor}
 \newcommand{\blue}[1]{{\textcolor{blue}{#1}}}

\newtheorem{theorem}{Theorem} 
\newtheorem{definition}{Definition}[theorem]

\title{EuLearn: A 3D database for learning Euler characteristics}

\usepackage{fnpct}

\usepackage[foot]{amsaddr}
\usepackage{natbib}

\author{Rodrigo Fritz$^\ast$, Pablo Suárez-Serrato$^\ast$}
\address{$^\ast$Instituto de Matemáticas, UNAM.}
\author{Victor Mijangos$^\dagger$, Anayanzi D. Martinez-Hernandez$^\dagger$}
\address{$^\dagger$Facultad de Ciencias, UNAM.}
\author{Eduardo Ivan Velazquez Richards$^\ddagger$}
\address{$^\ddagger$IIMAS, UNAM.}

\date{}
\usepackage[verbose=true,letterpaper]{geometry}
\AtBeginDocument{
  \newgeometry{
    textheight=9in,
    textwidth=5.5in,
    top=1in,
    headheight=12pt,
    headsep=25pt,
    footskip=30pt
  }}

\begin{document}

\maketitle

\begin{abstract}
We present EuLearn\footnotemark, the first surface datasets equitably representing a diversity of topological types. 
We designed our embedded surfaces of uniformly varying genera relying on random knots, thus allowing our surfaces to knot with themselves. 
EuLearn contributes new topological datasets of meshes, point clouds, and scalar fields in 3D. 
We aim to facilitate the training of machine learning systems that can discern topological features. 
We experimented with specific emblematic 3D neural network architectures, finding that their vanilla implementations perform poorly on genus classification.
To enhance performance, we developed a novel, non-Euclidean, statistical sampling method adapted to graph and manifold data.
We also introduce adjacency-informed adaptations of PointNet and Transformer architectures that rely on our non-Euclidean sampling strategy. 
Our results demonstrate that incorporating topological information into deep learning workflows significantly improves performance on these otherwise challenging EuLearn datasets.
 
\end{abstract}
\footnotetext{The dataset and source code are publicly available at \url{https://huggingface.co/datasets/appliedgeometry/EuLearn} and \url{https://github.com/appliedgeometry/EuLearn_db}, respectively.}

\section{Introduction}
\label{Introduction}

\begin{figure}[ht]
    \renewcommand{\thesubfigure}{} 
    \centering
    \subfigure[$g = 0$]{\includegraphics[width=0.15\columnwidth]{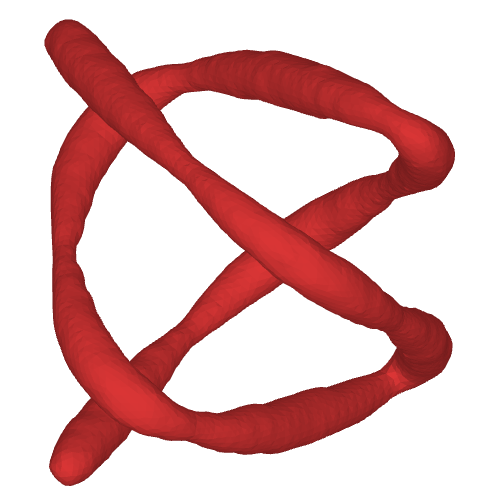}}
    \subfigure[$g = 1$]{\includegraphics[width=0.15\columnwidth]{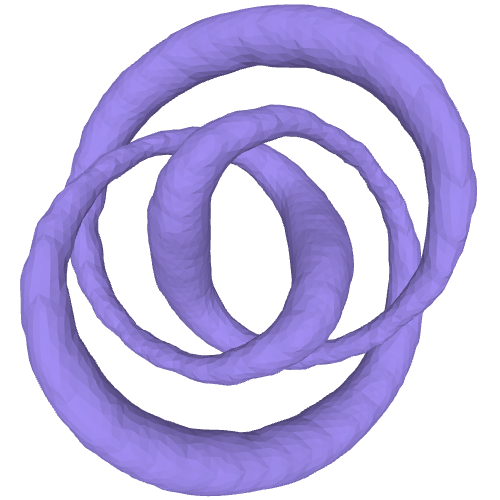}}
    \subfigure[$g = 2$]{\includegraphics[width=0.15\columnwidth]{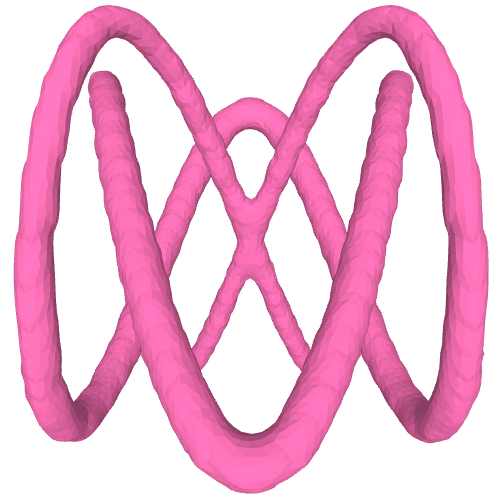}}
    \subfigure[$g = 3$]{\includegraphics[width=0.15\columnwidth]{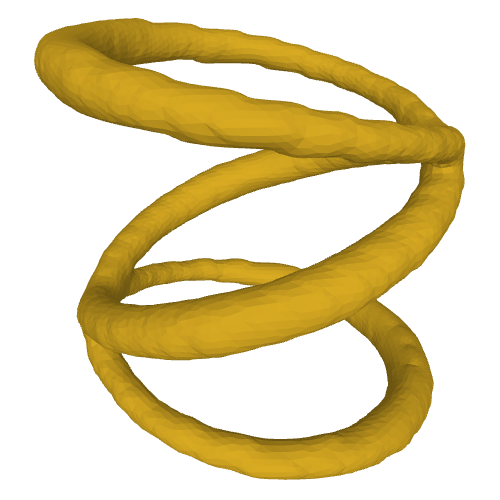}}
    \subfigure[$g = 4$]{\includegraphics[width=0.15\columnwidth]{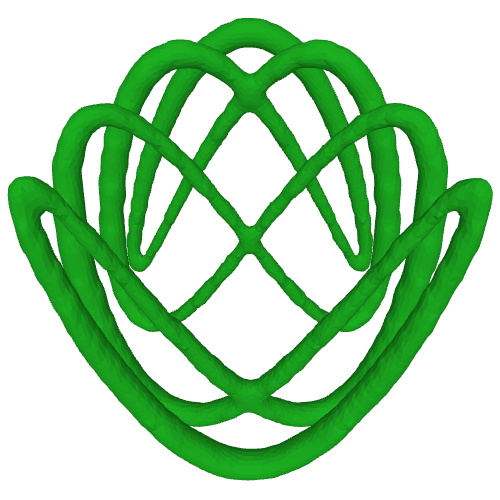}}
    \hspace{0mm}
    \subfigure[$g = 5$]{\includegraphics[width=0.15\columnwidth]{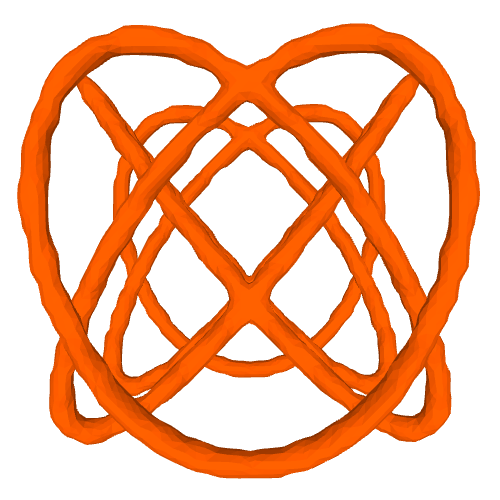}}
    \subfigure[$g = 6$]{\includegraphics[width=0.15\columnwidth]{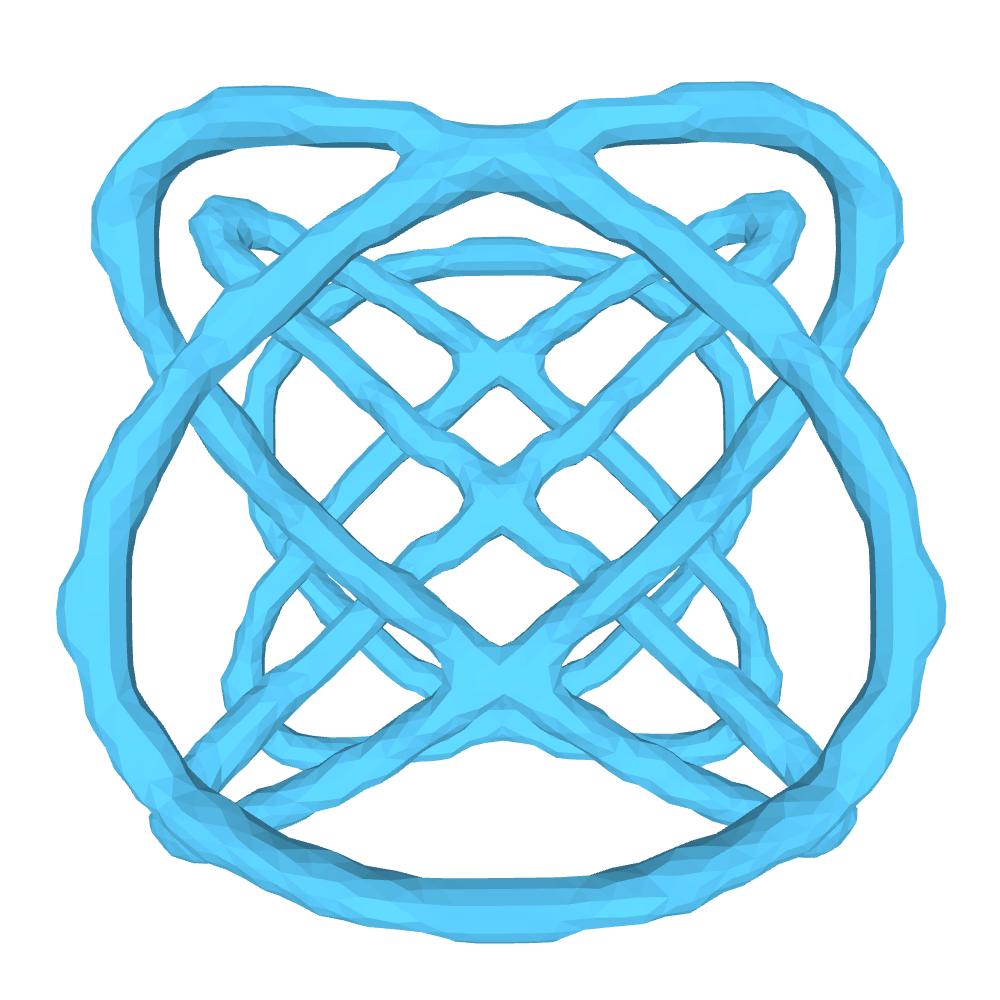}}
    \subfigure[$g = 7$]{\includegraphics[width=0.15\columnwidth]{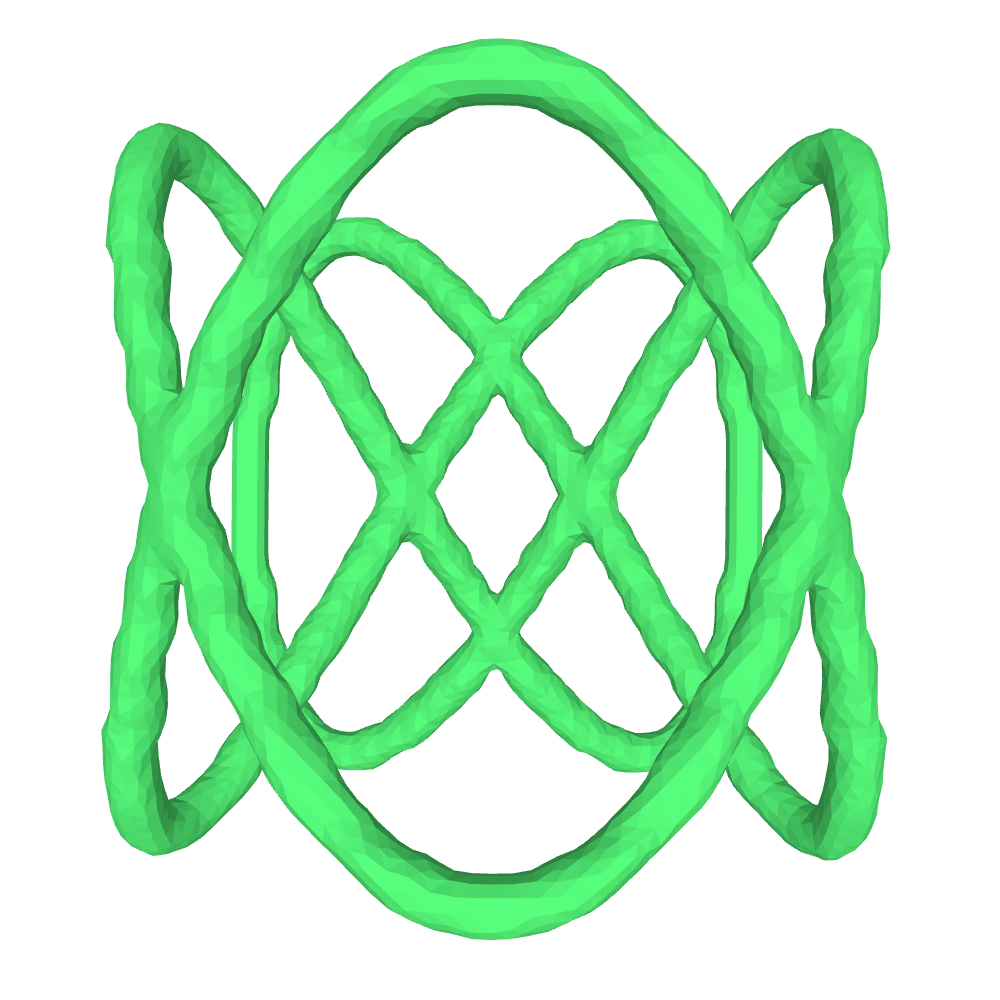}}
    \subfigure[$g = 8$]{\includegraphics[width=0.15\columnwidth]{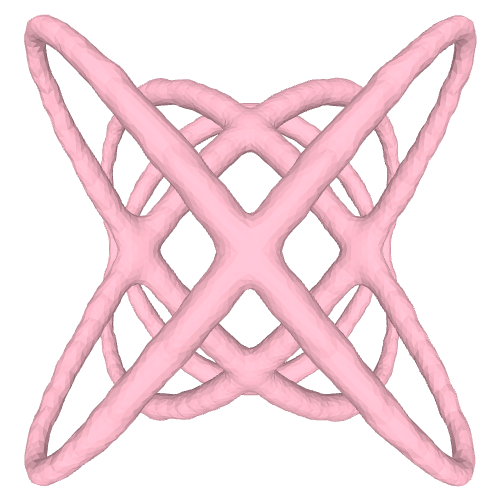}}
    \hspace{0mm}
    \subfigure[$g = 9$]{\includegraphics[width=0.15 \columnwidth, trim=0 15mm 0 20mm,clip]{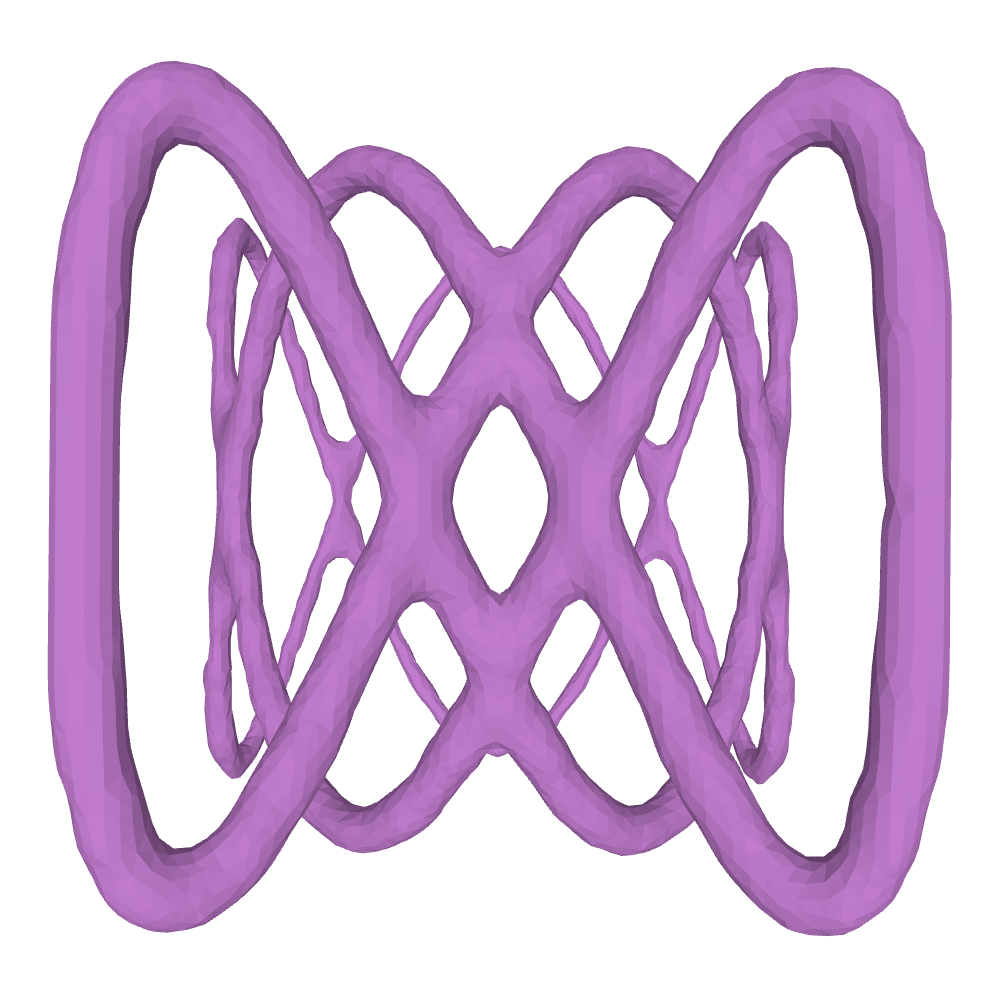}}
    \hspace{1mm}
    \subfigure[$g = 10$]{\includegraphics[width=0.15\columnwidth]{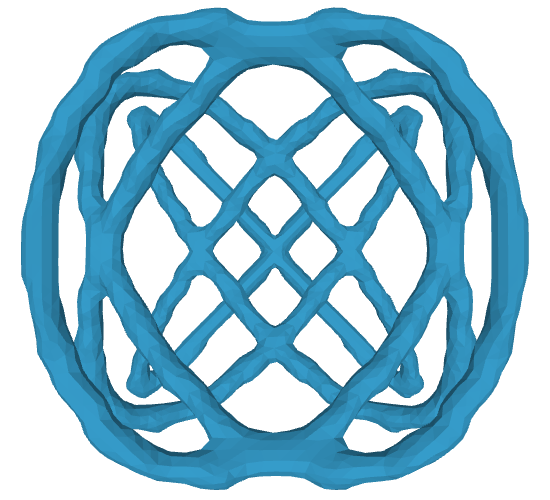}}
    \caption{Examples of embedded, knotted, surfaces for each genus in our EuLearn database.}
    \label{fig:Eulearn_sample}
\end{figure}
Advances in machine learning across various domains have been driven by the introduction of new datasets that integrate previously overlooked or underexplored information.
In computer vision, shape analysis, and geometric processing, the crucial contributions of FAUST \cite{Bogo:CVPR:2014} and ShapeNet \cite{shapenet2015} led to significant breakthroughs.
Their broader impact may be measured by the hundreds of patents that refer to these datasets. 
However, as far as we know, all current 3D point-cloud databases supported on surfaces lack uniform topological diversity.
The homeomorphism type of the associated underlying 2-dimensional surfaces is either fixed or not distributed in a known or controlled way.
For example, consider the following theoretical results (proven in \cref{App:Proofs}):
\begin{theorem}\label{thm:FAUST-S2}
    Every file in the FAUST dataset is homeomorphic to a $2$-dimensional sphere.
    Therefore, the FAUST dataset only has a single topological type.
\end{theorem}

\begin{theorem}\label{thm:not-unif-dist-data}
    The distribution of topological types of surfaces (when well defined) is not uniform in the following datasets: ShapeNet \cite{shapenet2015}, Thingi10K \cite{zhou2016thingi10kdataset100003dprinting}, ABC \cite{Koch_2019_CVPR}, SimJEB \cite{Whalen_2021}.
\end{theorem}
Therefore, even when restricted to the watertight subset of files in these datasets a machine learning regime trained on them will reflect the unbalanced nature of its topological distribution.

\begin{table}[!ht]
    \small
    \centering
    \begin{tabular}{c|c c c c} \hline
        \textbf{Model} & \textbf{Precision} & \textbf{Recall} & $\mathbf{F_1}$ & \textbf{Acc.} \\ \hline
         Attention (classic) & 0.01 & 0.10 & 0.02 & 0.10\\
         FNO & 0.02 & 0.11 & 0.04 & 0.20 \\
         DGCNN & 0.07 & 0.14 & 0.07 & 0.16 \\
         PointNet (classic) & 0.50 & 0.49 & 0.43 & 0.49 \\
         PointNet++ & 0.50 & 0.54 & 0.52 & 0.63 \\
         GS PointNet* (ours) & \textbf{0.82} & 0.79 & 0.78 & 0.79 \\
         GS Attention* (ours) & 0.81 & \textbf{0.81} & \textbf{0.79} & \textbf{0.81} \\ \hline
    \end{tabular}
    \vspace{2mm}
    \caption{Summary of the results obtained with different methods. The training set consisted of 1925 surfaces, while the test set consisted in 825 surfaces. *Graph Sampled (GS) PointNet and GS Attention are tailored architectures for our dataset.}
    \label{tab:AttentionResults}
\end{table}

The topological genus of a surface measures its ``holes'' or ``handles,'' characterizing the number of cuts required to transform the surface into simpler shapes, such as a flat plane or a sphere, without tearing it.
For instance, a sphere has genus 0 (no handles), a torus (doughnut shape) has genus 1, a double torus (doughnut with 2 holes) has genus 2, and so on.
The genus is a fundamental topological invariant useful for classifying surfaces.

Despite significant progress in geometric processing, the field of topological analysis and its integration with machine learning and deep learning methods still face critical challenges. 
For example, by \cref{thm:FAUST-S2} all surfaces in the FAUST dataset have genus 0, rendering them diffeomorphic to a two-dimensional sphere and devoid of genus diversity. 
In contrast, ShapeNet exhibits some genus diversity; however, this is poorly defined due to irregularly connected meshes where the components of an object are attached together rather than forming a single continuous mesh (see \cref{appendix:ShapeNet}).
Moreover, even when it is well-defined, by \cref{thm:not-unif-dist-data} the genus is not uniformly distributed, so it is a topologically unbalanced dataset.

Our EuLearn dataset, addressed these shortcoming, as it is constructed so that all components of an object are perfectly merged into branches of a unified tubular structure. 
This ensures the final object has a well-defined genus and Euler characteristic. 
 The key to achieving these well-defined structures was the application of the Marching Cubes algorithm, guided by a scalar field acting as a signed distance function to a zero-level surface, ensuring smooth transitions and accurate topology.
 We therefore obtain the following result (see \cref{App:Proofs}):
 \vspace{2mm}
 \begin{theorem}\label{thm:EuLearn-unif-dist}
     The distribution of topological types of surfaces in the EuLearn dataset is uniform.
 \end{theorem}

Here, we contribute databases whose underlying objects are compact and orientable surfaces randomly embedded in Euclidean space, and that by \cref{thm:EuLearn-unif-dist} have uniformly distributed genera (see \cref{fig:Eulearn_sample}).  
For our experiments, we construct a database of point-clouds, each supported on a single surface.
In this way, we contribute the first 3-dimensional dataset that takes into account varying topological features in a controlled manner (since the genus is distributed uniformly across the dataset).
Our dataset contains 3300 mathematical surfaces classified by genera, ranging from 0 to 10, with 300 examples of each genus.
Each surface results from thickening a closed curve in $\mathbf{R}^3$, with its genus determined by the number of self-intersections of the curve.

We thus address the problem of topologically unbalanced datasets, which has so far obstructed progress of spatial, and more generally, topological, learning.
Our work here contributes to ameliorating the blindness of current models to topological features, as they have not yet been exposed to datasets with adequately ordered topological data.

Each surface is provided with three files: a mesh, a discrete level set on a $100^3$ resolution grid, and a smoothed version of the mesh.
The full dataset occupies 15.2GB. \cref{sec:HigherGenusSurfaces} details the technical process used to construct these surfaces.
To construct our EuLearn dataset, we had to overcome the following key challenges: the numerical computation of self-intersections of parameterized knots, setting bounds for the thickening interval of the tubular neighborhoods using the reach, and discretizing these neighborhoods as level sets.
These tasks were handled using a modular workflow that switched between parallel programming for scalar field calculations, concurrent programming for self-intersection detection, and sequential programming for mesh generation and smoothing.

\begin{figure}
    \centering
    \includegraphics[width=0.8\linewidth]{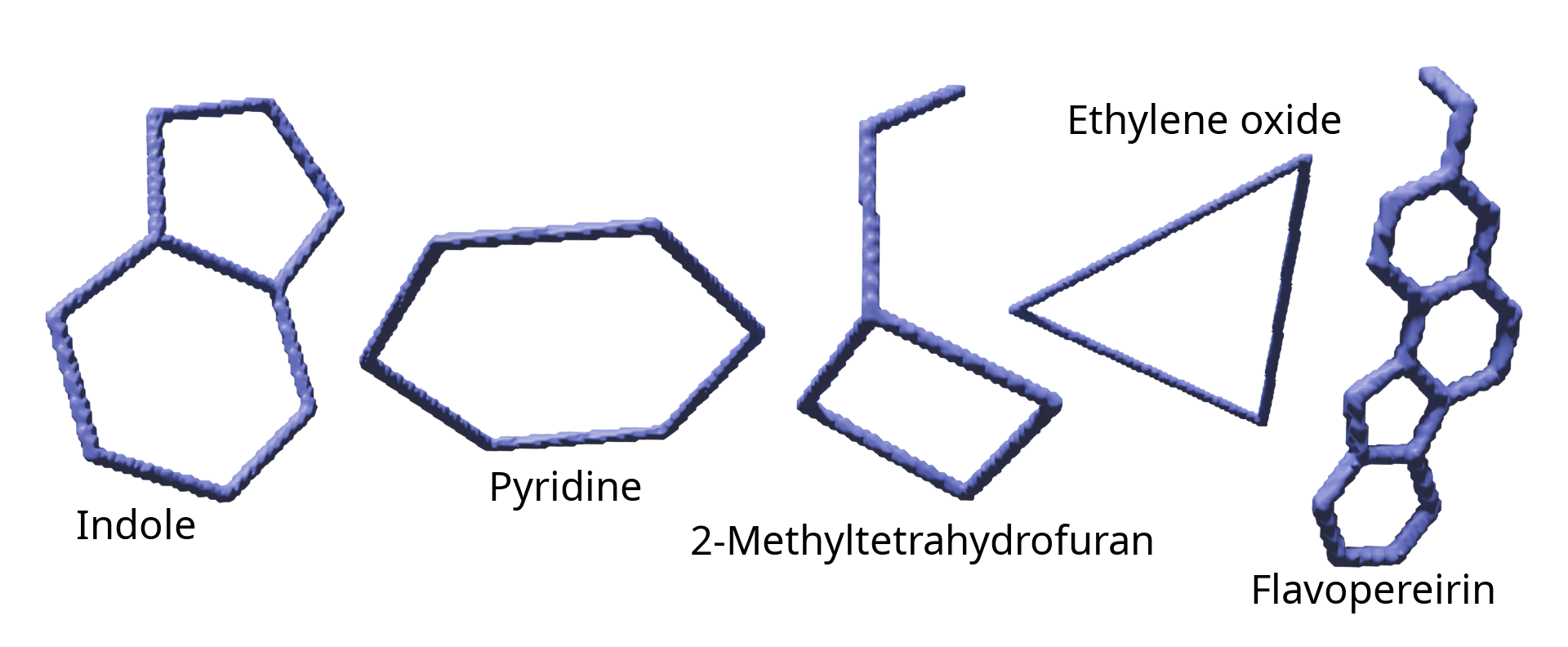}
    \caption{Examples of meshes generated from tubular neighborhoods around the following molecules: indole ($C_6H_4CCNH_3$), pyridine ($C_5H_5N$), 2-methyltetrahydrofuran ($C_5H_{10}O$), ethylene oxide ($C_2H_4O$), flavopereirin ($C_{17}H_{15}N_{2}$), respectively from left to right. QSAR molecular analysis \cite{Drugdesign} models and predicts chemical properties using descriptors such as ring count (topologically related to genus), molecular shape, area, and interatomic distances \cite{toporingcount, drugdiscovery}.}
    \label{fig:molecules}
\end{figure}

Our aim is for these datasets to be used in geometry processing and shape analysis tasks, particularly for deep learning architectures.
More broadly, we raise the question of how capable are neural networks in understanding topological features.
We hope they will be used as evaluation tools and benchmarks, to gauge the ability and performance of a learning system to recognize varying topological features.
We found that the certain popular methods, which perform very well on other tasks, fail to adequately identify and categorize our EuLearn dataset by genus (see \cref{tab:AttentionResults}).
Once a mesh is constructed, computing the Euler characteristic is accessible by counting the vertices, edges, and faces, and applying Euler's formula. 
While the Euler characteristic of a (well-defined, connected) mesh is readily computable from the combinatorial data of a mesh object, the point of our work is to understand how individual machine learning architectures achieve topological classification tasks, such as identifying the genus of an object.
The ability to recognize topologically distinct features is essential in scientific domains such as polymer analysis \cite{Panagiotou2019}, DNA entanglement \cite{Arsuaga2005}, creation of meshes in computer vision \cite{gkioxari2019mesh}, and protein folding research \cite{Qu2024}, to name a few.
Our curve thickening and database construction methods may enhance results in QSAR molecular analysis\cite{Drugdesign}, where topological features (e.g., ring count) \cite{toporingcount} and geometric descriptors (e.g., shape, area, interatomic distances) \cite{prediction} are linked to chemical properties (see Figure~\ref{fig:molecules}). 
Such analysis have been widely used in drug discovery and molecular property prediction \cite{drugdiscovery}.
 
Observe that the task of accurately generating a mesh from a point cloud presents several difficulties. 
Current methods can fail for objects that are not spheres or not given a geometric or topological prior.
We anticipate that our EuLearn databases will aid in the evaluation of these methods, as they eventually adapt to different surface topologies. 
\cref{tab:AttentionResults} shows the baseline performance of three emblematic methods 
(over 1925 training and 825 testing instances). 
We propose two key performance enhancing improvements. 
First, recognizing the importance of adjacency information from the original mesh, we introduce a sampling strategy (see \cref{sec:sampling}) that preserves this data while reducing the point count and processing time. 
Second, we adapt the PointNet and attention layers (see \cref{sec:GNNs}) to explicitly incorporate this adjacency data, simplifying their original operations.

 We experimented on the EuLearn dataset using classical PointNet \cite{qi2017pointnet}, PointNet++ \cite{qi2017pointnetdeephierarchicalfeature}, Fourier Neural Operators (FNOs) \cite{andradeneural}, Dynamic Graph Convolutional Neural Networks (DGCNNs) \cite{dgcnn} and a version of a 3D transformer \cite{lu2022transformers3dpointclouds}. As mentioned earlier, to improve performance, we need to incorporate structural information into these architectures (see \cref{sec:GNNs} for further details). 
 
To guarantee the statistical significance and generalizability of our experimental results, we conducted an ablation study and a sensitivity analysis (see Appendix~\ref{app:Significance}). These analyses revealed that incorporating adjacency information has a significant impact on performance (Appendix~\ref{sec:ablation}), and we confirmed the reproducibility of our results, as variations in hyperparameters did not substantially affect the overall outcomes (Appendix~\ref{sec:sensitivity}).

A dataset of uniformly distributed genera on surfaces is relevant for the pretraining of foundation models. While these surfaces may be abstract in nature, their inherent topological characteristics enable foundation models to better generalize high-genus instances observed in real-world applications, as explained above.

\subsection{Previous Dataset Work}

The FAUST (Fine Alignment Using Scan Texture) dataset is a collection of high-resolution 3D scans of human subjects in various poses. 
It was designed to evaluate algorithms for dense correspondence and shape analysis. 
FAUST is made up of 300 meshes representing 10 subjects in 30 different poses. 
The dataset provides ground-truth correspondences between shapes, making it useful for tasks such as surface registration, deformation learning, and geometric feature extraction \cite{Bogo:CVPR:2014}.  

The ShapeNet dataset is one of the most extensive repositories of 3D objects, offering a structured and richly annotated collection of models.
There are 2 subsets: ShapeNetCore, with about 51~300 models across 55 categories, and ShapeNetSem, which is a smaller, more densely annotated subset consisting of 12~000 models spread over a broader set of
270 categories.
The categories of both subsets include objects such as vehicles, tools, everyday household items, and others.
ShapeNet’s hierarchical labeling system allows researchers to perform tasks ranging from coarse-grained classification to fine-grained segmentation. Its diversity in geometry and semantics has established ShapeNet as a benchmark for advancing 3D computer vision, enabling studies in object recognition, segmentation, and generative modeling \cite{shapenet2015}.

MANTRA \cite{ballester2024mantramanifoldtriangulationsassemblage} is a new dataset of abstract surface triangulations and three-dimensional manifolds. 
Using this dataset, three topological classification tasks were evaluated on models based on simplicial complexes and graph-based models. 
The results showed that while neural networks based on simplicial complexes outperform graph-based networks, they also face challenges when handling topological classification tasks.
However, these abstract objects are not embedded.

{Other relevant datasets include ABC \cite{Koch_2019_CVPR}, Thingi10K \cite{zhou2016thingi10kdataset100003dprinting}, and SimJEB \cite{Whalen_2021}, which we further analyze in \cref{App:ABC}.}

The structure of this work is as follows.
\cref{sec:HigherGenusSurfaces} describes the dataset generation process, while \cref{Experiments} details the design and implementation of the neural network architectures. The main conclusions are presented in \cref{Conclusions}.
We include ten appendices for supplementary material:     \cref{App:Proofs} provides the proofs of our three theorems.
\cref{App:EulearnDataset} offers additional information about the EuLearn dataset.
\cref{App:PreviousArchitectures} and \cref{app:LayersDescription} detail the architectures used in our experiments and the specifics of their layer designs.
\cref{App:Performance} contains tables summarizing experimental results.
\cref{appendix:sampling} presents a graphical comparison between the sampling process and the original point cloud.
\cref{appendix:ShapeNet} discusses challenges faced by large-scale datasets such as ShapeNet in ensuring robustness and reliability.
\cref{App:ABC} shows the genus distribution across the analyzed datasets.
\cref{app:Significance} provides a statistical analysis of our experimental outcomes.
Finally, \cref{App:Datasheet} includes a datasheet for the EuLearn dataset, following the guidelines proposed by Gebru et al.~\cite{10.1145/3458723}.

\section{Construction of our EuLearn Dataset}
\label{sec:HigherGenusSurfaces}

We will now explain our construction of surfaces with varying genera. To achieve this, we proceeded by thickening closed, self-intersecting curves into precisely controlled tubular neighborhoods. 
For example, thickening a unit circle $S\subset {\mathbf{R}^2}\subset {\mathbf{R}^3}$, we may construct the surface of a torus, which has one handle.
Controlling the number of self-intersections in parametrized families of \textbf{singular knots} (knots with permanent self-intersections), we will construct surfaces of higher genus.
The knots we utilize are drawn from two primary families, each defined by a set of parameters. 
Each self-intersection of a singular knot curve adds one handle to the resulting surface, so $n$ self-intersections result in a surface with genus $n+1$.

We can summarize the construction process with the following four key steps: (1) selection of curve parameters, (2) curve thickening, (3) surface discretization and (4) smoothing. In the following paragraphs, we explain step 1, while the other steps are detailed in the  \cref{App:EulearnDataset}.

\subsection{Selection of Curve Parameters}
\label{subsec:selection}

The two curve families that our database relies on are subsets of a larger, overarching family known as Fourier Knots. Mathematically, this family consists of 1-dimensional manifolds embedded in 3-dimensional space (knots).
They can be parameterized by a function $\mathcal{K}:S^1 \to \mathbf{R}^3$, where $S^1$ is the unit circle.

\begin{definition}
    A \textbf{Fourier knot} is a knot whose coordinates are parameterized by \cite{kauffman1997fourier,lamm2012fourier}:
    \begin{align*}
         x(t) = \sum_{l=1}^i A_{x,k}\cos(n_{x,l} t + \phi_{x,l}), \quad
         y(t) = \sum_{l=1}^j A_{y,k}\cos(n_{y,l} t + \phi_{y,l}), \quad
         z(t) = \sum_{l=1}^k A_{z,k}\cos(n_{z,l} t + \phi_{z,l}).
    \end{align*}
    Here, $n_{x,l}, n_{y,l}, n_{z,l} \in \mathbf{Z}$, $\phi_{x,l}, \phi_{y,l}, \phi_{z,l} \in \mathbf{R}$, and $0\leq t \leq 2\pi$.
    Every type of knot can be represented by a Fourier knot \cite{kauffman1997fourier,boocher2009sampling}.
\end{definition}

From this family, we define the two specific families, Lissajous and Fibonacci Knots, on which we based our dataset.
\begin{definition}
    A \textbf{Lissajous knot} $\mathcal{K}(t)$ is parameterized by \cite{Bogle,kauffman1997fourier}:
    \vspace{-2pt}
    \begin{align*}\label{knot}
        x(t) = \cos(n_x t + \phi_x), \qquad
        y(t) = \cos(n_y t + \phi_y), \qquad 
        z(t) = \cos(n_z t + \phi_z), \nonumber
    \end{align*}
    where now $n_x<n_y<n_z \in \mathbf{Z}$ are relatively prime to avoid self-intersections, and also the phases must NOT be some of the following:
    \begin{equation*}
        \phi_x = \dfrac{m}{n_z}\pi,\qquad 
        \phi_y = \dfrac{m}{n_z}\pi,\qquad
        \phi_x = \frac{n_x}{n_y}\phi_y + \dfrac{m}{n_y}\pi,
    \end{equation*}
    where $m \in \mathbf{Z}$ is such that $0 < m < n_z$ and, without loss of generality, $\phi_z = 0$ \cite{lamm97}.
\end{definition}

Additionally, we used singular Fibonacci knots, slightly modifying the usual parameterization \citep{kauffman1997fourier}.
The frequencies of these knots follow the Fibonacci sequence, starting with the $n$-th term of the sequence, so $F_1(t)$ has $n_x = 1$, $n_y = 1$, $n_z = 2$, then $F_2(t)$ has $n_x = 1$, $n_y = 2$, $n_z = 3$, then $F_3(t)$ has $n_x = 2$, $n_y = 3$, $n_z = 5$, and so on.

\begin{definition}
A \textbf{Fibonacci knot} $F_n(t)$ is parameterized by: 
\begin{align*}
    x(t)  = \cos(n_x t + \phi_x), \quad
    y(t) = \cos(n_y t + \phi_y),  \quad
    z(t)  = \dfrac{1}{2}\cos(n_z t + \phi_z) + \dfrac{1}{2}\sin(n_y t + \phi_y)
\end{align*}
where $n_x,n_y,n_z$ follow the Fibonacci sequence starting with the $n$-th term of the sequence.
\end{definition}

For instance, $F_3(t)$ with $\phi_x = 0$, $\phi_y = \pi/2$, $\phi_z = 0$ has 6 self-intersections which produces a genus 7 surface, and $F_4(t)$ with $\phi_x = 0$, $\phi_y = 1/2$, $\phi_z = 1/2$ has 1 self-intersection which produces a genus 2 surface.

Initially, we manually curated the knot parameters and identified the corresponding number of self-intersections for each knot using the symmetry properties of the parametrization. 
To this end we generate sets of six parameters $(n_x, n_y, n_z, \phi_x, \phi_y, \phi_z)$ and determine the number of self-intersections on the resulting curve. 
To achieve this, we generated all combinations of integer triads within the range $[0,12]$ and performed the Cartesian product with triads of phase angles $(\phi_x, \phi_y, \phi_z)$. 
In these phase triads, only one phase was nonzero and chosen from the set ${\pi/2,\pi/3,\pi/5,\pi/7}$.
 We then automated the discovery of self-intersection points for a given knot, producing a balanced amount of number of singular knots per genus.
Given a singular knot and its combination of six parameters, this automatization starts by creating a polygonal representation of the curve. 
The chosen discrete representation consisted of the minimum number of uniformly distributed points that approximate the curve length within an error lower than 2.5$\%$.

\begin{algorithm}
    \small
    \caption{Detection of self-intersections in curves}\label{alg:self-intersections}
    \begin{algorithmic}[1]
    \STATE $\textbf{Procedure:}$ Find self-intersections of a closed curve $\mathcal{K}$.
    \STATE $n \leftarrow$ Number of points approximating the curve.
    \STATE $\mathcal{K}\leftarrow$ Curve parameterization.
    \STATE $\mathcal{I}=\{\,\}\leftarrow$ Empty set to store self-intersection points.
    \STATE $\mathcal{S}=\{ \mbox{seg}_i\}_{i=0}^{n-1}\leftarrow$ Set of polygonal segments for the curve.
    \FOR{$(\mbox{seg}_i,\mbox{seg}_j)\in \mathcal{S}\times\mathcal{S},\, i<j$}{}
      {\STATE Let $K_i$ and $K_j$ be the arcs with endpoints matching those of $\mbox{seg}_i$ and $\mbox{seg}_j$, respectively.}
      {\STATE Reparameterize $K_i$ and $K_j$ on the domain $[0,1]$.}
      {\STATE Use \emph{Differential Evolution Algorithm} to minimize the distance $d(K_i,K_j)$ and denote the minimum as $(\xi,\zeta)$.}
      \IF{$|\mathcal{K}_i(\xi)-\mathcal{K}_j(\zeta)|<10^{-6}$}
      \STATE $\mathcal{I}=\mathcal{I}\cup{\mathcal{K}(\xi)}$
      \ENDIF
    \ENDFOR
    \STATE number of intersections $=\,\#\mathcal{I}$ 
    \end{algorithmic}
\end{algorithm}

To detect self-intersections, we computed the distances between segment pairs in each polygonal curve using the Differential Evolution algorithm \cite{differentialevolution}. 
Self-intersections were detected if that distance was below a threshold of $10^{-6}$ and its coordinates had not been previously computed for another segment pair. 
Algorithm \ref{alg:self-intersections} summarizes this procedure.

\section{Experiments}
\label{Experiments}

We worked with a subset of the EuLearn dataset consisting of 2750 surfaces, spanning 11 distinct genera (from genus 0 to genus 10).
Each genus class had 250 surfaces, generated using the aforementioned method. 
This dataset is conformed by the original surfaces and a smooth version of them that are more visually appealing. The smooth surfaces contribute to the regularization of the models by serving as a data augmentation strategy, enhancing the robustness and generalization capabilities of the models.
 
 In the following sections, we explain the experiments we conducted to classify surfaces according to genus.
Initially, we introduce the sampling of points from a surface to ensure that each surface contains a representative set of points that preserve the connectivity of the original mesh. 
Next, we outline the general architecture employed in the experiments, and including the description of the specific layers used. 
Finally, we discuss the results obtained.

\subsection{Sampling} \label{sec:sampling}

Starting from our mesh dataset, we construct a dataset of point clouds with a reduced number of points.
To this end, we developed a sampling algorithm that we describe here.
The dataset consists of point clouds of varying sizes, with the size of each point cloud influenced by factors such as genus, generation parameters, and voxel size. The variable sizes and potentially large number of points in some point clouds posed challenges for processing the dataset using neural models, primarily due to two reasons: (i) certain surfaces required substantial memory capacities to accommodate their size, and (ii) the different sizes do not allow performing uniform batches for training purposes.
One of the first challenges in processing the surface dataset was to reduce the size of the surfaces while retaining their essential information. To this end, we prioritized preserving the adjacency information encoded in the triangulation of the surfaces, with the aim of leveraging neural models that exploit neighboring relationships in the surface. To achieve this, we employed a sampling procedure designed to reduce the number of points in the surface while maintaining the connectivity information of the original mesh, thereby facilitating the application of graph-based neural networks.

The dataset is composed of a set of point clouds and a graph $G=(V,E)$ that represents the triangulation of the surface, where each node $ v\in V$ is associated with a point in the point cloud and the edges $e \in E$ encode the adjacency relations between points, thus defining the polygonal structure of the surface.
The sampling procedure consists of iteratively traversing the nodes of $G$, then finding the neighboring nodes of each examined node.
A subset of these neighboring nodes is sampled, while the remaining nodes are removed from the surface. To preserve the topological structure of the surface, the original node is reconnected to the immediate neighbors of the deleted nodes, thereby maintaining the connectivity of the graph. This iterative process is repeated until the desired number of sampling points is reached, yielding a reduced representation of the surface that retains its essential topological properties. 
This procedure may be seen as a fine grained coarsening of the graph itself.
The pseudocode for this procedure is presented in Algorithm~\ref{alg:sampling}.

\begin{algorithm}
    \small
    \caption{ Graph Sampling (GS)}
    \label{alg:sampling}
    \begin{algorithmic}[1]
        \STATE $\textbf{Procedure:}$ Sampling $X$, $G = (V,E)$, $sample\_size$
            \STATE $N \leftarrow \textsc{Sorted}(V)$
            \STATE $k \leftarrow 0$ 
            \WHILE{$k < \text{sample\_size}$}
                \STATE $actual\_node \leftarrow \textsc{Top}(N)$
                \STATE $neighbors \leftarrow \textsc{GetNeighbors}\big(actual\_node\big)$
                \STATE $sampled\_neighbors \leftarrow \textsc{Sample}(neighbors)$
                \STATE Remove from $N$, $X$ and $V$ the not sampled nodes.
                \STATE Reconnect points in $E$
                \STATE $k \leftarrow k + |sampled\_neighbors| + 1$
            \ENDWHILE
        \STATE $\textbf{end Procedure}$
    \end{algorithmic}
\end{algorithm}

First we sort the vertices of $G$. 
The order given by this sorting determines how each node is traversed, functioning as the positioning for the case of Transformer architectures. 
The sorting itself is obtained applying the \textsc{Sorted} method. 
This ordering was achieved via the construction of a spanning tree, but other methods that enable the imposition of a node ordering (including random ordering) could in principle be applied. 
The sorted nodes are then stored in a stack data structure $N$, allowing for selection of the top node as the current node. 
We use a Depth-first Search (DFS) method to select the top nodes. 
To examine the neighbors of the current node, the method \textsc{GetNeighbors} utilizes the adjacency information encoded in the original graph, while ignoring previously deleted points. 
The sampling procedure (the method \textsc{Sample}) is performed through a randomized selection, where a subset of the neighbors is selected. Inspired by PointNett++ \cite{qi2017pointnetdeephierarchicalfeature}, we sample points within a ball of radius $r$, including only those points that are neighbors nodes in the original mesh.
Finally, the method removes the unselected nodes and performs a reconnection of the new graph.
When the size of sample is reached, the method halts.

In our experiments, we extracted 3000 points from the mesh of each surface.
This reduction process involved an augmentation of the degree of each vertex representing a sampled point. Initially, the average degree for each mesh was 6, but after sampling, the average degree increased to 37. 
This increase is attributed to the reconnection performed in the sampling algorithm, which reconnected the neighbors of deleted points to adjacent points. This reconnection not only made more efficient use of memory but also resulted in smaller and more densely populated adjacency matrices. 
The sampling process enabled us to manage the dataset more effectively, as each data instance now has a uniform size. Additionally, the memory required to load the point clouds and their adjacency information was significantly reduced.

\subsection{Deep Learning Models} \label{sec:GNNs}

In our experiments, we evaluated the performance of several graph neural models in classifying a dataset of meshes based on their genus. 
The objective was to assess the models' ability to classify the data according to its Euler characteristic. While there are existing algorithms to determine the Euler characteristic \cite{hacquard2024euler}, our purpose is to assess the performance of neural models using minimal topological information. We evaluated three models: Fourier Neural Operators (FNO), PointNet, and Transformers, along with graph-informed variations of PointNet and Transformers. Each model was trained and evaluated using a consistent schedule, with standardized hyperparameters and training protocols to ensure a fair comparison. The dataset comprised 1925 training examples (175 per class) and 825 test examples (75 per class). 

%
%
The layers used in the architecture of the deep neural network are described in detail in Appendix~\ref{app:LayersDescription}. The architecture comprises the following layers:
\vspace{-5pt}
\begin{enumerate}
    \item A graphical input layer that processes the data points (a point cloud) along with the adjacency matrix. %
    The selected models were Fourier Neural Operator (FNO), DGCNN, PointNet++, PointNet, a 3D Transformer, and a graph-informed variant of the latter two models.
    \item Following the representation of the input data by the graph layers, a batch normalization layer \cite{ioffe2015batchnormalizationacceleratingdeep} is applied to mitigate internal covariate shift. Subsequently, a dropout layer with a probability of 0.3 is utilized.
    \item The subsequent layers involve global max pooling and flattening to transform the representations of the points into a single vector. In each case, the resulting vector is of dimension 512, with the exception of the FNO, where the vector dimension is 100. A dropout layer with a probability of 0.3 is then applied.
    \item For classification, we used a multi-layer perceptron (MLP) with a hidden layer and a ReLU activation function, $\psi(h) = W\ \text{ReLU}(W_1h+b_1) + b$.
    Then, the final classification is achieved through a softmax activation function. 
\end{enumerate}

The main variants on our experiments where the graph layers utilized. 
We focused on three models, representative of the surface classification paradigms: Fourier Neural Operators (FNO), PointNet and a 3D transformer. 
The FNO implementation is based in \citealp{andradeneural} (see \cref{sec:FNO}). The implementations of PointNet and attention layers are based on \citealp{qi2017pointnet} (see \cref{app:pointnet}) and \citealp{zhao2021pointtransformer} (see \cref{app:Attention}). 
It is worth noting that these layers do not require adjacency information $A$, relying only on the point cloud or scalar field information (in the case of FNO).

We introduce adapted versions of PointNet and a transformer. 
Both using the sampling procedure described in \cref{alg:sampling} and incorporate the adjacency matrix $A$ for neighborhood aggregation.
The graph sampled (GS) PointNet apply the aggregation function on a PointNet layer over the neighbors in the sampled mesh. 
Formally:
$h _i = \max_{i\in \mathcal{N}_i} \phi\big(\psi(x_i), p_i - p_j \big)$. \cref{fig:Layers} b) shows the layer diagram.
In contrast with classical PointNet implementations the position vectors are the original input data points in every layer, and not the previous layer representations. 
(see \cref{sec:PointNet}).
Finally, the GS attention layer applies the aggregation sum over the neighbors in the sampled mesh as well. 
Then it applies the softmax estimation over the direct dot product of the points (in contrast with traditional implementations that apply a linear or affine transformation to the points previous to the dot product) \cref{fig:Layers} shows this layer's process, described formally as $h_i = \sum_{j\in \mathcal{N}_i} \alpha(x_i, x_j) \psi_v(x_j)$. 
Here $\alpha(x_i, x_j)$ is a softmax activation over the dot product, and $\psi_v(x_j) = xV$, with $V$ a matrix of learnable parameters. 
It is worth noting that GS attention does not apply multi-head mechanisms, but sequentially processes the attention layers (see \cref{sec:attention}).

The main difference between each experiment lies in the variation of the graph neural network layers. 
For model training, we optimized the cross-entropy loss function. 
The chosen optimizer was the Noam optimizer \cite{vaswani2017attention}, which is known for its adaptive learning rate schedule, particularly beneficial in the context of attention-based models. Additionally, we incorporated weight decay regularization. 
Each model was trained over 10,000 epochs with identical hyperparameters (see \cref{app:hyperparemeter}).
These hyperparameters optimized convergence and stability during the training process.

%

\subsection{Results} \label{sec:results}

We show our experimental results across the six different models in \cref{tab:AttentionResults}. 
We evaluated the models using precision, recall, $F_1$ score, and accuracy metrics. 
The average over the classes is the precision, recall, and $F_1$ scores reported. The macro and weighted averages are identical since the test dataset is balanced (75 examples per class).
Our graph sampled attention transformer model achieved the best performance, while the FNO and classic attention transformer models exhibit the lowest results. 
Models that rely solely on point cloud information, such as the PointNet model, which does not incorporate adjacency information, achieve low accuracy (0.49) due to the lack of relational information between mesh points. 
In contrast, models that integrate adjacency information achieve high accuracies and $F_1$ scores close to 0.8, highlighting the importance of structural information provided by the adjacency matrix for genus classification. 
This structural information is crucial for enhancing model performance, and its absence significantly reduces classification accuracy. 
The modifications, using our graph sampling algorithm, not only improved performance but also required fewer computational resources due to the sampling and simplification of the computations.
\subsection{Hardware Used}

 When constructing the database and conducting our experiments, our primary server has 4 nodes, each equipped with 2 Intel Xeon Gold 5122 processors running at 3.6 GHz, with 4 cores each. 
 Each node is supported by 64GB of RAM and 2 NVIDIA Tesla V100 GPUs. 
 We also utilized a secondary server equipped with a GeForce GTX 980 video card, with 4GB of dedicated VRAM, and 32GB of RAM.

\section{Conclusions}
\label{Conclusions}
In this paper, we introduce EuLearn, a pioneering dataset of 3D surfaces. It is the first of its kind to feature varying topological types and uniformly distributed genera, marking a significant advancement in the available datasets. 
We conducted experiments on emblematic architectures, representing several methods of approaching 3D data.
We found significant improvements in model performance by employing an adjacency-preserving sampling method, which we also developed here.
Moreover, our strategy reduces the processing time of classical methods. 

Our findings underscore the practical need for new deep learning architectures that can identify specific topological information, such as higher genera in point clouds. This has significant implications for a wide range of applications in 3D data analysis. 
To contribute to this effort, we provide a methodology for creating similar datasets. 

We hope that our EuLearn datasets will be a transformative resource for testing and benchmarking in related research communities, inspiring new directions and breakthroughs in the field.
They could also be used as a pretraining step to guarantee that a proposed deep learning architecture understands topological information, learning the uniformly varying Euler characteristics in our EuLearn dataset. 

We highlighted a topological blind spot in the available deep learning tools and methods, caused by the lack of topological diversity in datasets. 
We contributed to understanding this problem and ameliorating it by creating a carefully crafted and curated dataset with uniform topological diversity.
After conducting experiments showing the limitations of classical deep learning architectures we proposed a sampling strategy that helps them improve performance. 

We invite DL practitioners to experiment with their own designs, test their performance on the genus classification task using our EuLearn dataset, and incorporate our sampling strategy if needed to improve results.
We believe that our EuLearn datasets will serve as a valuable resource for testing and benchmarking in allied communities, potentially reshaping the future of deep learning in 3D data analysis.

\section{Acknowledgments}
The authors thank DGTIC-UNAM for access to the Miztli HPC resources, grant LANCAD-UNAM-DGTIC-430.
RF thanks CONAHCyT for a graduate fellowship. 
This work was supported by Universidad Nacional Autónoma de México Postdoctoral
Program (POSDOC) for author EIVR, who also acknowledges the postdoctoral fellowship received during the production of this work. VM acknowledges the support from project PAPIIT TA100924 ``Investigación de sesgos inductivos en aprendizaje profundo y sus aplicaciones''.

\bibliography{bibliography}
\bibliographystyle{plain}

\newpage

\appendix
%

\section{Proofs}
\label{App:Proofs}

\subsection{Proof of \cref{thm:FAUST-S2}}

\begin{proof}
    By computation. Each file from FAUST is a watertight mesh, representing a compact orientable and triangulated manifold of dimension $2$.
    So its Euler characteristic can be directly computed, and from this we conclude that every file is homeomorphic to a $2$-dimensional sphere.
\end{proof}

\subsection{Proof of \cref{thm:not-unif-dist-data}}

\begin{proof}
    Many of the files in these datasets are either not water-tight, or are not even a single mesh, as the objects they represent are constructed as scenes with multiple meshes.
    When the object is represented by a single mesh, and it is possible to compute the Euler characteristic from this mesh, the computations reveal that the Euler characteristic is not uniformly distributed. 
    For example, the ABC and Thigie10K datasets concentrate highly around 0, 
    while the SimJEB dataset concentrates on 6.
    See Figures \ref{fig:ABC}, \ref{fig:Thingi10K}, and \ref{fig:SimJEB}, which illustrate how these distributions skew away from uniformity.  
    In any case, the result follows by computation.
\end{proof}

\subsection{Proof of \cref{thm:EuLearn-unif-dist}}

\begin{proof}
    By construction. 
    We explain in the main text and with further details in the appendices how our EuLearn dataset is constructed so that each object is watertight, is a $2$-dimensional compact connected triangulated manifold, and is a single mesh. 
    With these properties, we may compute their Euler characteristic, and in doing so we may verify that, by the detailed construction explained in this paper, the Euler characteristic is uniformly distributed.
\end{proof}

\section{EuLearn Dataset}
\label{App:EulearnDataset}

\begin{figure}[H]
    \renewcommand{\thesubfigure}{} 
    \centering
    \subfigure[$g=0$]{\includegraphics[width=0.2\linewidth]{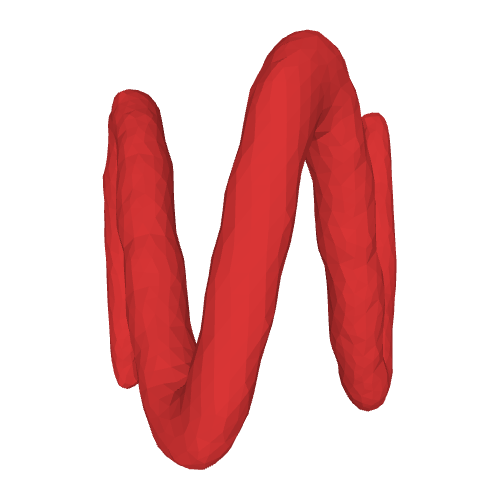}}
    \subfigure[$g=0$]{\includegraphics[width=0.2\linewidth]{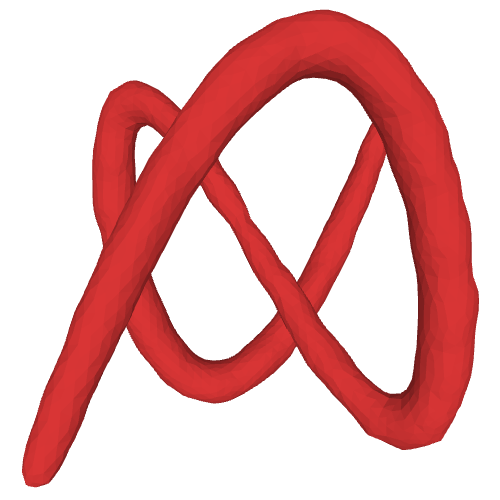}}
    \subfigure[$g=0$]{\includegraphics[width=0.2\linewidth]{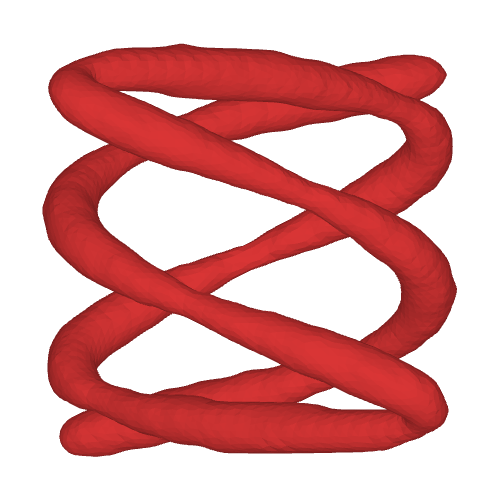}}
    \subfigure[$g=0$]{\includegraphics[width=0.2\linewidth]{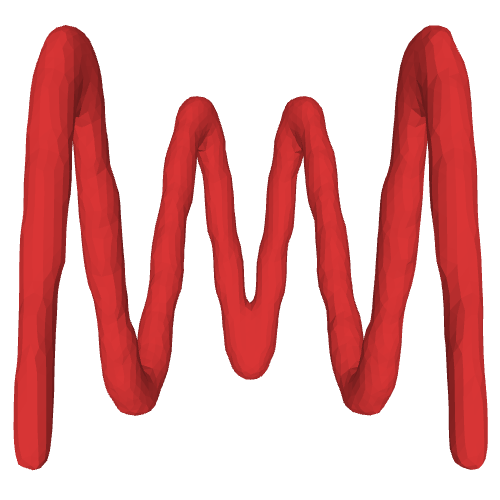}}

    \subfigure[$g=1$]{\includegraphics[width=0.2\linewidth]{g1_eight_f4_9785E3.png}}
    \subfigure[$g=1$]{\includegraphics[width=0.2\linewidth]{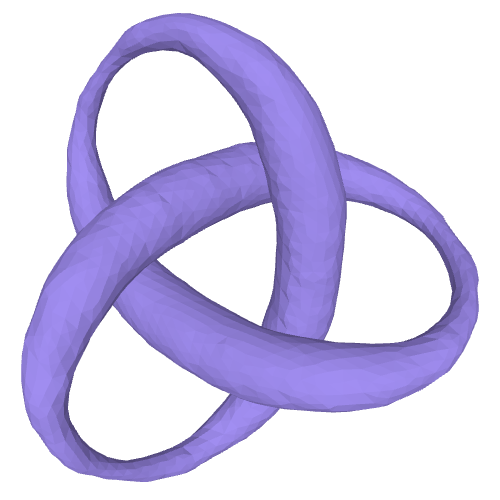}}
    \subfigure[$g=1$]{\includegraphics[width=0.2\linewidth]{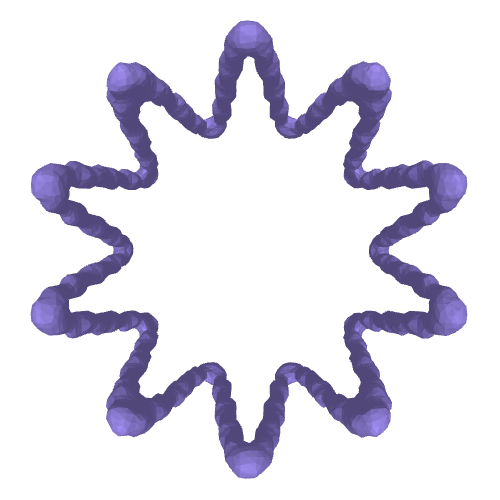}}
    \subfigure[$g=1$]{\includegraphics[width=0.2\linewidth]{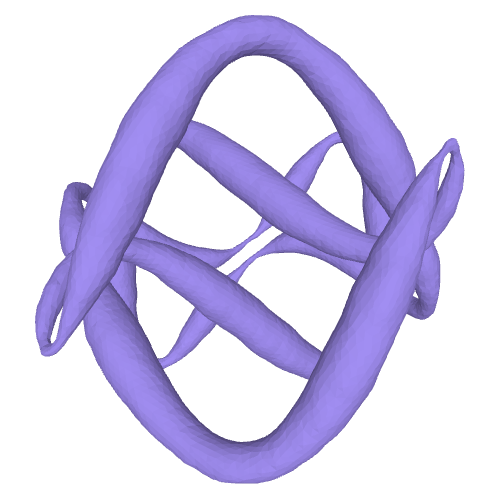}}
    
    \subfigure[$g=2$]{\includegraphics[width=0.2\linewidth]{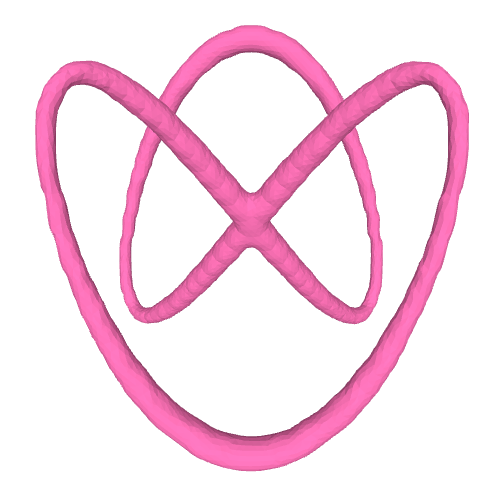}}
    \subfigure[$g=2$]{\includegraphics[width=0.2\linewidth]{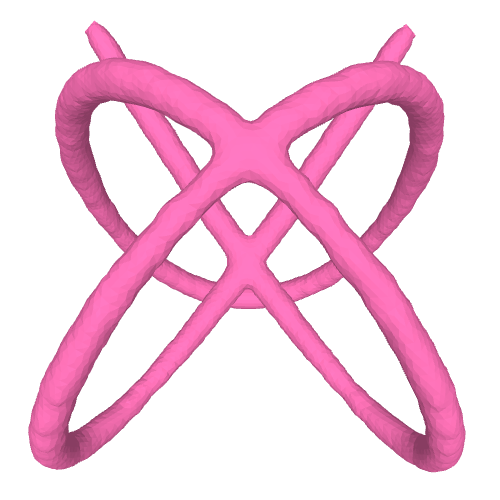}}
    \subfigure[$g=2$]{\includegraphics[width=0.2\linewidth]{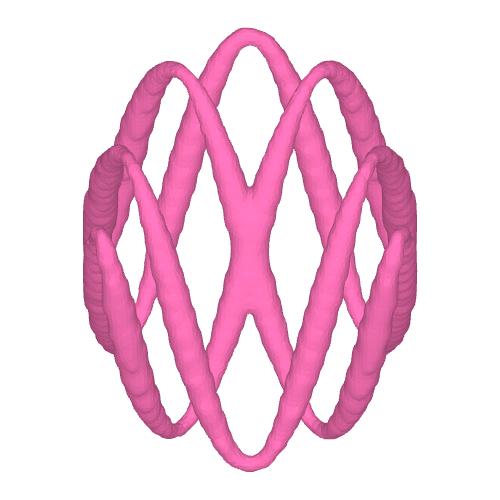}}
    \subfigure[$g=2$]{\includegraphics[width=0.2\linewidth]{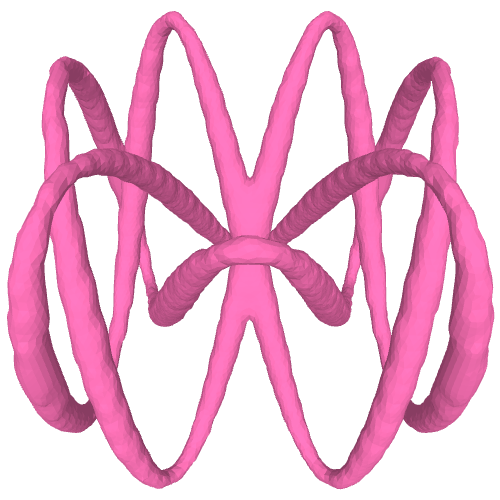}}
    
    \subfigure[$g=3$]{\includegraphics[width=0.2\linewidth]{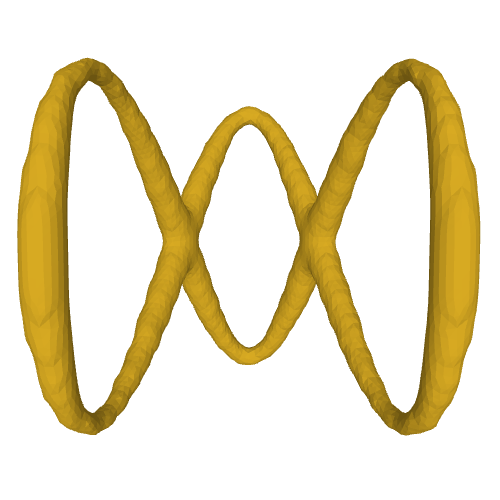}}
    \subfigure[$g=3$]{\includegraphics[width=0.2\linewidth]{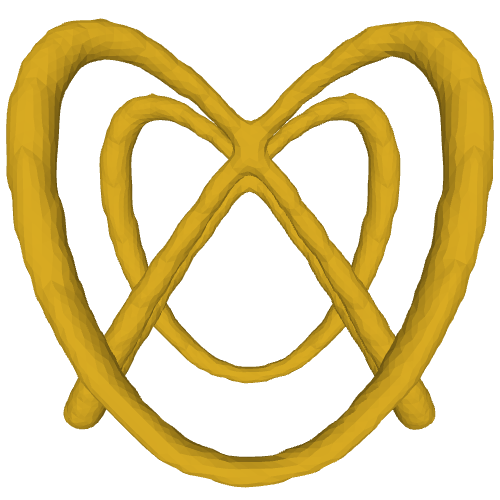}}
    \subfigure[$g=3$]{\includegraphics[width=0.2\linewidth]{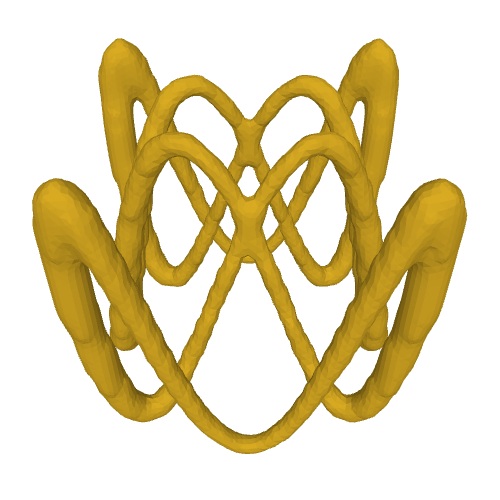}}
    \subfigure[$g=3$]{\includegraphics[width=0.2\linewidth]{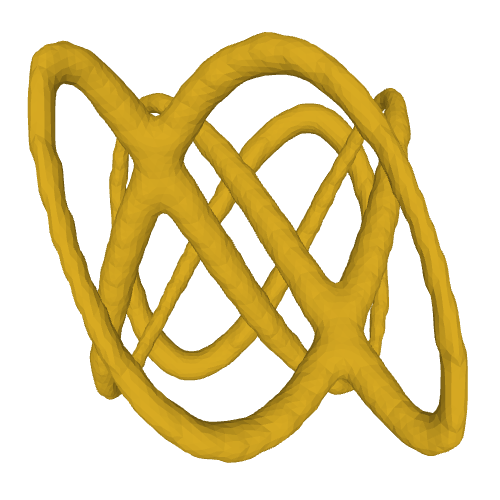}}

    \subfigure[$g=4$]{\includegraphics[width=0.2\linewidth]{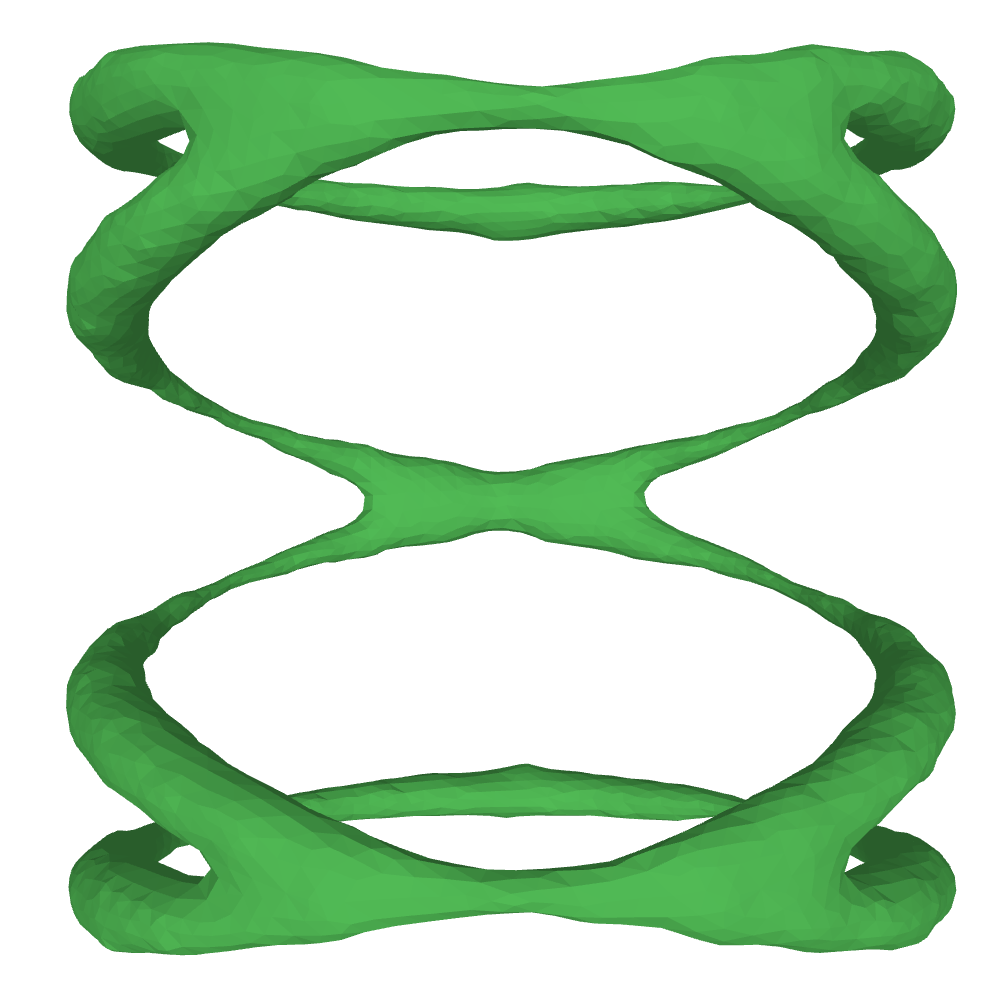}}
    \subfigure[$g=4$]{\includegraphics[width=0.2\linewidth]{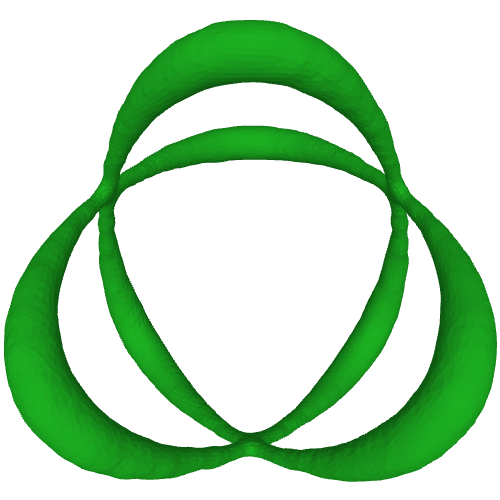}}
    \subfigure[$g=4$]{\includegraphics[width=0.2\linewidth]{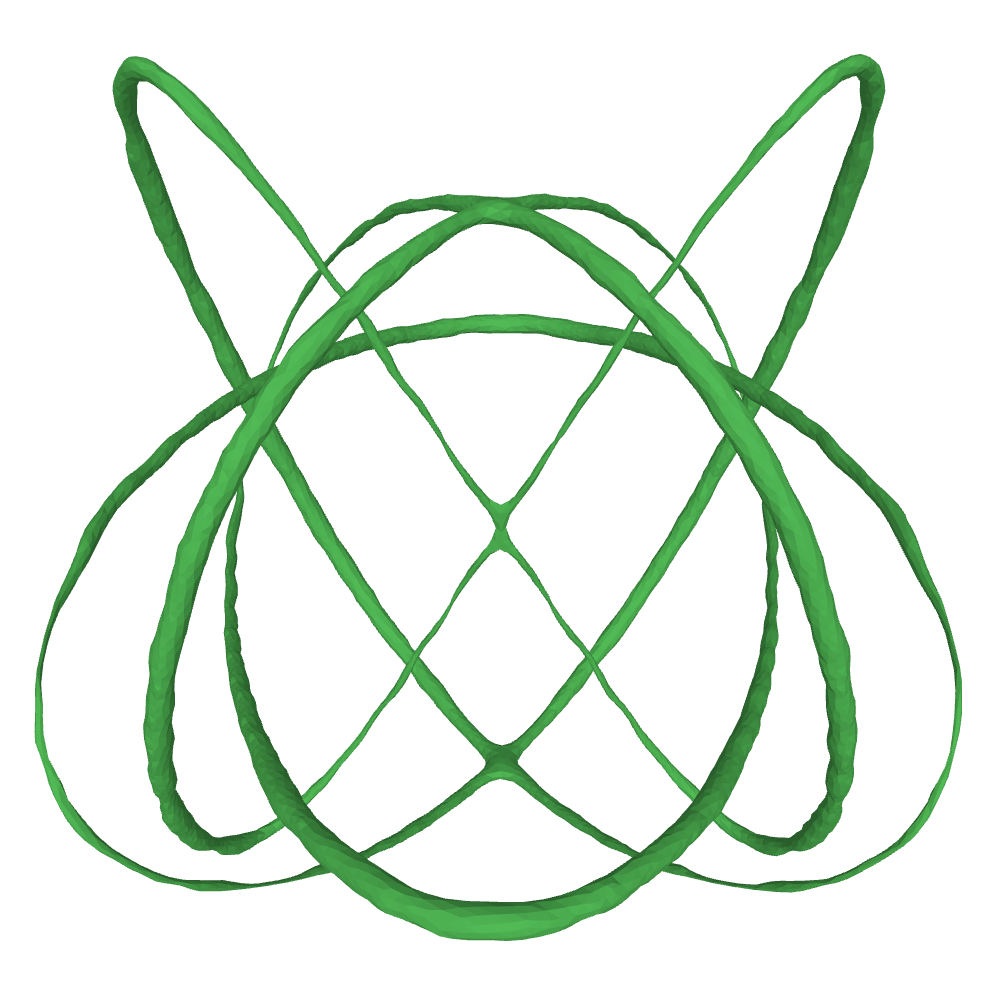}}
    \subfigure[$g=4$]{\includegraphics[width=0.2\linewidth]{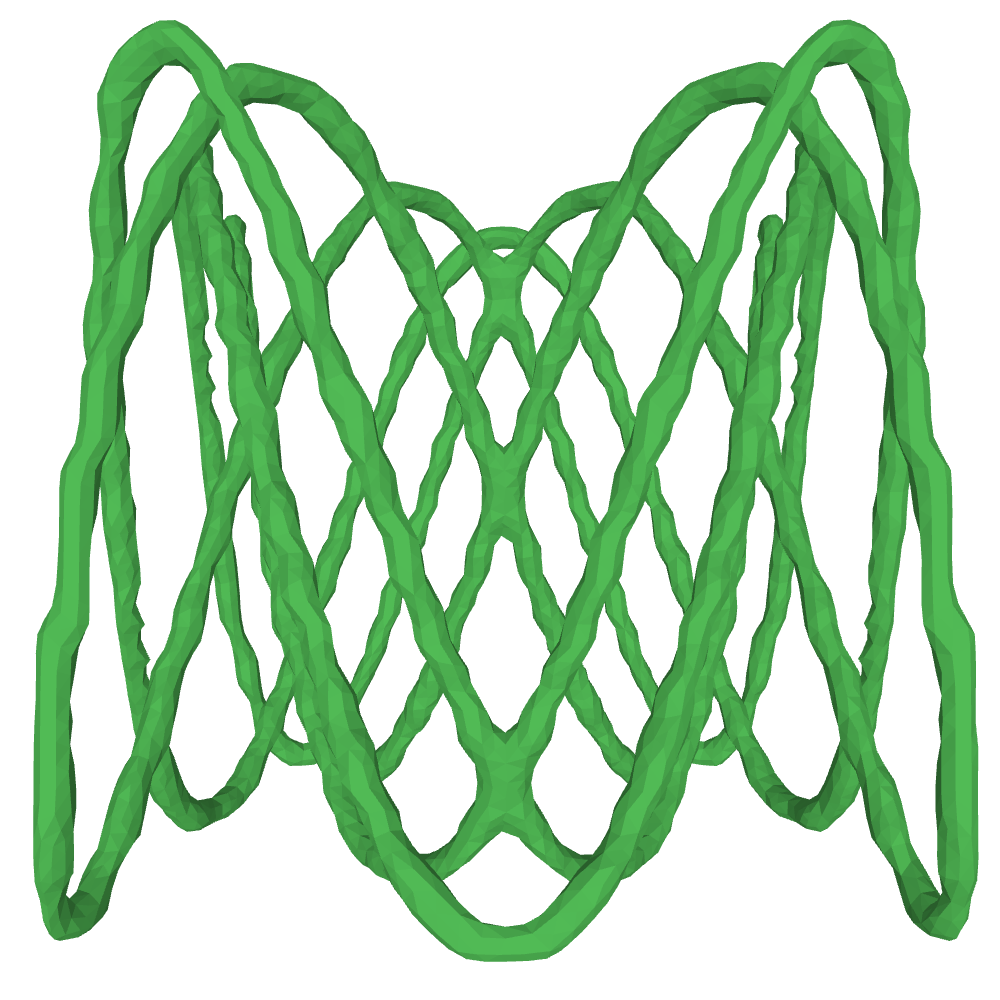}}
    
    \subfigure[$g=5$]{\includegraphics[width=0.2\linewidth]{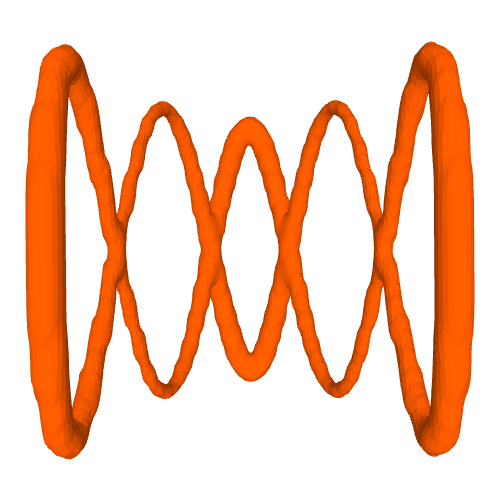}}
    \subfigure[$g=5$]{\includegraphics[width=0.2\linewidth]{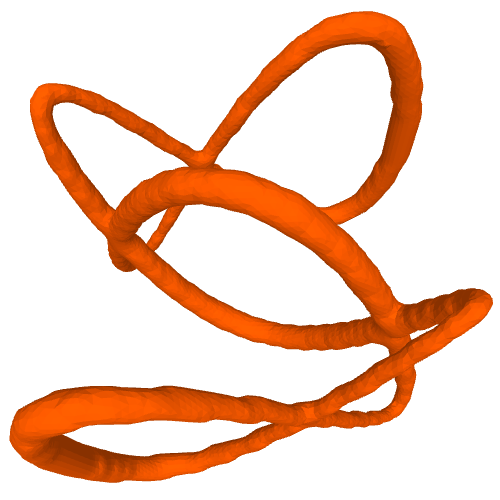}}
    \subfigure[$g=5$]{\includegraphics[width=0.2\linewidth]{g5_156_pi200_FF5B00.png}}
    \subfigure[$g=5$]{\includegraphics[width=0.2\linewidth]{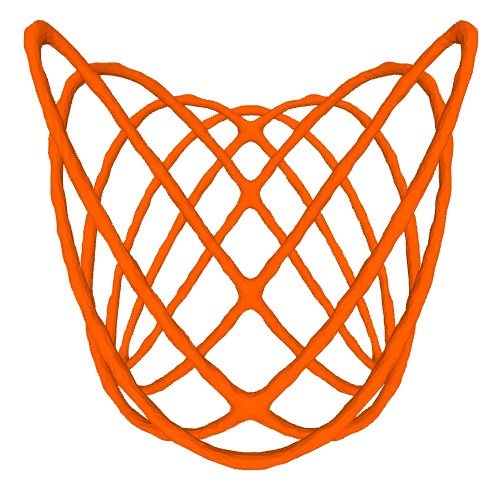}}
    
    \caption{A sample of 4 examples for each genus from 0 to 5 from the EuLearn dataset.}
    \label{fig:Eulearn_dataset1}
\end{figure}

\begin{figure}[H]
    \renewcommand{\thesubfigure}{} 
    \centering
    \subfigure[$g=6$]{\includegraphics[width=0.2\linewidth,trim=0 -5mm 0 0,clip]{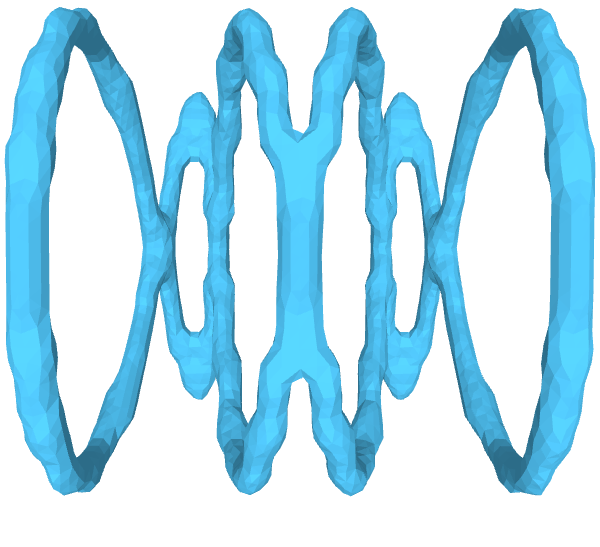}}
    \subfigure[$g=6$]{\includegraphics[width=0.2\linewidth,trim=0 2cm 0 2cm,clip]{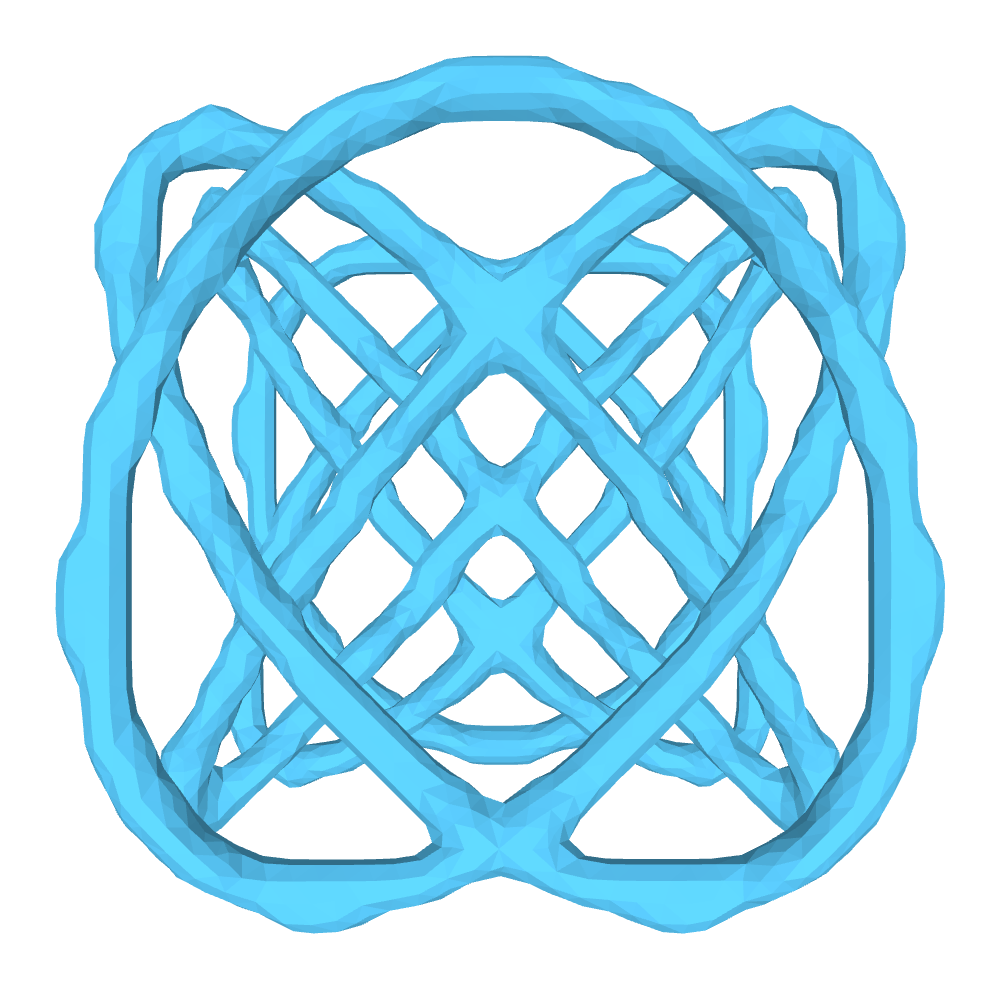}}
    \subfigure[$g=6$]{\includegraphics[width=0.2\linewidth,trim=0 16mm 0 2cm,clip]{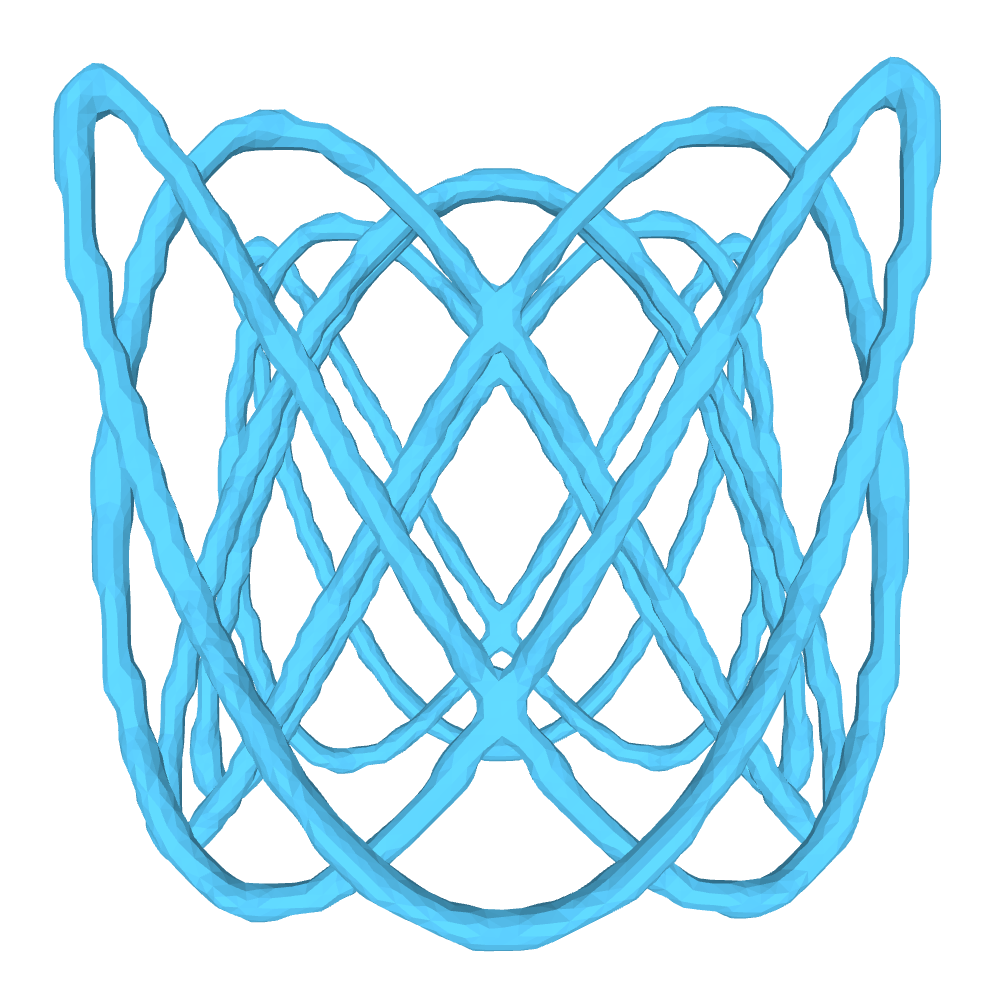}}
    \subfigure[$g=6$]{\includegraphics[width=0.2\linewidth]{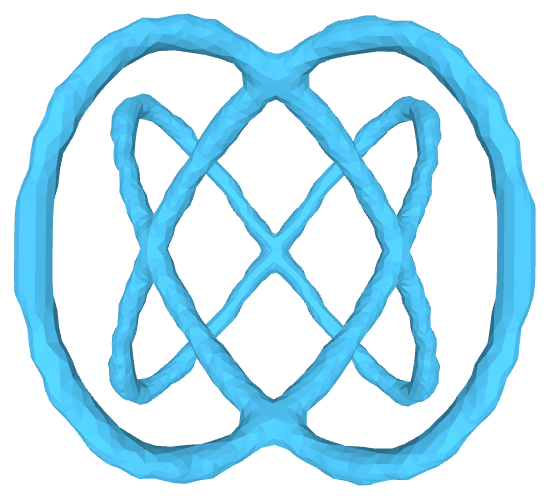}}
    
    \subfigure[$g=7$]{\includegraphics[width=0.2\linewidth]{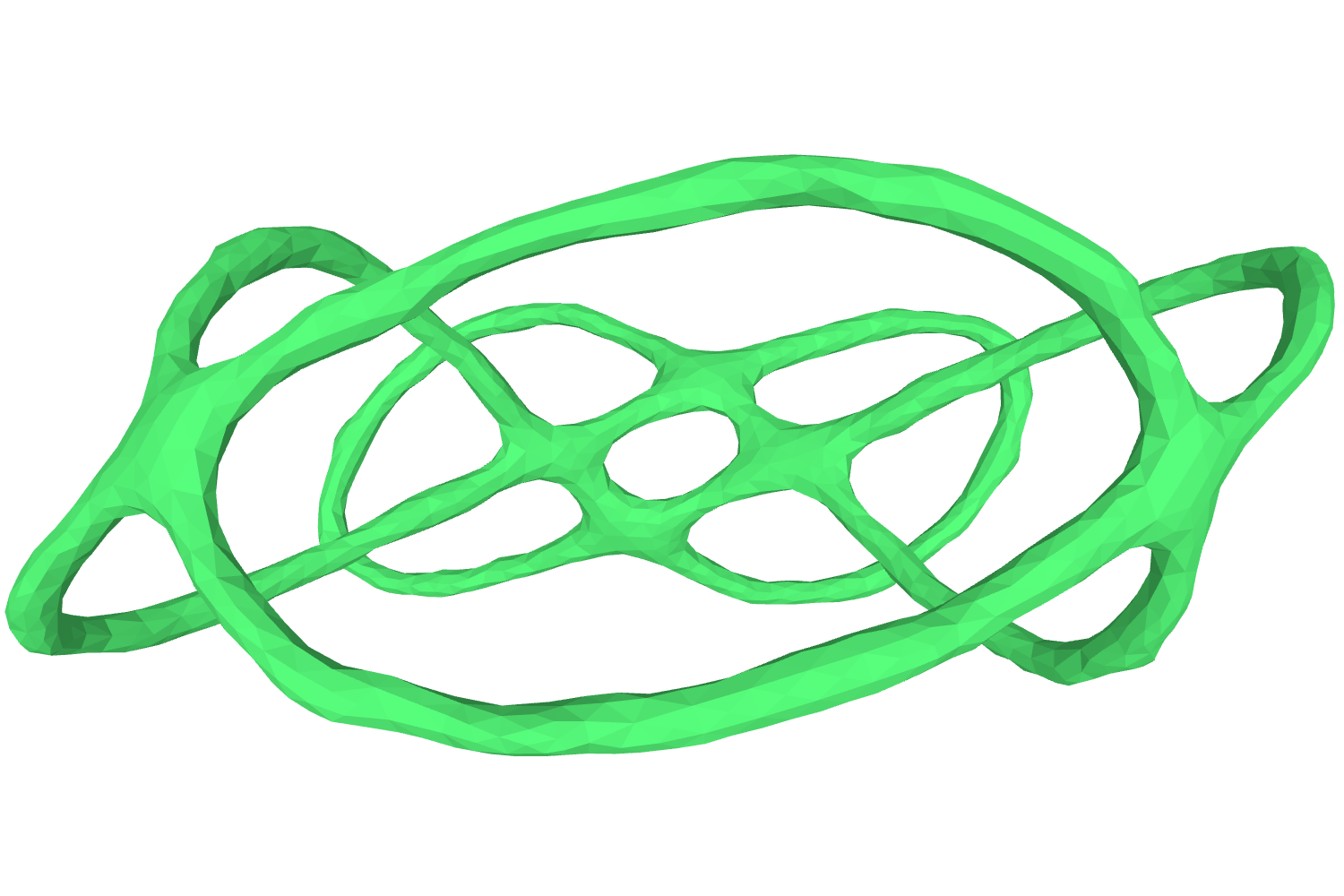}}
    \subfigure[$g=7$]{\includegraphics[width=0.2\linewidth]{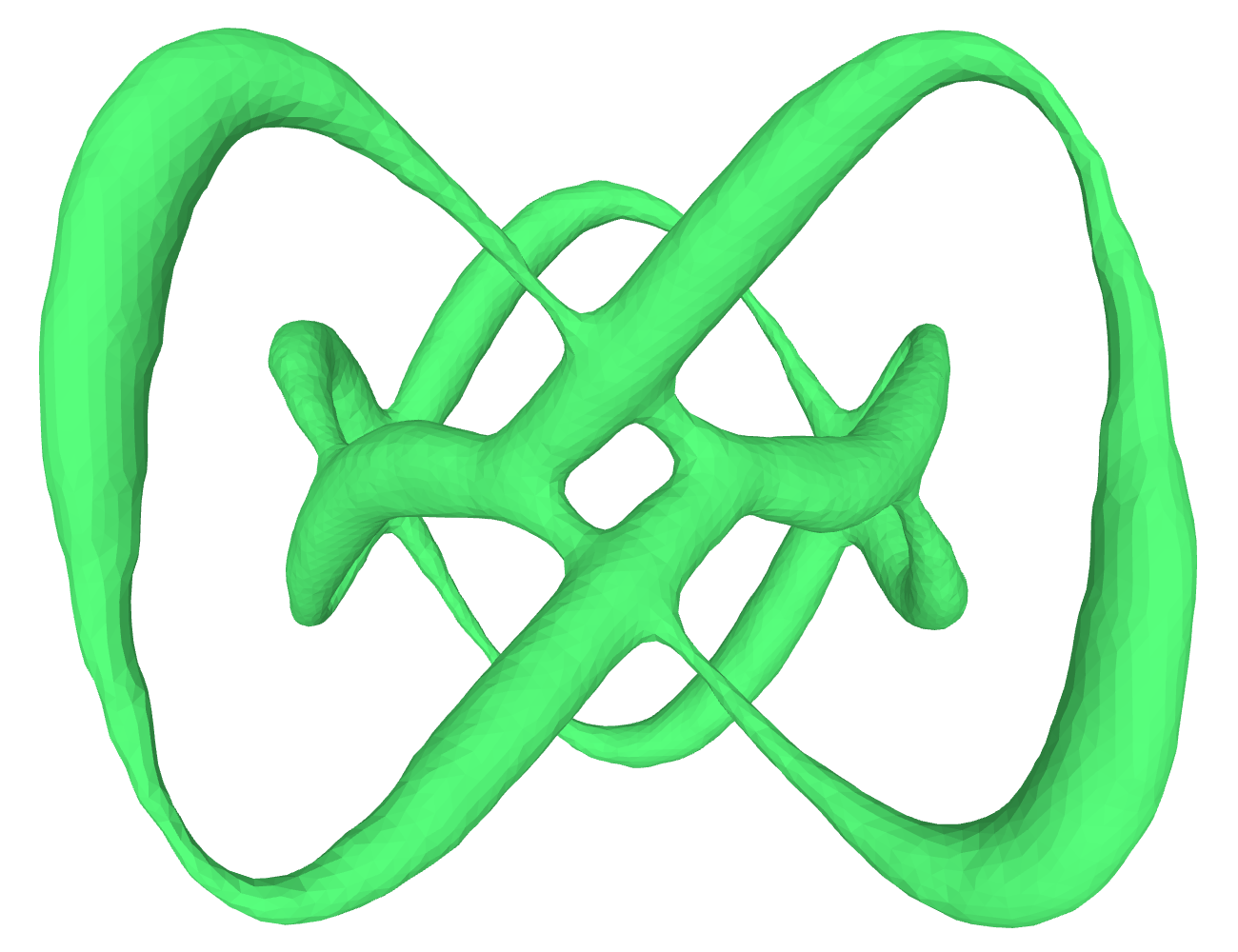}}
    \subfigure[$g=7$]{\includegraphics[width=0.2\linewidth,trim=-2cm 2cm -2cm 0,clip]{g7_liss235.png}}
    \subfigure[$g=7$]{\includegraphics[width=0.2\linewidth]{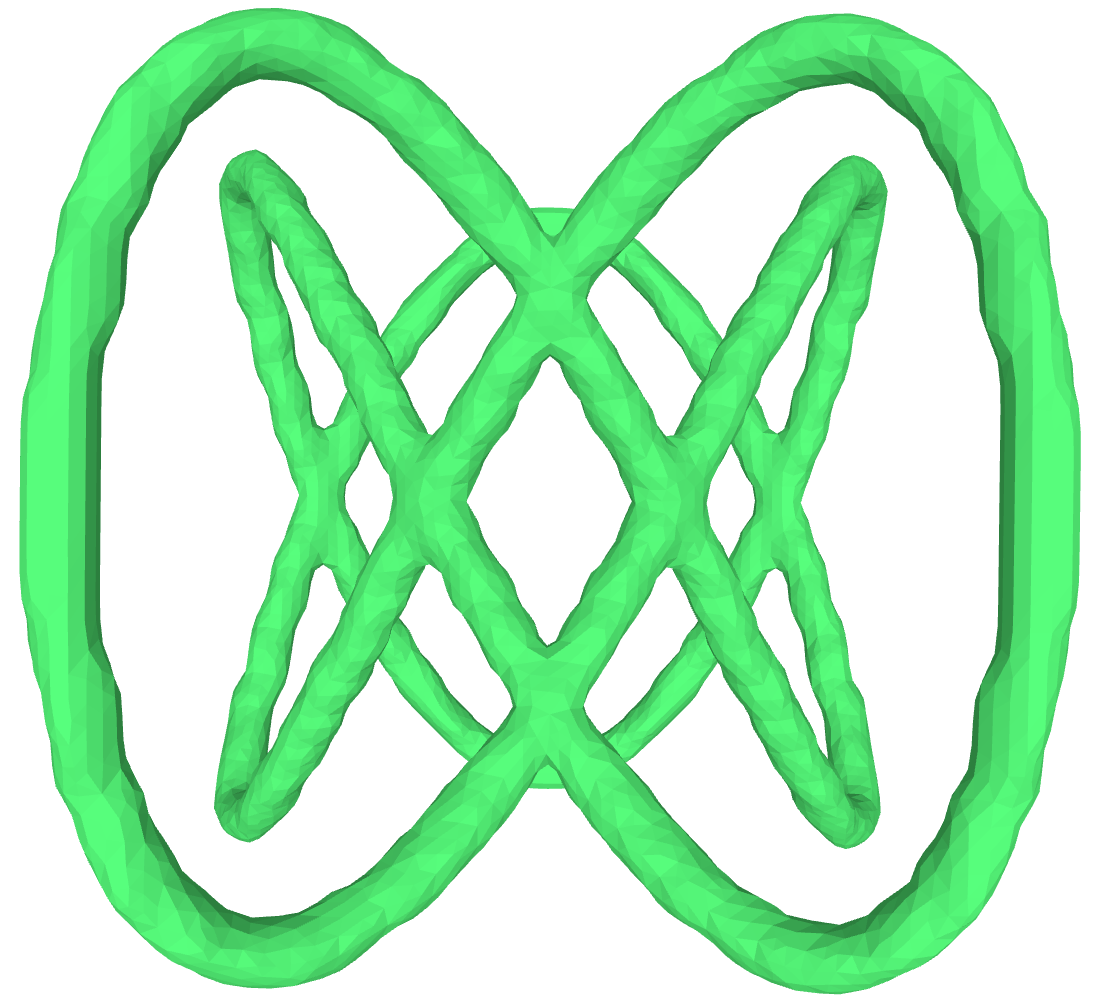}}
    
    \subfigure[$g=8$]{\includegraphics[width=0.2\linewidth]{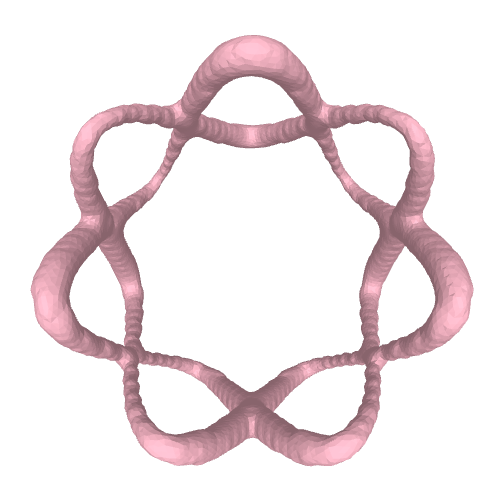}}
    \subfigure[$g=8$]{\includegraphics[width=0.2\linewidth]{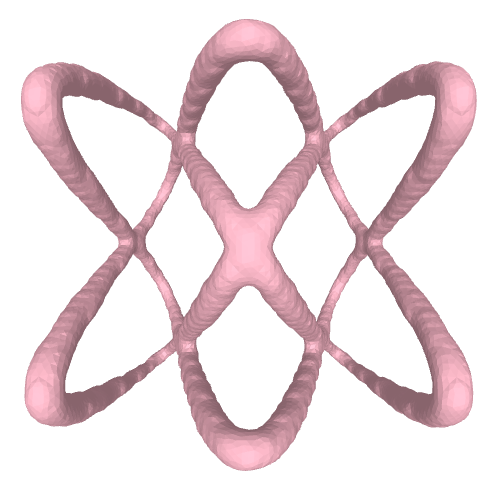}}
    \subfigure[$g=8$]{\includegraphics[width=0.2\linewidth]{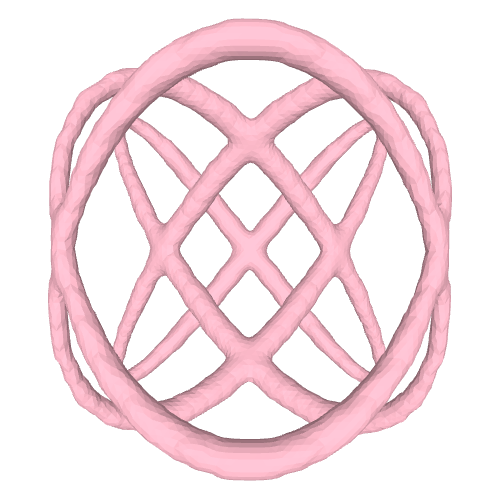}}
    \subfigure[$g=8$]{\includegraphics[width=0.2\linewidth]{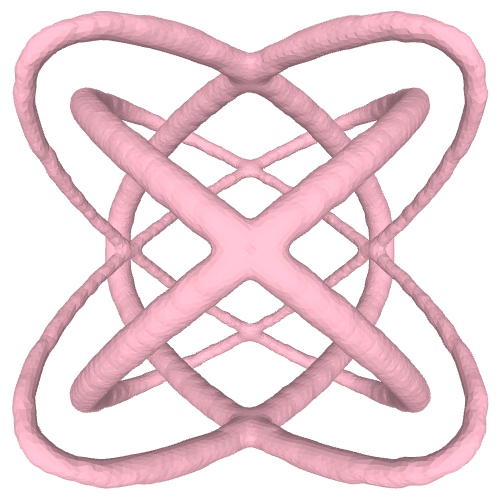}}

    \subfigure[$g=9$]{\includegraphics[width=0.2\linewidth]{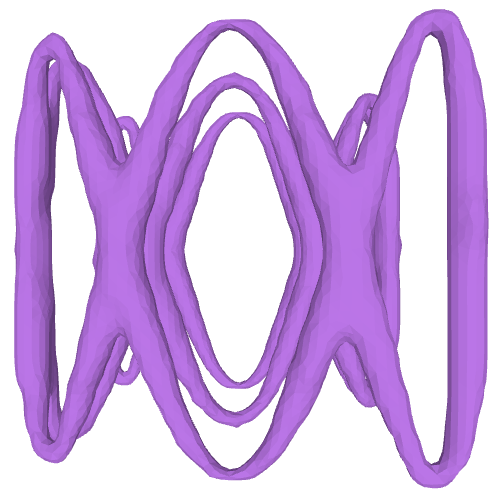}}
    \subfigure[$g=9$]{\includegraphics[width=0.2\linewidth]{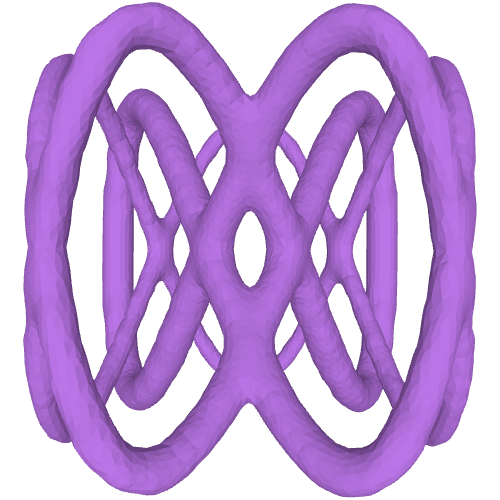}}
    \subfigure[$g=9$]{\includegraphics[width=0.2\linewidth]{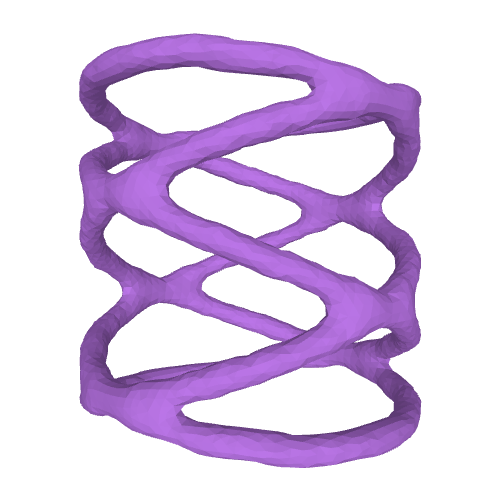}}
    \subfigure[$g=9$]{\includegraphics[width=0.2\linewidth]{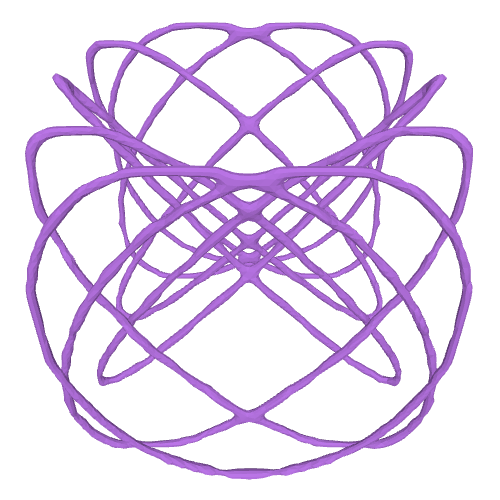}}
    
    \subfigure[$g=1$0]{\includegraphics[width=0.2\linewidth]{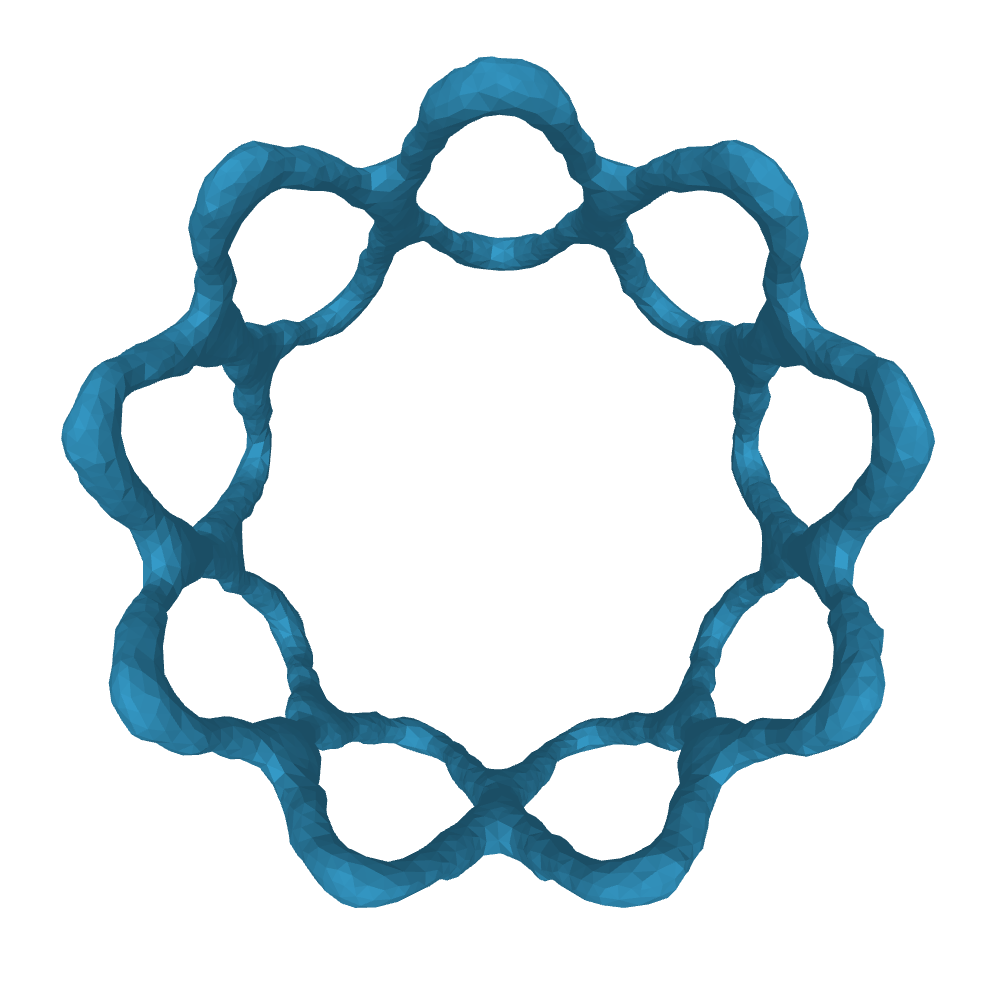}}
    \subfigure[$g=1$0]{\includegraphics[width=0.2\linewidth]{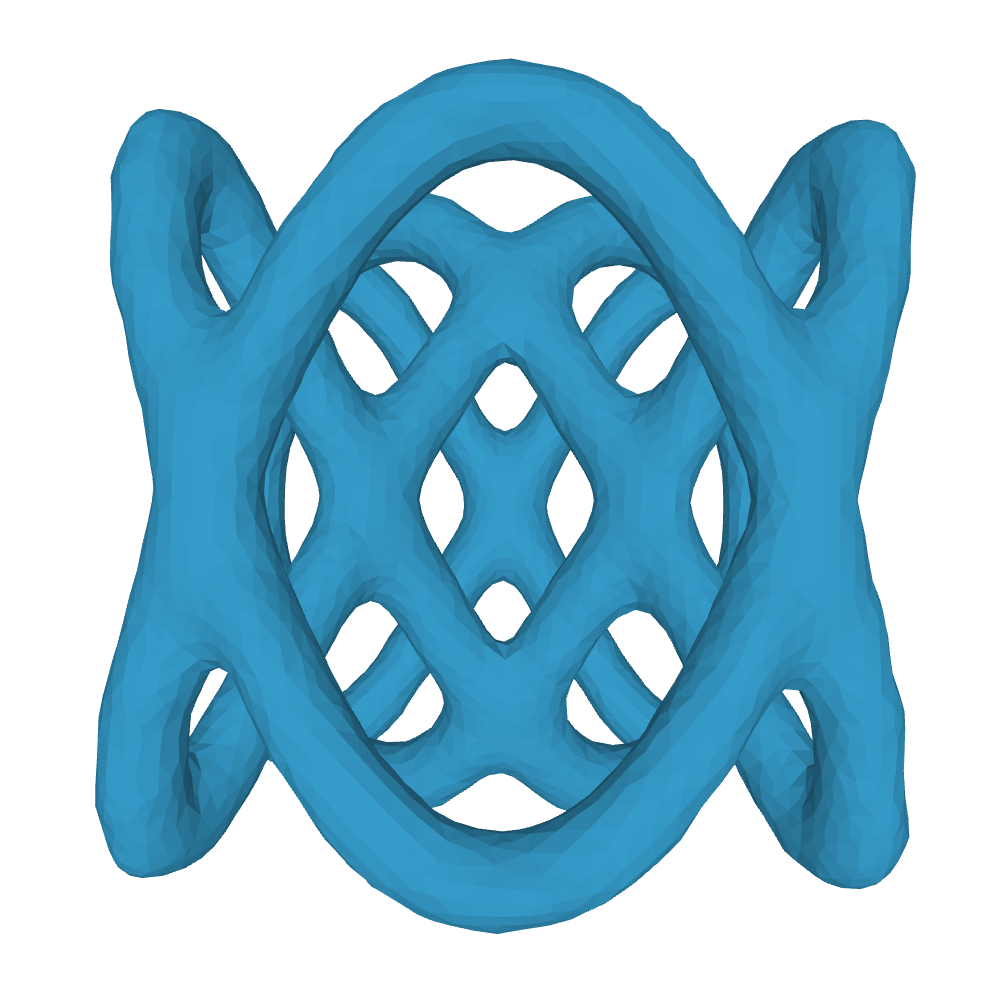}}
    \subfigure[$g=1$0]{\includegraphics[width=0.2\linewidth]{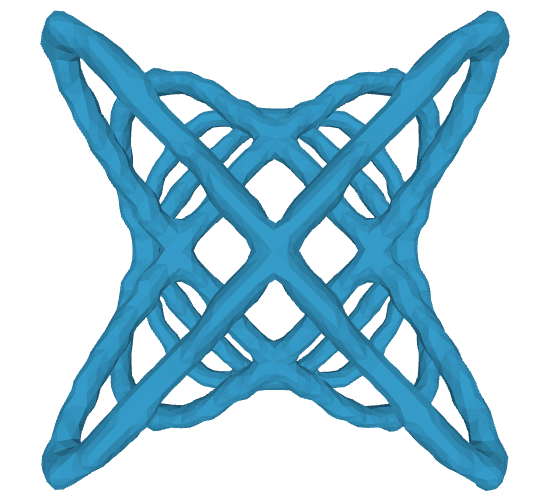}}
    \subfigure[$g=1$0]{\includegraphics[width=0.2\linewidth]{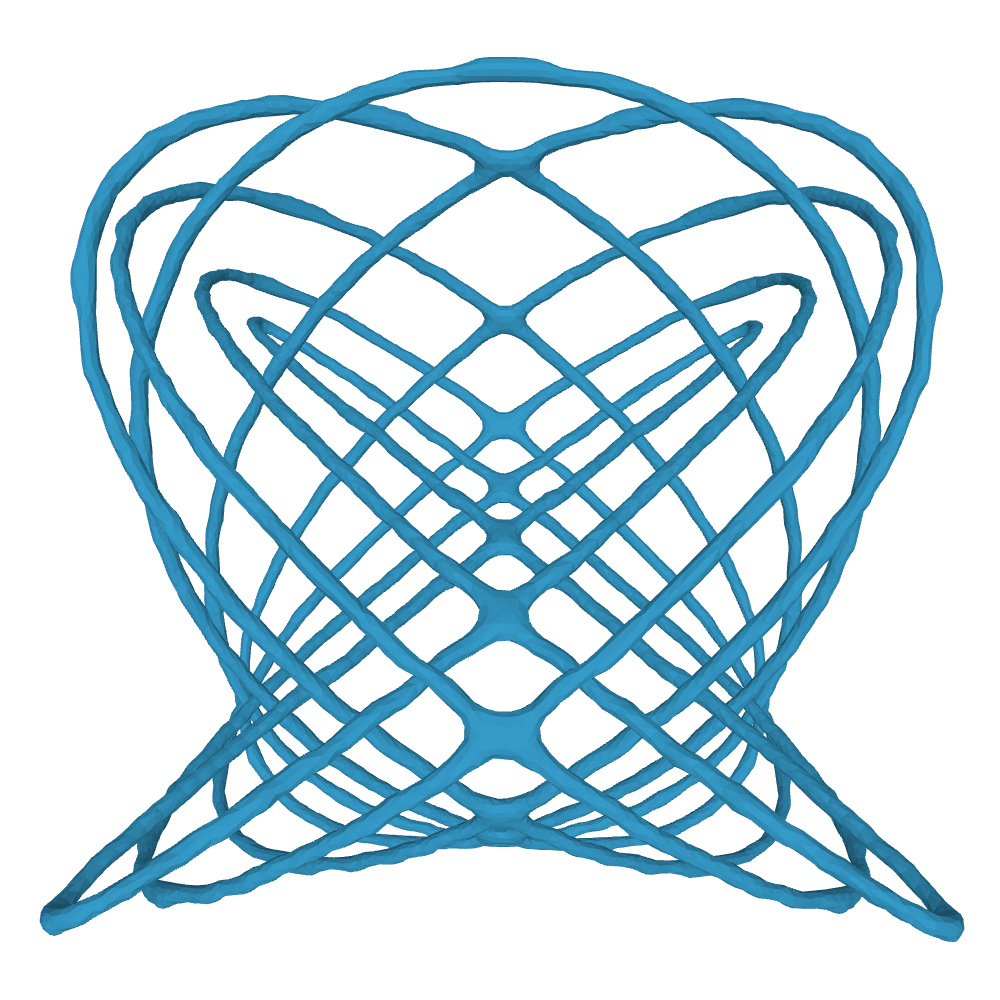}}
    \caption{Another sample of 4 examples for each genus from 6 to 10 from the EuLearn dataset.}
    \label{fig:Eulearn_dataset2}
\end{figure}

Our EuLearn dataset contains 3,300 surfaces.
As the Marching Cubes algorithm produces sharp surfaces, we applied smoothing techniques.
The EuLearn dataset includes 15 distinct initial singular knot types for each of the 11 genera, ranging from genus 0 to genus 10.
For each genus, there are 300 surfaces generated by applying sinusoidal variations to the 15 seed singular knots, with frequencies ranging from 1 to 20. 
In Figures \ref{fig:Eulearn_dataset1} and \ref{fig:Eulearn_dataset2}, we present a sample of 4 examples (out of 15) for each genus.

\subsection*{Dataset Construction Steps 2 through 4}

We developed Step 1 in \cref{subsec:selection} and left pending Steps 2 through 4, which now explain in the following subsections.

\subsection{Step 2: Curve Thickening}
To generate a surface by thickening a polygonal curve, we need to construct a structure on a neighborhood surrounding the curve that represents the final surface.
Various data structures, such as point clouds, meshes, level sets, and others, can be used for this purpose.

\begin{figure}[H]
    \centering
    \subfigure[A Lissajous singular knot (red) enclosed in its bounding box (blue), with its triangulated mesh (green) generated by the Marching Cubes algorithm. Only the voxels intersecting the tubular neighborhood around the knot are shown.]{\includegraphics[width=0.35\linewidth]{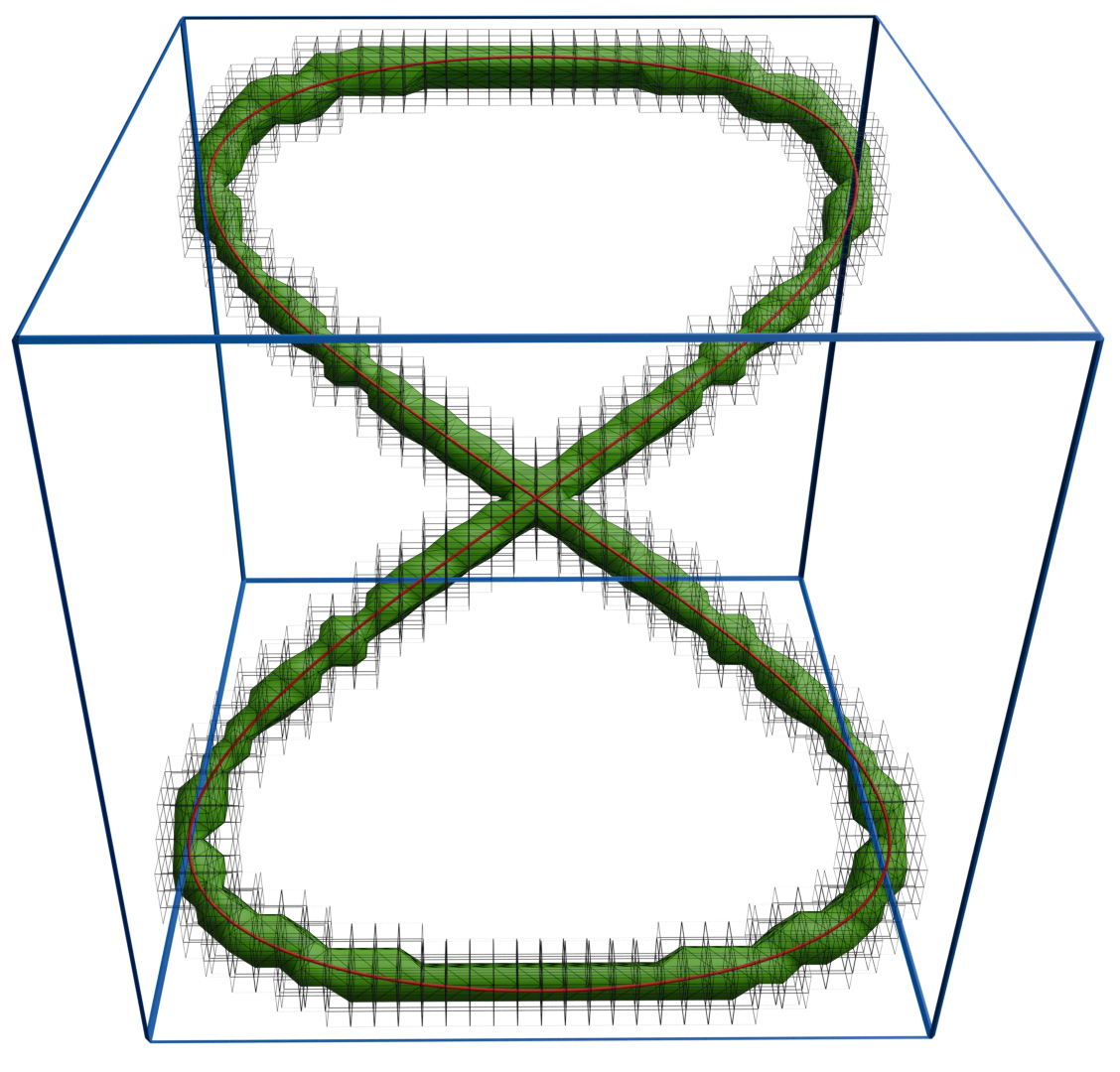}
    \label{fig:voxels}
    }
    \hspace{2mm}
    \subfigure[Detail of the interior of a singular knot's tubular neighborhood. Note that the thickening process does not inherit the self-intersection singularity of the 1D original curve, establishing an embedded regular surface without self-overlapping.]{\includegraphics[width=0.58\linewidth]{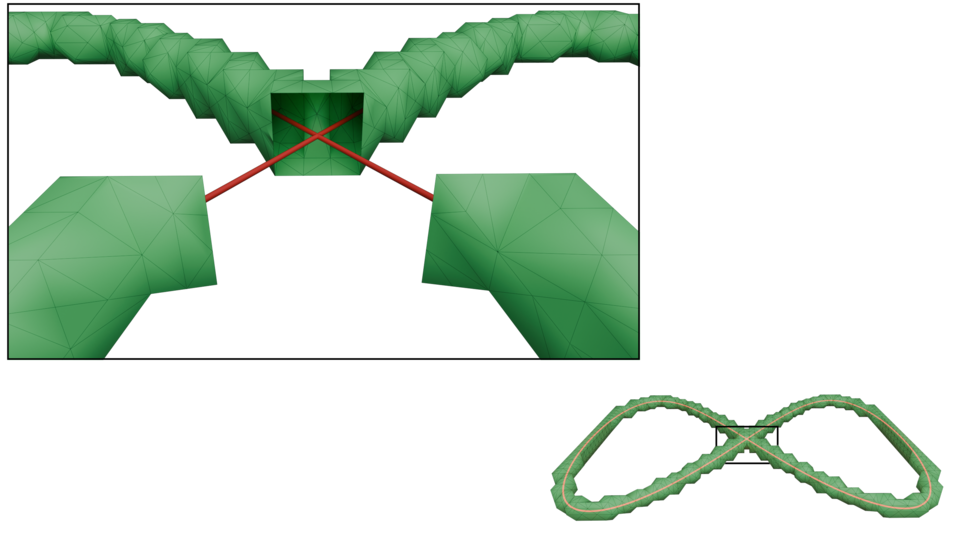}
    \label{fig:inside}
    }
    \caption{Detail of the curve thickening process.}
\end{figure}

Thickening a self-intersecting curve becomes more challenging with structures other than level sets, which rely on discretizing a scalar field on a grid and applying the Marching Cubes algorithm~\cite{WengerMarchingC} to create a surface mesh consisting of vertices, edges, and faces.
We discretize the curve by generating a scalar field for a neighborhood of radius $\delta>0$ around it, and assigning each spatial point a value equal to its distance from the curve minus $\delta$. 
In this representation, the zero-level set corresponds to the surface.
To achieve this goal, we developed a modified approach that constrains the face sizes produced by the Marching Cubes algorithm above a fixed size. 
This restriction prevents small faces from being lost due to rounding errors, and it was crucial for the creation of our EuLearn database.
We will explain this method in detail below.
The discrete representation of the scalar field consists of a grid of $100^3$ voxels around the curve. 
On the nodes we assign the integer value $1$ if they are inside a neighborhood of radius $\delta$ around the curve and $-1$ otherwise.
In the next section, we detail the selection of the neighborhood of size $\delta$.
To determine if a point in the space is inside or outside the neighborhood, we compute its distance function from the polygonal curve and compare this value with the radius $\delta$ of the neighborhood.
This scalar field classifies the grid nodes into inside and outside nodes, thus constraining the zero-level set to traverse the midpoints of the voxels edges.
Consequently, after applying the Marching Cubes algorithm, the size of the surface faces is also determined by the size of the voxels and not exclusively by the distance between the curve and the grid nodes.
In \cref{fig:voxels} we show the process of thickening curves, and then generating surfaces from their tubular neighborhoods.

\cref{fig:inside} illustrates that the non-injective nature of singular knot curves introduces no ambiguity when generating a surface from the curve. 
This is because the singular point of the 1D curve expands into an entire circumference on the resulting 2D surface. 
The increase in dimensionality leads to a more structured and manageable geometric object, with well-defined topology, smooth geometry, consistent normals, and compatible with standard surface-processing algorithms.

\subsection{Step 3: Surface Discretization}
\label{subsec:discretization}

The curve-thickening process discussed in the previous section results in a genus $g$ surface $S$ formed by a triangulation with $V$ vertices, $E$ edges, and $F$ faces.
This representation allows for the genus of the surface to be determined using the Euler characteristic formula, 
$\chi(S) = V - E + F = 2 - 2g$.
However, when the thickening of the curve overreaches a critical limit, surface regions that are initially disjoint may merge and alter the genus.
The following definition of \emph{reach} determines this critical limit.%

\begin{definition}(Federer ~\cite{Federer})
    The \textbf{reach} $\rho$ of an embedded manifold $M$ is the largest number such that any point at a distance less than $\rho$ from $M$ has a unique nearest point on $M$.
\end{definition}

An alternative way to think of reach is via the notions of the medial axis of $M$ and its local feature size, notions which have been developed in the Computational Geometry community \cite{Amenta1999}. 
In fact, Federer’s original definition of reach for the purely Euclidean case is given implicitly in terms of the medial axis.

\begin{definition}
    The \textbf{medial axis} of a manifold $M$ embedded in $\mathbf{R}^d$ is the closure of the set of points in $\mathbf{R}^d$ with more than one nearest neighbor on $M$.    
\end{definition}

By determining the medial axis of the manifold, we can calculate the reach, and thus determine the largest radius that a singular knot can be thickened into, without the resulting boundary surface touching itself.
Curves with fewer self-intersections and shorter arc lengths typically have a larger reach.
To thicken the curves consistently regarding their genus or length, we considered a non-constant neighborhood determined by a sine function along the curve (see \cref{subsec:sinusoidal}), which is always bounded by the voxel size and the surfaces' reach.
The sine function that defines the non-constant neighborhood enables artificial augmentation of the database by tailoring its frequency, amplitude, and phase shift.

\subsection{Step 4: Smoothing}
As stated in \cref{subsec:discretization}, the surfaces in our database were generated by representing the tubular neighborhood around a curve by a scalar field and applying the Marching Cubes algorithm to this field discretization.
The output of the Marching Cubes algorithm is a mesh whose resolution depends on the voxel size.
While smaller voxels provide finer mesh detail, they also increase the processing time and may limit the database handling efficiency.
Using a relatively small constant number of voxels reduces the mesh resolution.
To enhance the smoothness of the mesh while keeping the number of faces constant, we additionally generate a smoothed version of the mesh by averaging the angle between the faces of the mesh.
This smoothing procedure was performed by integrating the Blender command-line \textit{Smooth Vertices} tool into our workflow.

\subsection{Data Augmentation}
\label{subsec:sinusoidal}

\begin{figure}
    \centering
    \subfigure[Front view.]{\includegraphics[width=0.28\linewidth]{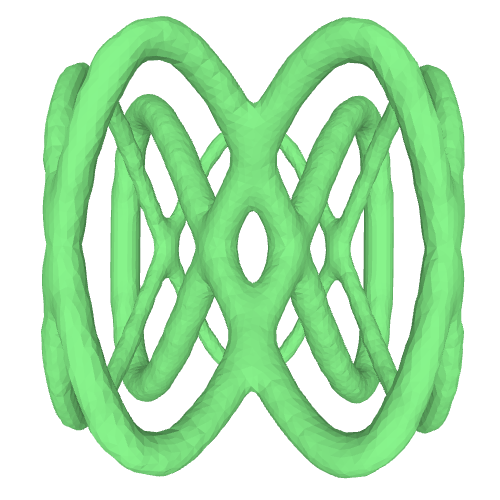}}
    \subfigure[Top view.]{\includegraphics[width=0.28\linewidth]{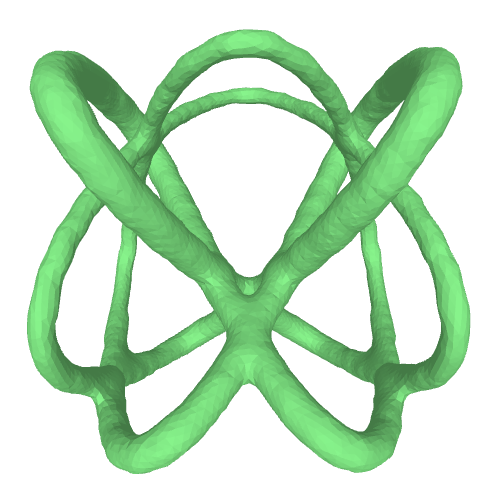}}
    \subfigure[Back view.]{\includegraphics[width=0.28\linewidth]{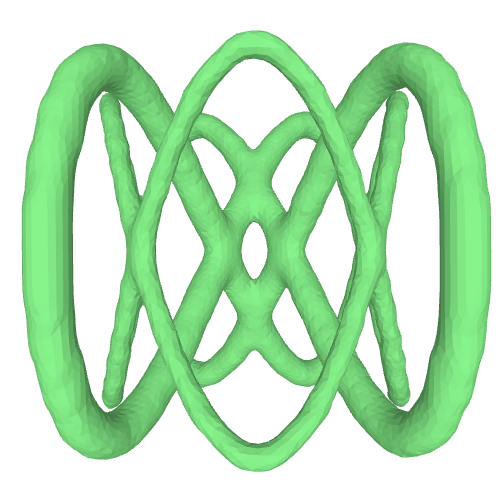}}
    \caption{A genus 9 surface generated with a cosine frequency of 2 for the sinusoidal extension. This surface was built upon a singular Lissajous knot with frequencies $3,4,7$ and $\phi_z = \pi/2$, which has 8 self-intersections.}
    \label{Fig:Freq2surface}
\end{figure}

To extend our dataset, we resorted to a sinusoidal variation of the tubular radius of each thickened (singular) knot.
This sinusoidal variation is actually given by a cosine function, since it gives a more symmetric thickening than the sine function. 
The frequency of this cosine function allows us to construct a variable tubular radius. 
We chose frequencies from 1 to 20, and thus our dataset is extended to 20 different surfaces for the same (singular) knot.
Observe that these 20 surfaces, by construction, have the same isotopy type. 
Figure \ref{Fig:Freq2surface} shows an example of such a surface with frequency equal to 2 in this sinusoidal variation of the thickness.

\section{Related Previous Architectures}
\label{App:PreviousArchitectures}

PointNet, introduced by Qi et al. \cite{qi2017pointnet}, is a pioneering architecture for 3D point cloud classification that aggregates both local and global information from point cloud data by analyzing the Euclidean neighborhood of each point. 
PointNet++ \cite{qi2017pointnetdeephierarchicalfeature} extends this concept by recursively applying the aggregation layer to partitions of the input data. %
Meanwhile, MeshCNN \cite{Hanocka_2019, wang2018} introduces a convolution operator over the edges of a mesh, where the spatial support is defined by the set of neighboring edges. %
Similarly, by focusing on edges rather than vertices, MeshCNN \cite{feng2019meshnet} exploits the connectivity structure of the mesh to model local relationships and enhance geometric feature representation.

Edge convolutions \cite{lee2022edgeconv} are a class of methods that operate on the edge information of a graph, performing convolutions over the points that define an edge. Unlike traditional graph convolutional methods that focus on the vertices (or nodes) of the graph, edge convolutions represent the edges, capturing the relationships and interactions between nodes. The Dynamic Graph Convolutional Neural Network (DGCNN) proposed by \cite{wang2018} implements the concept of edge convolutions, where the edge features are dynamically updated during the training process. This dynamic updating of edges allows the model to capture the evolving relationships between nodes and adapt to the changing graph structure, enabling more effective learning and representation of complex geometric and topological features.

Fourier Neural Operators (FNOs) \cite{li2020fourier} were originally introduced for solving partial differential equations. %
Subsequently, these operators were extended to image classification tasks \cite{johnny2022fourier}, showing a satisfactory performance.
More recently, an adaptation of Fourier Neural Operators for surface reconstruction was proposed \cite{andradeneural}, achieving a good performance, particularly in low-resource scenarios. 
Motivated by the success of transformers \cite{vaswani2017attention} for different tasks, a range of 3D transformer-based architectures have been proposed for processing 3D point clouds and meshes, primarily focusing on classification and segmentation tasks \cite{zhao2021pointtransformer, guo2021pct, yu2022point}. These architectures typically employ transformer encoders, which rely on self-attention mechanism. %

\begin{figure}[H]
    \centering
    \begin{minipage}[b]{0.33\linewidth}
        \centering
        \includegraphics[width=0.55\linewidth]{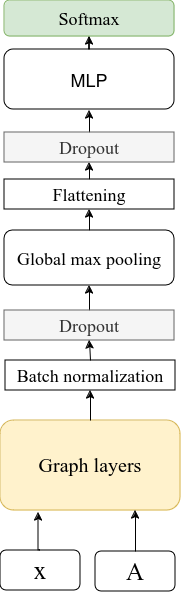} \label{fig:LayersA}
        \\ a) General Neural Architecture.
    \end{minipage}%
    \hspace{7mm} 
    \begin{minipage}[b]{0.43\linewidth} 
        \centering
        \hspace{0mm}\includegraphics[width=0.4\linewidth]{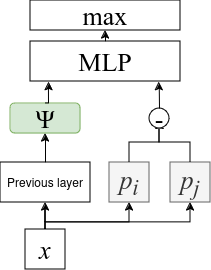} \label{fig:LayersB}
        \\ b) GS PointNet Layer.
        \vspace{8mm}\\ 
        \includegraphics[width=0.4\linewidth]{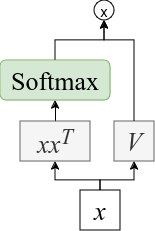} \label{fig:LayersC}
        \\ c) GS Attention Layer.
    \end{minipage}
    \caption{a) Neural architectures for genus classification comprising: 1. graph layer taking an input data $x$ and adjacency matrix $A$, our modifications to PointNet and Attention layers are shown in subfigures b) and c), 2. batch normalization and dropout of 0.3, 3. global max pooling with flattening and dropout of 0.3, 4. output MLP with softmax activation; b) Our GS PointNet layer uses as position the difference of the input vector and applies a MLP combining this position and a previous layer vector to finally apply a max aggregation; c) Our GS Attention layer weights the value vectors $V$ with attention weights and applies a direct dot product to the input vectors without applying any previous transformation.}
    \label{fig:Layers}
\end{figure}

\section{Description of the layers} \label{app:LayersDescription}

The general architecture of the deep models was described in \cref{sec:GNNs}. This architecture consists of the layers: 1) a graphical input layer; 2) batch normalization; 3) dropout regularization; 4) max pooling and flattening; and 5) multilayer perceptron for classification with softmax. A diagram describing the general architecture and the two propoed layers can be seen in \cref{fig:Layers}a. 
This appendix presents the description of the graphical input layers used for our experiments: we describe Fourier Neural Operators (FNO), Dynamic Graph Convolution Neural Networks (DGCNN), PointNet, PointNet++, GS PointNet, attention and GS attention layers.

\subsection{Fourier Neural Operators} \label{sec:FNO}

Fourier neural operators (FNOs) \cite{li2020fourier, johnny2022fourier,andradeneural} focus on leveraging the Fourier space for processing data, applying the Fast Fourier Transform (FFT) to the input, followed by a weighted sum with learnable parameters, and then transforming the data back to the original spatial domain using the inverse FFT. This information is subsequently integrated with the original input through a neural network architecture, enabling FNOs to effectively capture and model interactions at multiple spatial scales. A FNO block is defined as follows:

\begin{definition}[Fourier Neural Operator]
A Fourier Neural Operator Block is given by the layers:
$$h_{i+i} = \psi\Big( Wh_i + F^{-1} \big(  R \cdot F(h_i) \big) \Big)$$
Here $\psi$ is a MLP, $R \in \mathbf{C}^{d_{max} \times d\times d}$ is a weight tensor  and $F$ is the Fourier transform.
\end{definition}
The function $\psi$ in the previous formula is a MLP; specifically, we use an MLP with one hidden layer and a Gaussian Error Linear Unit (GeLU) activation function \cite{hendrycks2023gaussianerrorlinearunits}.
For the complete FNO layer, the input data $x$ is first transformed as $h_1 = P(x)$, where $P$ is an initial MLP. Subsequently, multiple FNO blocks are applied to obtain an output representation of the data $h_t$. In our experiments, we use three FNO layers. Finally, this representation is fed to an MLP to produce $Q(h_T)$.

In our experiments, the FNO architecture differs from other models in the nature of the input data. Specifically, FNO takes as input a scalar field rather than the point cloud of the mesh. The scalar field is incorporated into the dataset as additional surface information. The FFT is applied to the scalar field, rather than to the points of the mesh. This change in input data type necessitates specific modifications to the architecture. Notably, the scalar field is represented as a $100 \times 100 \times 100$ tensor. Besides this, the overall procedure and architecture remain consistent with those of the other models.

We used two FNO layers with three FNO blocks each. Each layer is of the same input and output dimension, i.e. $100 \times 100 \times 100$. We incorporate a dropout of 0.3 between each of these layers.

\subsection{Dynamic Graph Convolutiona Network}

Dynamic Graph Convolutional Neural Networks (DGCNNs) were proposed by \cite{wang2018} to process graph-structured data. They are specifically designed for processing 3D point clouds and meshes, but have applications in other graph-related tasks, such as social network analysis. The basic idea is to apply a graph convolution operation on the data graph $G = (V, E)$, where $V$ represents the set of vertices (or nodes) and $E$ represents the set of edges.
The basic idea is to use a graph convolution on the data graph $G$.

\begin{definition}[Edge convolution]
    Let $G=(V,E)$ be a graph representing the relations of the data $x_1,...,x_n$. A graph convolution is a layer defined as:
    \begin{equation}
    h_i = \bigoplus_{(i,j) \in E} \psi_\Theta(x_i, x_j)
\end{equation}
Here $\bigoplus$ is a comutative operator $\sum$ or $\max$ and $\psi_\Theta: \mathbf{R}^d \times \mathbf{R}^d \to \mathbf{R}^d$ is a non linear learnable function with parameters $\Theta$.
\end{definition}

The operator and the function $\psi_\Theta$ in our implementation are based on \cite{wang2018} and represent a convolutional operation that aggregates information along the edges of the graph, rather than relying on a fixed neighborhood defined by a kernel.
The graph construction, similar to the common PointNet layer, is based on $k$-nearest neighbors graphs, as proposed by the authors \cite{wang2018}. This approach relies on the metric of the embedded space rather than the surface itself. Our implementation also incorporates a dynamic graph update mechanism, which consists of 5 layers. For further details, we refer the reader to \cite{wang2018}.

\subsection{PointNet and PointNet++} \label{app:pointnet}

PointNet was first proposed by \cite{qi2017pointnet} for point cloud classification. This architecture is permutation invariant. In the standard implementation of the PointNet architecture, a point cloud input is processed by an aggregation function based on the maximum value of the set of processed neighbors. The neighbors are typically obtained using a $k$-nearest neighbor procedure, and these neighbors are then fed into a neural network along with positional information.
\\\\
\textbf{Definition} (PointNet layer) A PointNet Layer is a neural network layer that takes as input a set of point $x_1,...,x_n$ and information about their neighbors, and estimates a hidden representation as:
\begin{equation}\label{eq:pointnet}
    h_i = \phi\Big( \max_{j \in \mathcal{N}_i} \psi(x_i, p_i - p_j) \Big)
\end{equation}
for each $i \in \{1,...,n\}$. Here, $\phi$ and $\psi$ are feedforward networks, while $p_i, p_j$ define the position of the points $x_i$ and $x_j$, respectively.
\\\\
Generally, $\phi$ and $\psi$ are ReLU networks with one single hidden layer. The position $p_i$ of a point is commonly the original input vector, while $x_i$ are the previous layer representation. 
As we mentioned before, the neighbors of a point $x_i$, $x_j \in \mathcal{N}_i$, are the nearest element to the point $x_i$ using a metric in the embedded space (such like an euclidean metric).

We also use PoitNet++ \cite{qi2017pointnetdeephierarchicalfeature}, following an implementation based on the work of Yan \cite{Pytorch_Pointnet_Pointnet2}.
We remark that we found a notable performance improvement when considering having an intrinsic sampling versus the standard extrinsic one.

\subsection{GS PointNet} \label{sec:PointNet}

The PointNet architecture \cite{qi2017pointnet} is based on metric information in the embedded space to determine the neighbors in the mesh (see Appendix~\ref{app:pointnet}). However, as the objective of the experiment task relies on topological information, the performance of classical PointNet is suboptimal (above 0.5 as can be seen in Table~\ref{tab:AttentionResults}).
In order to address this limitation, we adapt the PointNet architecture to incorporate adjacency information through the adjacency matrix. Specifically, the aggregation method is modified to consider the mesh neighbors of each point on the original surface. The Graph Sampled (GS) PointNet adaptation we made is the following:
\\\\
\textbf{Definition} (GS PointNet Layer) A GS PointNet Layer is a neural network layer that takes as input a set of point $x_1,...,x_n$ and information about their neighbors, and estimates a hidden representation as:
\begin{equation}\label{eq:pointnetAdapt}
    h_i =  \max_{j \in \mathcal{N}_i} \phi\Big( \psi(x_i), p_i - p_j \Big)
\end{equation}
where $\psi(x_i) = Wx_i$ represents a linear projection without bias, and $\phi$ is an MLP with a hidden layer and a SiLU activation function \cite{elfwing2017sigmoidweightedlinearunitsneural}.
\\\\
We assume that the position information is better represented in the input points; therefore, the position vectors $p_i$ and $p_j$ are always derived from the original input data points, while the vector $x_i$ corresponds to the output of the previous layer (see Figure~\ref{fig:Layers}).
These modifications enable the model to better capture local surface information, thereby enhancing its performance in the classification task.

For both the classical implementation of PointNet and the GS PointNet, we employ two layers in a sequential configuration. This choice of layer configuration was found to yield the best performance for our specific task. We used two PoinNet layers in every model each with dimensions 256.

\subsection{Attention} \label{app:Attention}

Attention layers are commonly used in transformer architectures \cite{vaswani2017attention}. Originally used to process natural language, attention layers have been adapted to process other types of data such as images (with visual transformers \cite{liu2023survey}) and surfaces \cite{lu2022transformers}. 
The main idea of attention layers is to assign attention weights, based on softmax probabilities, to the elements of the input in order to make an addition aggregation. The classical definition of an attention layer is as follows:
\\\\
\textbf{Definition} (Attention layer) An attention layer is defined as:
\begin{equation} \label{eq:attentonClassic}
    h_{i} = \sum_j \alpha(x_i, x_j) \psi_v(x_j)
\end{equation}
where $\phi_v(x_j)$ is a projection (over value space) of the vector $x_j$ and $\alpha(x_i, x_j)$ are the attention weights estimated as below:
$$\alpha(x_i,x_j) = \text{softmax}\left(  \frac{ \phi_k(x_i)^T \phi_q(x_j) }{ \sqrt{d} } \right)$$
Here $\phi_q$ and $\phi_k$ are projections over the query and key spaces, respectively. Finally, $d$ is the model dimension.
\\\\
The original implementation of attention layers involves summing over all the other points in the input, thereby defining a fully connected graph structure. Additionally, classical attention implementations incorporate multi-head attention.
This feature helps explain why it is not readily capable of recognizing different topological types.
Multi-head attention operates by computing multiple independent attention heads, each of which is an attention representation as defined in Equation~\ref{eq:attentonClassic}. These attention heads are then concatenated, and a linear transformation is applied to project the concatenated representation back to the original model dimension, ensuring that the output and input dimensions match.

\subsection{GS Attention} \label{sec:attention}

The last models we used were attention based models (see \cref{app:Attention} for details). 
While transformers are state-of-the-art models for various tasks, they require a large number of examples for effective training. A simple 3D transformer performs poorly in our classification task (see \cref{tab:AttentionResults}). To address this, we propose integrating a graph attention layer into the neural network architecture, similar to those introduced by \cite{velickovic2018graphattentionnetworks}. This layer focuses on the relational information of mesh points by aggregating information from their neighbors, using both the point cloud and the adjacency matrix of the input mesh.
\\\\
\textbf{Definition} (GS attention layer) A GS attention layer is defined as:
\begin{equation} \label{eq:GraphAttention}
    h_{i} = \sum_{j \in \mathcal{N}_i } \alpha(x_i, x_j) \psi_v(x_j),
\end{equation}
where $\psi_v(x_j)$ is a projection (over value space) of the vector $x_j$ and $\alpha(x_i, x_j)$ are the attention weights over the neighbors of $x_i$. 
\\\\
The linear projection $\psi_v(x)$ is computed in the usual way as $xV$, where $V$ consists of learnable parameters of the model. Another modification included in the proposed layer is the simplification of the attention weight estimation as follows:
$$\alpha(x_i,x_j) = \text{softmax}_{j\in \mathcal{N}_i}\left(  \frac{ x_i^T x_j }{ \sqrt{d} } \right)$$
Here $d$ is the model dimension. 
In contrast with classic implementations, we eliminate the linear projections of the data (see Appendix~\ref{app:Attention}). 
This change is motivated by two key notions: (i) the orthogonality or co-linearity of the points in the mesh can be informative for representing its structure, particularly at points of contact where intersections are representative of the genus of the surface.; and (ii) since each layer has adjacency information, the multi-head attention scheme is not necessary and therefore there is no longeer a need for different projections. 
This last point is particularly relevant, as the removal of multi-head attention is motivated by the fact that multiple heads are known to capture different relationships among input instances \cite{clark2019does}. 
However, since the relational information of the mesh is already contained in the adjacency matrix, we opt for multiple layers arranged sequentially, similar to other common graph layers instead of multi-head.

In order to incorporate adjacency information, we applied masking. This involves multiplying a copy of the adjacency matrix, where the 0 values are replaced with large negative numbers (approaching $-\infty$), with the matrix $xx^T$. 
When the softmax function is applied, these large negative values transform to 0, effectively disconnecting the corresponding instances. 
Consequently, the implementation of a layer of this kind has a complexity similar to that of a classical attention layer.
The proposed layer structure is illustrated in \cref{fig:Layers}.

We use two sequential layer with a model dimension of 256 and with ReLU activation function, between layers a batch normalization and a dropout of 0.3 were applied. While for the classic attention layer we use similar parameters but instead of sequential layers we used two attention heads.

\subsection{Hyperparameter selection} \label{app:hyperparemeter}

For all experiments we used similar hyperparameters, with variations depending on the specific layer requirements, as detailed in \cref{app:LayersDescription}. We employed the Noam optimizer during training, as we observed that this optimization method has a significant impact on the correct convergence of the training process and exhibits low variation across different folds, thereby allowing for reproducibility of our experiments (see \cref{app:Significance}).

The hyperparameters of the Noam optimizer were carefully tuned to optimize the performance of our neural model in genus classification experiments. The warmup period was set to 40,000 as proposed by \cite{vaswani2017attention}. The initial learning rate was set to 0, which is a common choice for the Noam Optimizer \cite{vaswani2017attention}, as it allows the optimizer to dynamically adjust the learning rate during training. As is shown in \cref{app:Significance}, this selection does not affect the convergence of the treining procedure.

The $\epsilon$ value was set to 1e-9. The betas were set to (0.9, 0.99) as it is a standard selection \cite{kingma2017adammethodstochasticoptimization}. Finally, a weight decay of 1e-5 was applied to regularize the model as we see that this helps the model to not overfit on certain cases. These hyperparameter values, as shown in Table \ref{tab:hyperparameters}.

\begin{table}[H]
\centering
\begin{tabular}{ll}
\hline
\textbf{Hyperparameter} & \textbf{Value} \\
\midrule
Warmup & 40~000 \\
Initial Learning Rate & 0 \\
$\epsilon$ & 1e-9 \\
Betas & (0.9, 0.99) \\
Weight Decay & 1e-5 \\
\hline
\end{tabular}
\caption{Hyperparameters in our genus classification experiments. }
\label{tab:hyperparameters}
\end{table}


\section{Performance Results}
\label{App:Performance}

The next tables show the complete classification report for the four architectures used in the experiments. 

\begin{table}[H]
    \small
    \centering
    \begin{tabular}{c|c c c} \hline
        \textbf{Genus} & \textbf{Precision} &    \textbf{Recall} &  \textbf{F1-score} \\ \hline
0  &   0.32  &  1.00  &  0.49    \\
1  &   0.29  &  0.35  &  0.32    \\
2  &   0.58  &  0.95  &  0.72    \\
3  &   1.00  &  0.30  &  0.46    \\
4  &   0.00  &  0.00  &  0.00    \\
5  &   0.42  &  0.55  &  0.48    \\
6  &   0.00  &  0.00  &  0.00    \\
7  &   1.00  &  0.25  &  0.40    \\
8  &   0.61  &  0.85  &  0.71    \\
9  &   0.81  &  0.65  &  0.72    \\
10  &  0.47  &  0.49  &  0.48    \\ \hline
\textbf{Accuracy}  &    &    &     0.49  \\
\textbf{Macro avg}  &   0.50  &  0.49  &  0.43  \\
\textbf{Weighted avg}  &   0.50  &  0.49  &  0.43  \\ \hline
    \end{tabular}
    \caption{(Classic) PointNet results over 10 000 epochs for al 11 genera on the dataset, trainning with Noam optimizer}
    \label{tab:AppxPointNet}
\end{table}

Every experiment was run over 10 000 epochs, all following the same regime indicated in Section~\ref{sec:GNNs}, and using the Noam optimizer with the hyperparamenters of \cref{tab:hyperparameters}.

For the classical PointNet implementation, Table~\ref{tab:AppxPointNet} indicates that the overall results are low for all genera, particularly for genera 4 and 6. In the case of a classical transformer with a multi-head attention mechanism, the results are notably poor as can be seen in Table~\ref{tab:AppxAttention}. It is evident that almost every test example was classified under a single genus. This is a common issue in transformer architectures, which typically require a vast amount of data to learn effectively. However, our training set contained fewer than 2000 examples, which could potentially be insufficient for the attention mechanism to learn effectively.

\begin{table}[H]
    \small
    \centering
    \begin{tabular}{c|c c c} \hline
        \textbf{Genus} & \textbf{Precision} &    \textbf{Recall} &  \textbf{F1-score} \\ \hline
0 & 0.48 & 0.96 & 0.64  \\
1 & 0.00 & 0.00 & 0.00  \\
2 & 0.92 & 0.86 & 0.89  \\
3 & 0.62 & 0.76 & 0.68  \\
4 & 0.90 & 0.69 & 0.78  \\
5 & 0.00 & 0.00 & 0.00  \\
6 & 0.00 & 0.00 & 0.00  \\
7 & 0.73 & 1.00 & 0.84  \\
8 & 0.86 & 0.60 & 0.71  \\
9 & 0.54 & 0.50 & 0.52  \\
10 & 0.99 & 0.57 & 0.72	  \\ \hline
\textbf{Accuracy}  &    &    &     0.63  \\
\textbf{Macro avg}  &   0.50  &  0.54  &  0.52  \\
\textbf{Weighted avg}  &   0.50  &  0.54  &  0.52  \\ \hline
    \end{tabular}
    \caption{PointNet++ results over 10 000 epochs for al 11 genera on the dataset, trainning with Noam optimizer}
    \label{tab:AppxPointNetPlus}
\end{table}

\begin{table}[H]
    \small
    \centering
    \begin{tabular}{c|c c c} \hline
        \textbf{Genus} & \textbf{Precision} &    \textbf{Recall} &  \textbf{F1-score} \\ \hline
0  &   0.00  &  0.00  &  0.00    \\
1  &   0.00  &  0.00  &  0.00    \\
2  &   0.00  &  0.00  &  0.00    \\
3  &   0.00  &  0.00  &  0.00    \\
4  &   0.00  &  0.00  &  0.00    \\
5  &   0.00  &  0.00  &  0.00    \\
6  &   0.00  &  0.00  &  0.00    \\
7  &   0.00  &  0.00  &  0.00    \\
8  &   0.00  &  0.00  &  0.00    \\
9  &   0.10  &  1.00  &  0.18    \\
10  &  0.01  &  0.10  &  0.02    \\ \hline
Accuracy  &    &    &    0.10  \\
Macro avg  &   0.01  &  0.10  &  0.02  \\
Weighted avg  &   0.01  &  0.10  &  0.02  \\ \hline
    \end{tabular}
    \caption{(Classic) Attention results  over 10 000 epochs for al 11 genera on the dataset, trainning with Noam optimizer}
    \label{tab:AppxAttention}
\end{table}

The PointNet++ results are better than its classical implementation.
However, because this architecture still relies on the embedding metric space (extrinsic) instead of the metric of the surface (intrinsic), the results are not great.
We see in \cref{tab:AppxPointNetPlus} that for some genus the classification is adequate, however there are certain genera (1, 5 and 6) on which the classification is poor.

\cref{tab:AppxGraphPointNet} shows the results for our proposed adaptation of PointNet integrating adjacency information. 
\cref{tab:AppxGraphAttention} provides the results for our adaptation of attention layers. As can be seen, incorporating adjacency information significantly enhances the performance of the deep learning models. Both models achieve an accuracy around 0.8. \%,   significantly better than any model that does not incorporate information from the adjacency matrix.

\begin{table}[H]
    \small
    \centering
    \begin{tabular}{c|c c c} \hline
        \textbf{Genus} & \textbf{Precision} &    \textbf{Recall} &  \textbf{F1-score} \\ \hline
0  &    0.85  &   0.85  &   0.85    \\
1  &    0.77  &   0.50  &   0.61    \\
2  &    0.95  &   0.90  &   0.92    \\
3  &    0.83  &   0.95  &   0.88    \\
4  &    0.80  &   0.80  &   0.80    \\
5  &    1.00  &   0.30  &   0.46    \\
6  &    0.57  &   1.00  &   0.73    \\
7  &    0.72  &   0.90  &   0.80    \\
8  &    0.84  &   0.80  &   0.82    \\
9  &    0.90  &   0.90  &   0.90   \\
10  &   0.79  &   0.79  &   0.79    \\ \hline
    \textbf{Accuracy}  &     &     &     0.79  \\
   \textbf{Macro avg}  &    0.82  &   0.79  &   0.78   \\
\textbf{Weighted avg}  &    0.82  &   0.79  &   0.78  \\ \hline
    \end{tabular}
    \caption{GS PointNet Layer results  over 10 000 epochs for al 11 genera on the dataset, trainning with Noam optimizer}
    \label{tab:AppxGraphPointNet}
\end{table}

\begin{table}[H]
    \small
    \centering
    \begin{tabular}{c|c c c} \hline
        \textbf{Genus} & \textbf{Precision} &    \textbf{Recall} &  \textbf{F1-score} \\ \hline
0  &   0.71  &  1.00  &  0.83    \\
1  &   0.95  &  0.90  &  0.92    \\
2  &   0.90  &  0.90  &  0.90    \\
3  &   1.00  &  0.95  &  0.97    \\
4  &   1.00  &  0.35  &  0.52    \\
5  &   0.62  &  0.75  &  0.68    \\
6  &   0.83  &  1.00  &  0.91    \\
7  &   1.00  &  1.00  &  1.00    \\
8  &   0.29  &  0.20  &  0.24    \\
9  &   0.80  &  1.00  &  0.89    \\
10  &  0.81  &  0.86  &  0.83    \\ \hline
\textbf{Accuracy}  &    &    &     0.81 \\
\textbf{Macro avg}  &   0.81  &  0.81  &  0.79  \\
\textbf{Weighted avg}  &   0.81  &  0.81  &  0.79  \\ \hline
    \end{tabular}
    \caption{GS Attention Layer results  over 10 000 epochs for al 11 genera on the dataset, trainning with Noam optimizer}
    \label{tab:AppxGraphAttention}
\end{table}

\section{Sampling Process}
\label{appendix:sampling}

\begin{figure}[ht]
    \centering
    \subfigure[Point cloud of the 11~324 vertices that conform the original genus 7 mesh constructed from a Lissajous singular knot.]{\includegraphics[width=0.43\linewidth]{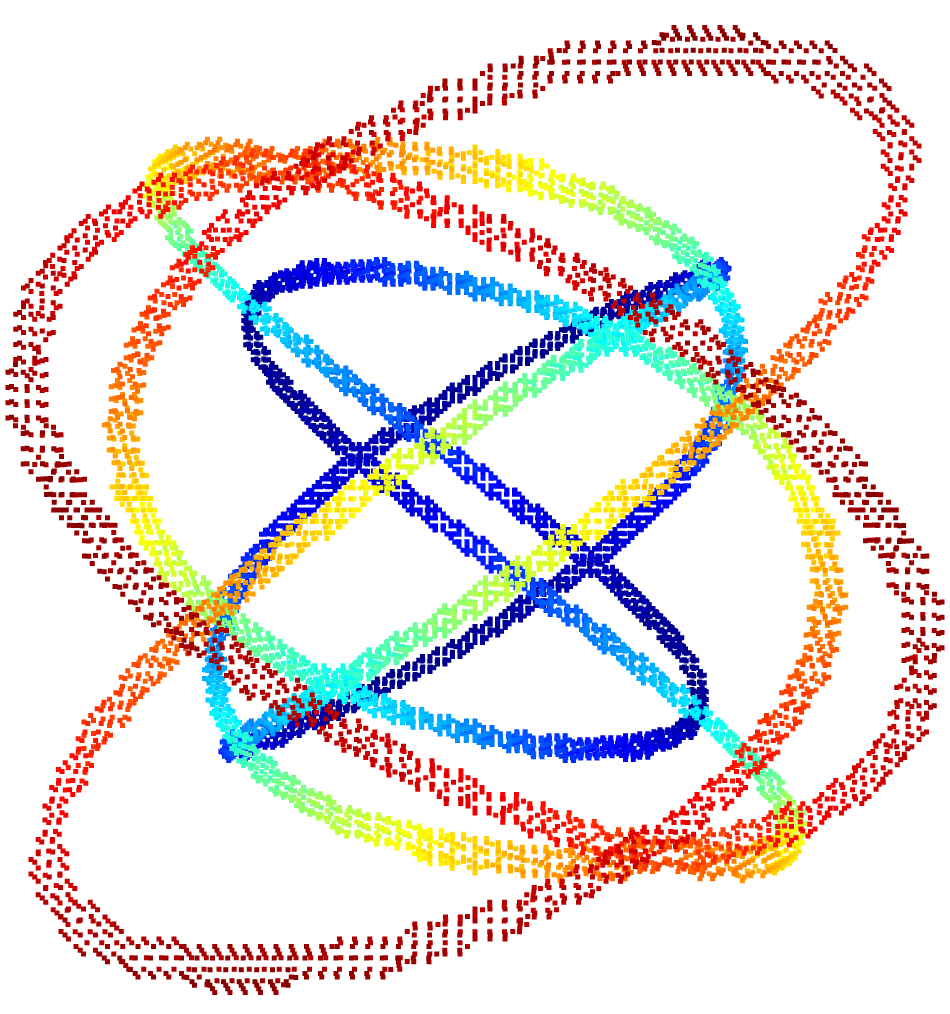}}
    \hspace{2mm}
    \subfigure[Sample of 3~000 vertices from the Lissajous mesh.]{\includegraphics[width=0.43\linewidth]{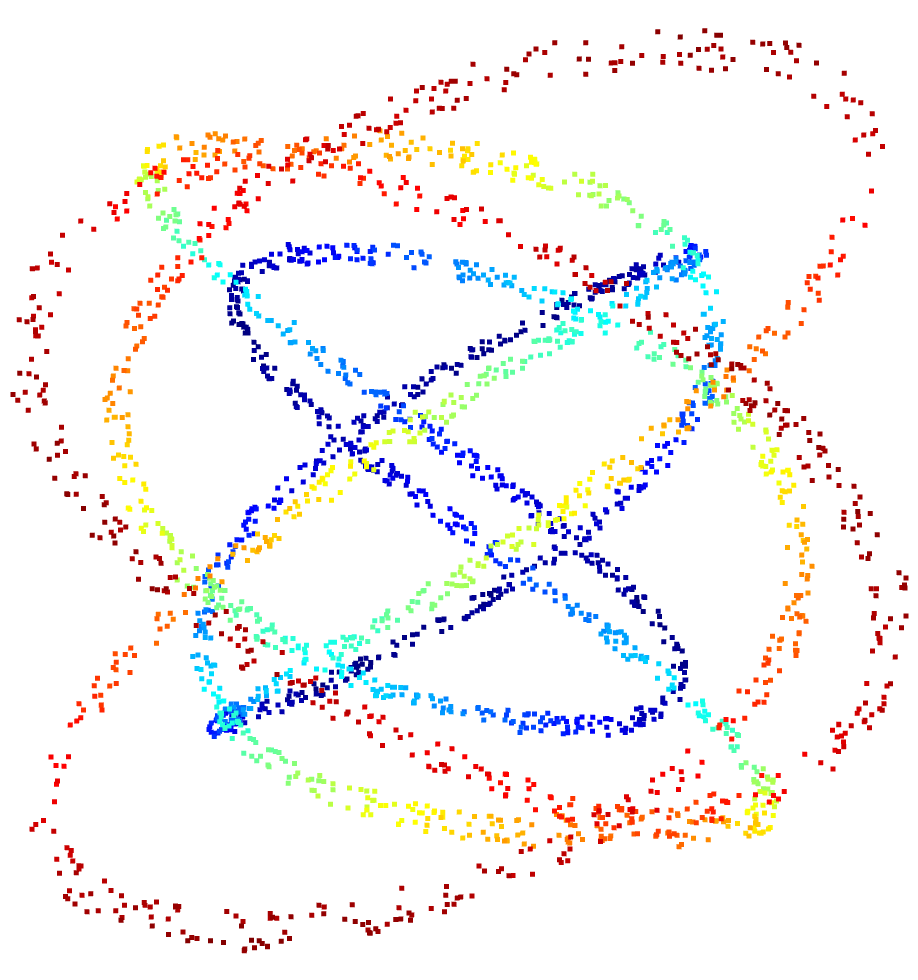}}

    \subfigure[Point cloud of the 43~726 vertices that conform the original genus 2 mesh constructed from a Fibonacci singular knot.]{\includegraphics[width=0.43\linewidth]{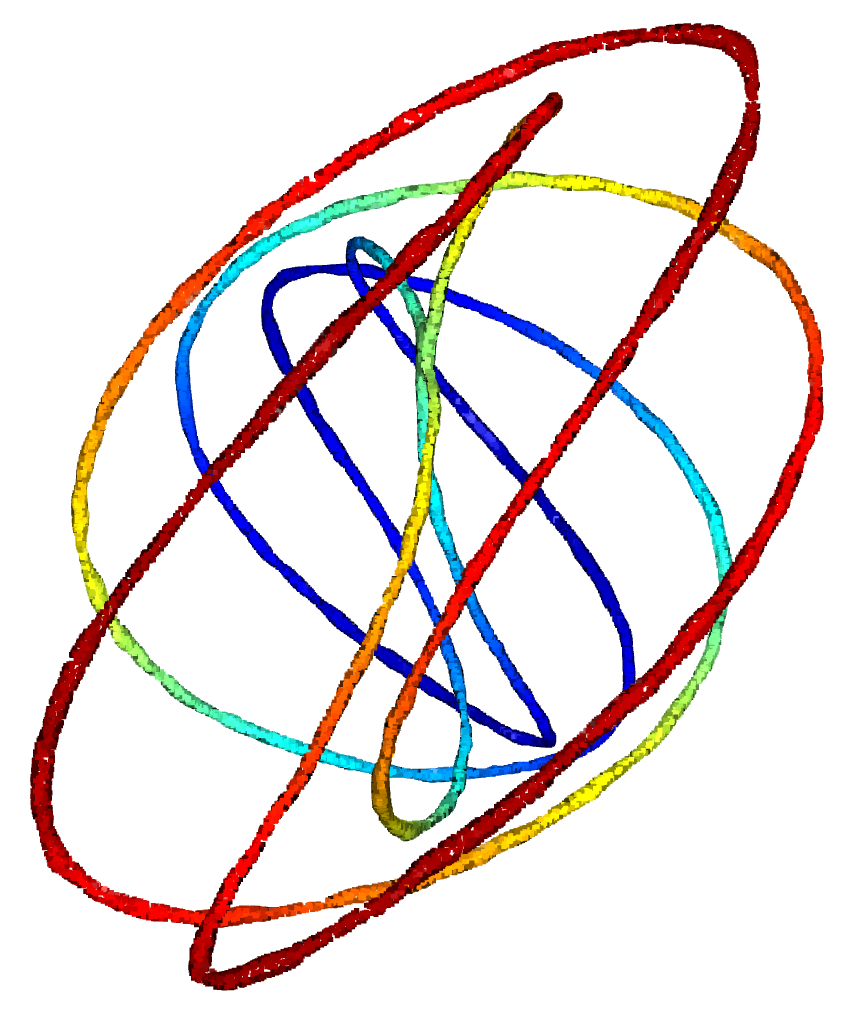}}
    \hspace{2mm}
    \subfigure[Sample of 3~000 vertices from the Fibonacci mesh.]{\includegraphics[width=0.43\linewidth]{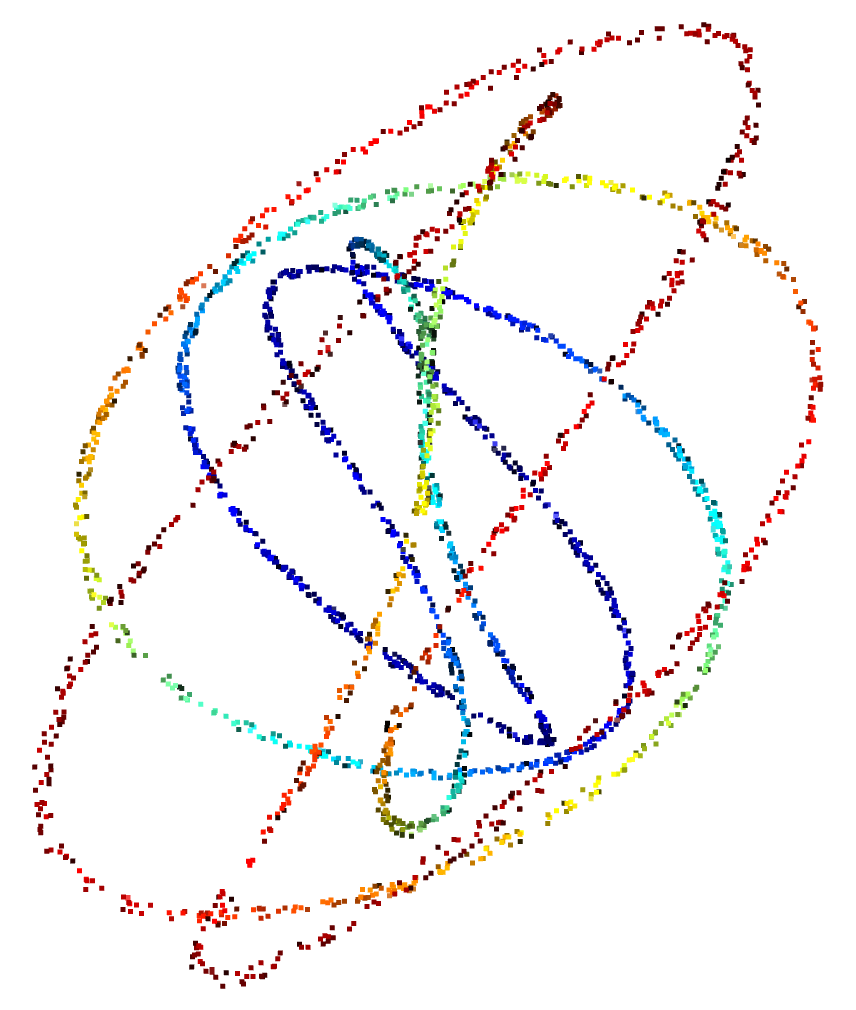}}    
    \caption{ Illustration of the sampling process for two meshes. The original point clouds (left) contain a large number of vertices, while the right-side images show the 3,000-vertex sampled subsets. The color gradient indicates relative depth, with red representing the closest points to the viewer and blue the farthest.}
    \label{fig:sampling}
\end{figure}

 In \cref{sec:sampling}, we describe the process of extracting 3,000 vertices from each mesh to create a standardized sample to train and test neural networks.
The meshes in our dataset contain varying numbers of vertices, so this sampling approach reduces memory consumption while enhancing the density of the adjacency matrix for graph-based representations.
\cref{fig:sampling} shows two examples of the point cloud obtained from the vertices of the triangulation compared to the sampled point cloud.

The extent of information loss depends on several factors, including the arc length of the original 1D curve, the thickness of the tubular neighborhood, and the resolution of the voxel grid.

\section{ShapeNet Examples with Higher Genera}
\label{appendix:ShapeNet}

In Section \ref{Introduction}, we mentioned that some of the meshes in the ShapeNet dataset are not well-defined manifolds. 
This is because most of the surfaces consist of multiple meshes, organized as scenes rather than single objects. 
To analyze this, we used the Trimesh package in Python, which provides the number of faces, edges, vertices, and, consequently, the Euler characteristic.

For our analysis of the ShapeNet dataset we considered two concepts: watertight mesh and manifold mesh. 
A watertight mesh is fully connected, forming a single continuous surface with no gaps.
A manifold mesh has edge consistency: every edge in the mesh is shared by exactly two faces.
Also, the immediate neighborhood around each vertex resembles a disk, this means that the local geometry is well-behaved and consistent with manifold theory.
However, a non-manifold mesh fails to satisfy key topological criteria, such as every edge being shared by exactly two faces or vertices being free of overlapping or ambiguity. 
Non-manifold properties often arise due to inconsistencies in mesh construction, especially in large datasets like ShapeNet, which aggregate models from diverse sources.

\begin{figure}
    \centering
    \subfigure[Bench from the ShapeNet dataset with 46 handles, although the Trimesh method computes the genus of this mesh to be 6.]{    \includegraphics[scale = 0.45]{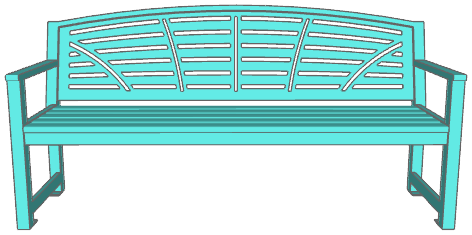}\label{fig:ShapenetBench1}}
    \hspace{4mm}
    \subfigure[ShapeNet mesh with genus $-4732$.]{\includegraphics[scale = 0.55]{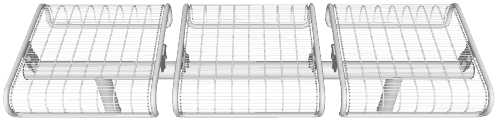}\label{fig:ShapenetBench2}}
    \caption{A significant number of the instances that have a well-defined Euler characteristic in the ShapeNet dataset exhibit negative genus or a non-negative genus that does not equal the number of handles in the surface.}
    \label{fig:ShapenetBench}
\end{figure}

In cases where the mesh is non-manifold, the Euler characteristic and genus are still defined, but the genus does not equal the number of visible holes of the surface. 
This discrepancy reveals the ill-defined topology of the ShapeNet dataset in general.
For example, the bench mesh of \cref{fig:ShapenetBench1} has a genus different from the number of holes in the surface.
Moreover, most of the ShapeNet meshes whose Euler characteristic is defined, Trimesh computes it to have a negative genus, such as the one in \cref{fig:ShapenetBench2}.

\begin{figure}
    \centering
    \subfigure[g=0]{\includegraphics[width=0.19\linewidth]{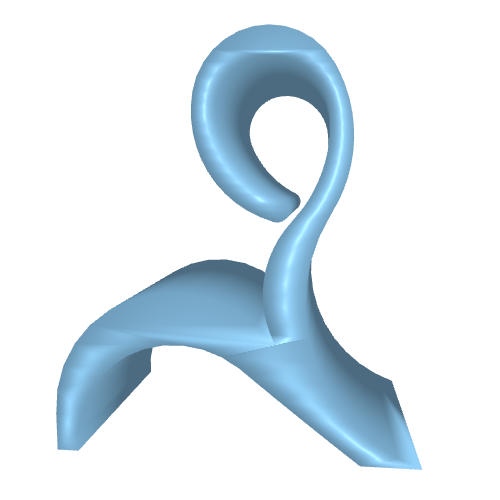}}
    \subfigure[g=0]{\includegraphics[width=0.19\linewidth]{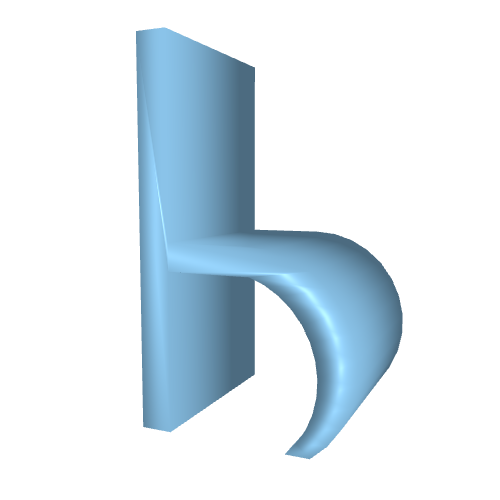}}
    \subfigure[g=0]{\includegraphics[width=0.19\linewidth]{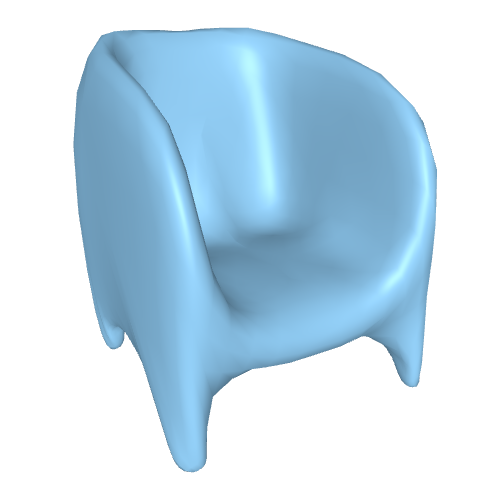}}
    \subfigure[g=1]{\includegraphics[width=0.19\linewidth]{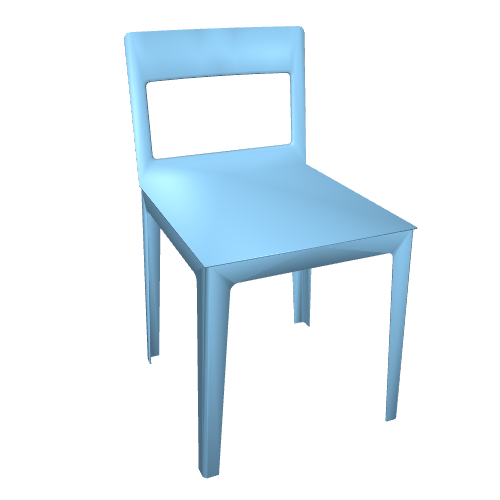}}
    \subfigure[g=1]{\includegraphics[width=0.19\linewidth]{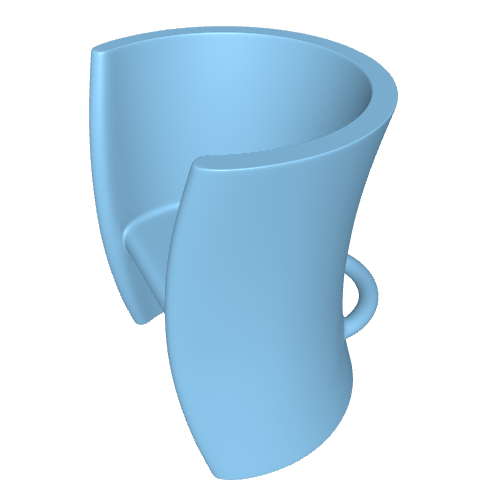}}
    \caption{The five ShapeNet chair meshes (out of 6778) that are watertight and have non-negative genus, which, consequently, matches the number of holes.}
    \label{fig:ShapeNetGenusHoles}
\end{figure}

Additionally, a subset of meshes in the dataset is watertight, meaning that they do not have boundary edges and are thus topologically valid (they form a closed surface). 
For these watertight meshes, only if the genus is non-negative, then it equals the number of holes, which means the mesh has a clear topological structure.
For instance, of the 6778 ShapeNet chair models, 61 (less than 1\%) have a genus greater than or equal to zero; 13 (0.2\%) are watertight, and 5 (0.074\%) are both watertight and have non-negative genus, only in this case their genus equals the number of holes (see \cref{fig:ShapeNetGenusHoles}).
Similarly, of 1813 bench models, 35 (less than 2\%) have non-negative genus; 4 of them (0.22\%) are watertight, and none of them are both.

\begin{figure}
    \centering
    
    \subfigure[Full view.]{\includegraphics[scale = 0.32]{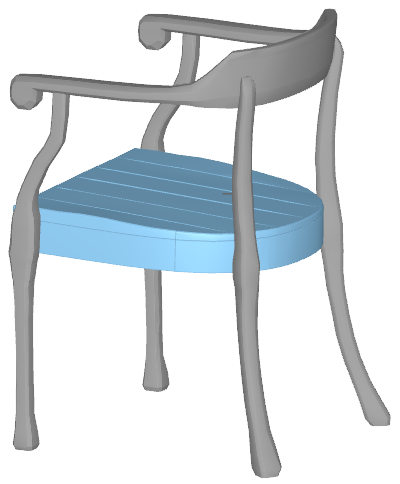}}
    \hspace{2mm}
    \subfigure[Zoom-In.]{\includegraphics[scale = 0.23]{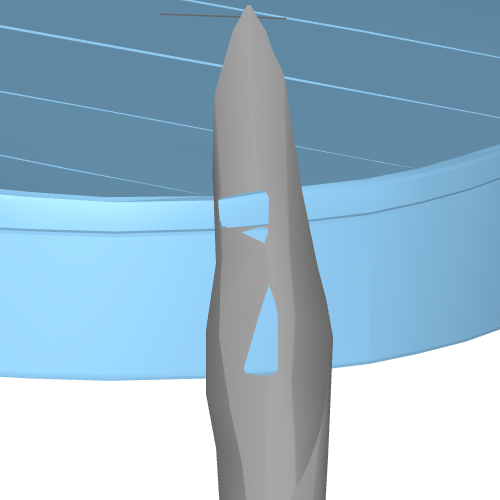}}
    \caption{Detail of the attachment of separate components into a single ShapeNet model consisting of an ill-defined mesh with genus 10, whereas the expected genus should be 3.}
    \label{fig:ShapeNetIllChair}
\end{figure}

\cref{fig:ShapeNetIllChair} illustrates how ShapeNet models are constructed by attaching parts together rather than emerging as a single piece. Our models have this single-piece condition guaranteed via the use of the Marching Cubes algorithm.

\section{ABC, Thingi10K, and SimJEB Datasets}
\label{App:ABC}

The ABC dataset contains more than a million CAD models \cite{Koch_2019_CVPR}, the Thingi10K dataset contains 10 thousand 3D printing models \cite{zhou2016thingi10kdataset100003dprinting}, and SimJEB contains 381 simulated jet engine bracket CAD models \cite{Whalen_2021}.

Although most of the models in these datasets have a well-defined watertight surface with a diverse genus distribution (the Thingi10K website even provides a filter to browse models by genus), the genera are not uniformly distributed, with the highest concentration at genus 0 for ABC and Thingi10K, with progressively fewer examples as the genus increases (see Figures \ref{fig:ABC} and \ref{fig:Thingi10K}), with such anomalous genera that, for Thingi10K, 2328 models are either negative genus surfaces or non-integer genus surfaces (387 are both), while the highest concentration of the SimJEB dataset is at genus 6 with 163 models, which equals to 42.8\% of the total (see \cref{fig:SimJEB}), with the lowest genus at $-80$ and the highest genus at 361.
The histograms of Figures \ref{fig:ABC}-\ref{fig:SimJEB} include a Kernel Density Estimation (KDE) curve in red, which is flat in the case of Thingi10K due to the highly negative (the most negative genus is $-190568.5$) and highly positive genera (the highest genus is 6781).
The lower limit for ABC is $-244$ and the upper limit is 94.

Furthermore, these datasets include models with negative (or even fractional, in the case of Thingi10K) genus values, which are topologically invalid.
Invalid genus values typically arise from inaccurate or inconsistent mesh processing, especially during automatic topology extraction. This can occur due to:

\begin{figure}[H]
    \centering
    \subfigure[Genus distribution for an ABC dataset sample of 2000 models.]{\includegraphics[width=0.48\linewidth]{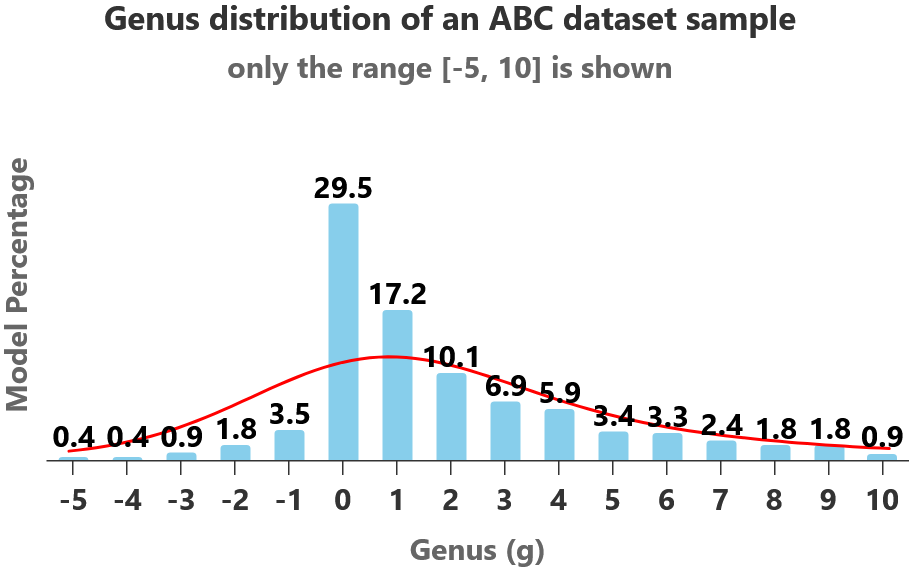}
    \label{fig:ABC}}
    \hspace{2mm}
    \subfigure[Genus distribution for the 9997 models with defined Euler characteristic in the Thingi10K dataset summary file.]{\includegraphics[width=0.48\linewidth]{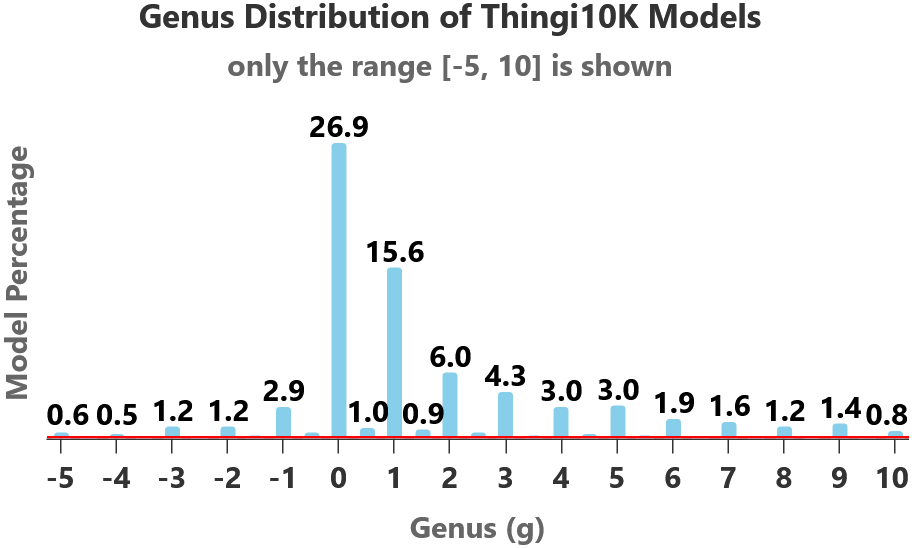}
    \label{fig:Thingi10K}}
    \linebreak
    \subfigure[Genus distribution for the 381 models in the SimJEB dataset.]{\includegraphics[width=0.48\linewidth]{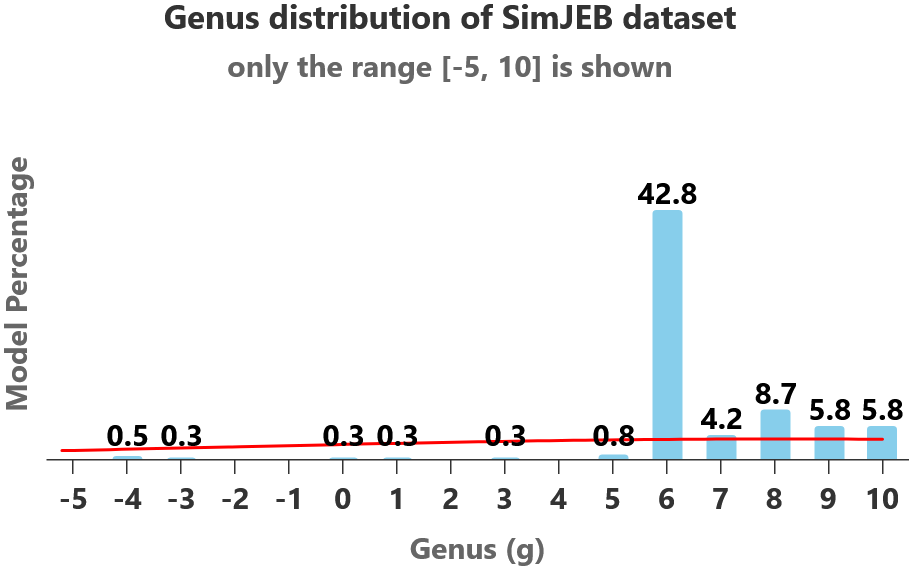}
    \label{fig:SimJEB}}
    \hspace{2mm}
    \subfigure[ShapeNet genus distribution for the 1086 (out of 6778) chair models with defined Euler characteristic. All are non-valid meshes, except for the three of genus 0 and the two of genus 1 of \cref{fig:ShapeNetGenusHoles}.]{\includegraphics[width=0.48\linewidth]{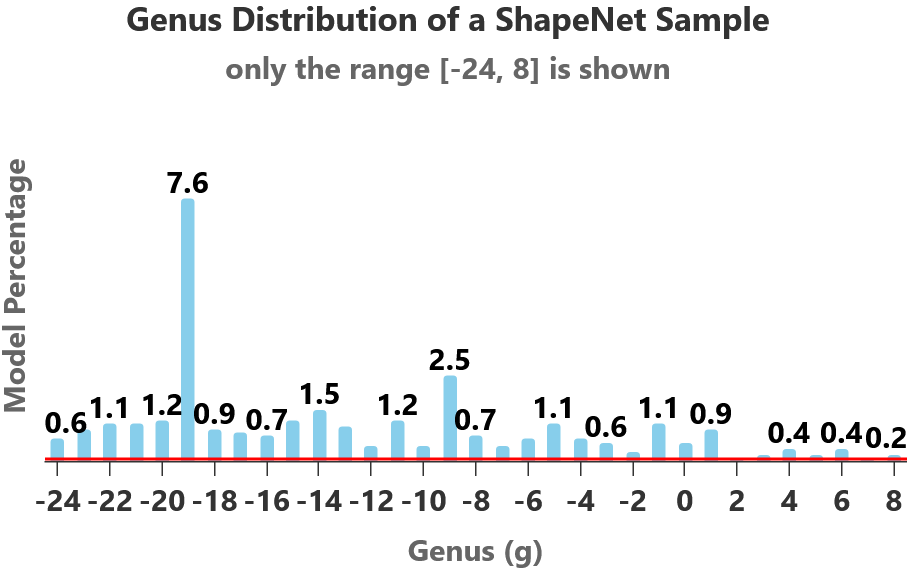}
    \label{fig:ShapeNetGenusDistr}}
    \caption{Genus distribution for the SimJEB and Thingi10K datasets, and for a sample of the ABC and ShapeNet datasets.}
\end{figure}

\begin{itemize}
    \item Non-manifold edges or vertices, which break Euler characteristic assumptions.
    \item Floating-point errors during genus computation on complex or noisy meshes.
    \item Incorrect handling of mesh holes or disconnected components.
    \item Use of heuristic-based genus estimation rather than strict topological invariants.
\end{itemize}

Such issues are more common in large-scale, crowd-sourced or automatically converted datasets.
These issues highlight the lack of a curated dataset that offers a regular and uniformly distributed range of genus-diverse surfaces, with clean and consistent topological properties.

\section{Analysis of Results} \label{app:Significance}

To guarantee the statistical significance and generality of our experimental results, we performed an ablation study (Section~\ref{sec:ablation}) and a sensitivity analysis (Section~\ref{sec:sensitivity}).

We performed a comprehensive evaluation through two complementary analyses to ensure the statistical significance and generality of our experimental results. 
Firstly, we conducted an ablation study (Section\ref{sec:ablation}). 
This analysis identifies the key elements that drive our results and shows how they interact. 
Secondly, we carried out a sensitivity analysis (Section\ref{sec:sensitivity}), which examines the robustness of our findings to variations in parameters, assumptions, and other factors that could affect the outcome. 
With these two analyses, we thoroughly validate our results, increase their reliability, and demonstrate their applicability.

\subsection{Ablation Study} \label{sec:ablation}

\begin{figure}
    \centering
    \includegraphics[width=0.9\linewidth]{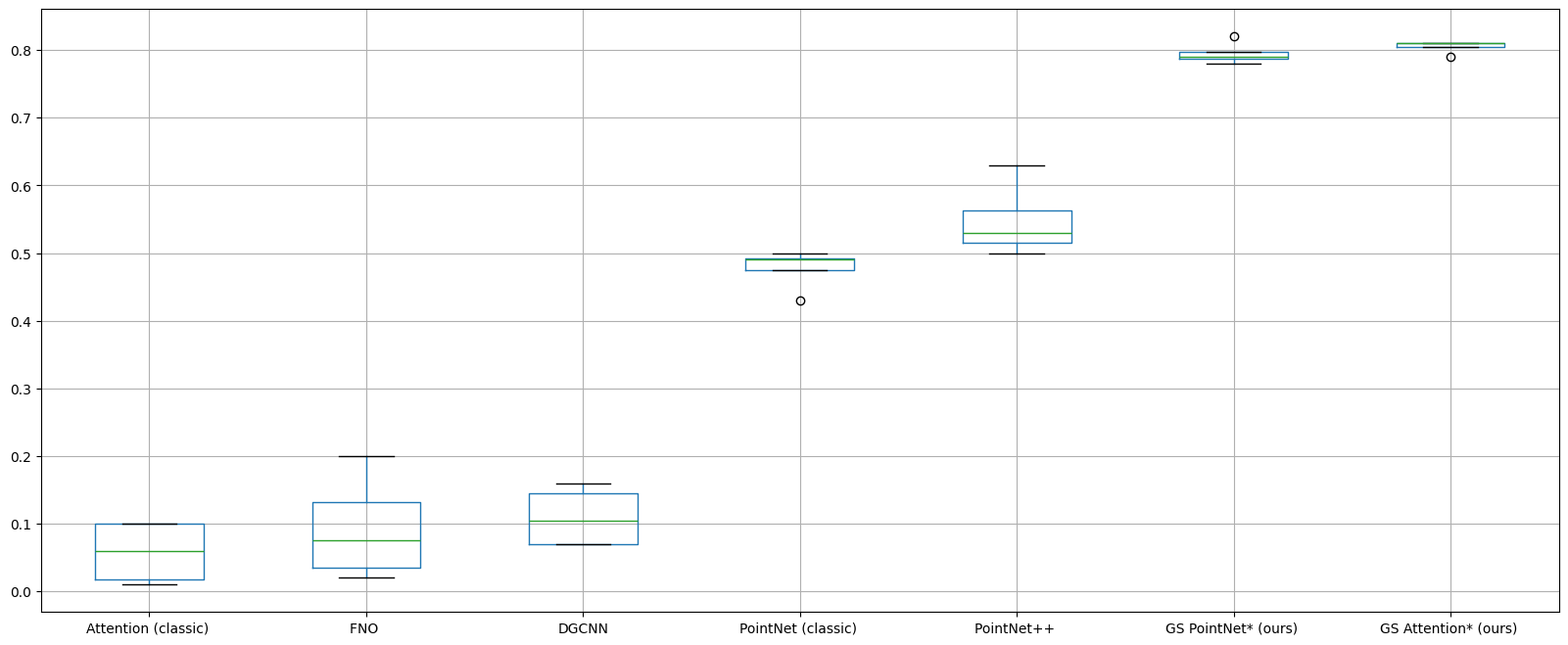}
    \caption{BoxPlot for the evaluation results for the different models tested.}
    \label{fig:BoxPlot}
\end{figure}

We conducted a quantitative ablation study to compare our graph-based sampling method's computational efficiency and classification performance with established techniques.
Figure~\ref{fig:BoxPlot} presents a box plot of the evaluation metrics for the models we tested on the genus classification task.
This figure illustrates that the classic attention transformer model, the FNO, and the DGCNN models exhibit lower evaluation scores, clustering below 0.1, and display higher variance.
In contrast, our proposed graph-informed methods demonstrate more concentrated distributions and lower variance across the evaluation metrics.
Both graph-based methods exhibit similar performance, surpassing the results obtained by other methods. 
Our study reveals that point cloud methods, including GCNN, FNO, and simple attention transformer, struggle with higher genus surfaces and exhibit poor performance.
In contrast, incorporating information about neighboring relations improves performance, as observed with PointNet and PointNet++ architectures, which yield similar distributed results. However, these methods and FNO, and GCNN are computationally expensive. Our proposed methods mitigate this complexity, with the simple PointNet method with adjacency information being more efficient than the original PointNet architecture. 
By leveraging adjacency information, our PointNet implementation reduces computational complexity by $O(N^2)$ steps, where N is the number of points. 
Furthermore, our attention implementation is more efficient, eliminating the need for attention heads, resulting in an execution time of $O(N^2)$ for this layer.

\begin{figure}
    \centering
    \includegraphics[width=\linewidth]{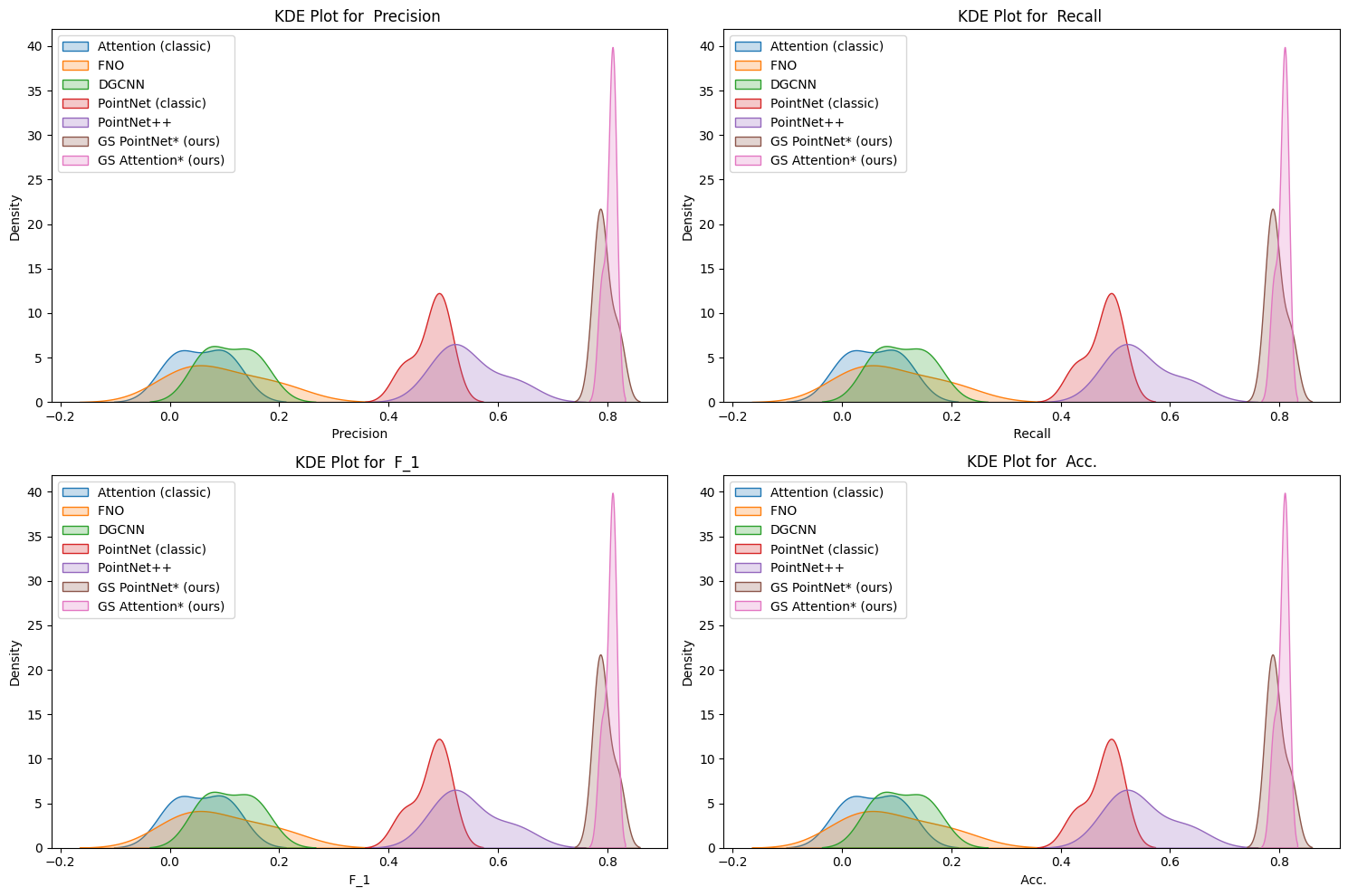}
    \caption{KDE plots for the evaluation results for the different models tested.}
    \label{fig:KDE}
\end{figure}

Figure~\ref{fig:KDE} displays the kernel density estimation (KDE) plots for the precision, recall, $F_1$, and accuracy metrics.
These plots illustrate the distribution of the models across regions based on their performance.
Notably, the proposed methods cluster together in the high-performance region, whereas the classic attention, FNO, and DGCNN models cluster in the low-performance region for all metrics.
Lastly, the PointNet and PointNet++ architectures occupy an intermediate position, clustering in the middle region.

These findings suggest that our graph-based sampling method offers a favorable trade-off between computational efficiency and classification performance, making it a promising approach for processing complex point cloud data. 
Future studies can build upon these results to further explore the benefits and limitations of graph-informed methods in various applications.

\begin{figure}
    \renewcommand{\thesubfigure}{} 
    \centering
    \subfigure[F-statistic]{\includegraphics[width=0.45\linewidth]{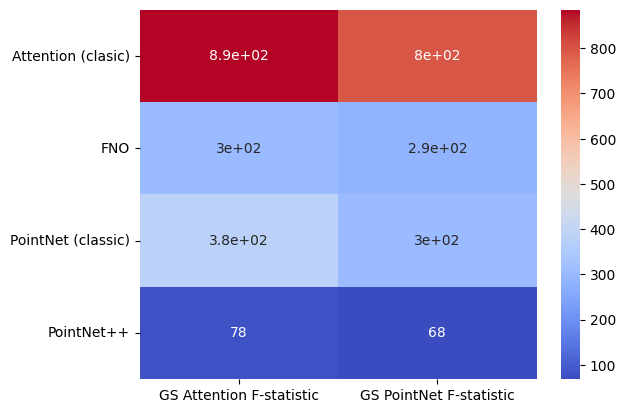}}
    \subfigure[$p$-values]{\includegraphics[width=0.45\linewidth]{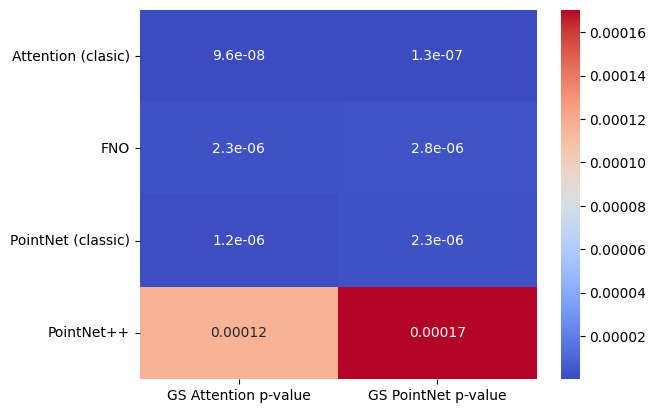}}
 \caption{F-satatistic analysis comparing graph-informed methods against classic methods}
    \label{fig:FStat}
\end{figure}

We performed an F-statistic analysis to assess the significance of the obtained results. 
Figure~\ref{fig:FStat} shows the outcomes of this analysis. 
The results indicate that the PointNet++ model yielded the lowest F-statistic values for GS Attention and GS PointNet. 
Furthermore, in all cases, the resulting p-values were below the threshold of 0.05, which suggests a statistically significant difference between the models.

The results of this study demonstrate that our graph-informed methods outperform other approaches in detecting topological patterns, leading to improved classification of genera.

\subsection{Sensitivity Analysis} \label{sec:sensitivity}

To optimize the performance of our best model for the task of genus classification, we conducted a sensitivity analysis to identify the optimal hyperparameters. 
We utilized Optuna to evaluate the effects of three key hyperparameters: 1) the learning rate, 2) the dimensionality of the vector representations, and 3) the random seed. 
Due to the computational requirements our models need, we limited our analysis to 10 trials, with the hyperparameters varied across a range of significant values.
For this sensitivity analysis, we use a subset of 50 data points per genus (from genus 0 to 5) in the training set, and 25 per genus in the validation set.

\begin{figure}
    \centering
    \includegraphics[width=0.9\linewidth]{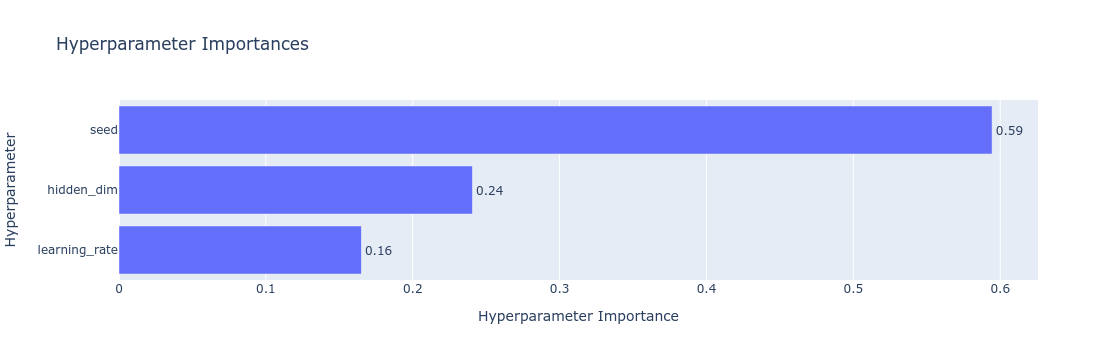}
    \caption{Relevance of the hyperparameters according to the sensitivity analysis over 10 trials}
    \label{fig:OptunaBestHyp}
\end{figure}

\begin{figure}
    \centering
    \includegraphics[width=0.9\linewidth]{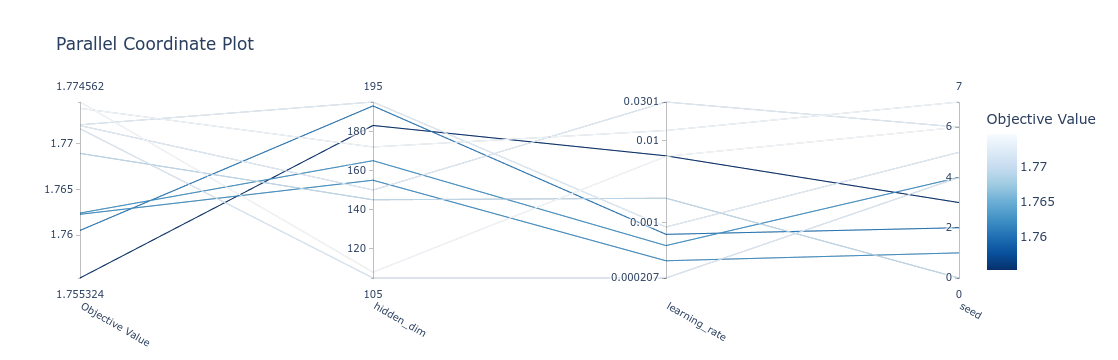}
    \caption{Parallel coordinate plot for the sensitivity analysis over 10 trials}
    \label{fig:Parallel}
\end{figure}

As illustrated in Figure\ref{fig:OptunaBestHyp}, we found the random seed to be the most significant hyperparameter, with a relative importance of 0.45. 
The dimensionality of the vector representation was the second most influential hyperparameter. 
Figure\ref{fig:Parallel} presernts a parallel coordinate plot of the results, which reveals that the optimal performance was achieved with a higher number of hidden dimensions and a relatively low initial learning rate. 
Furthermore, the results indicate that the random seed values between 2 and 4 yielded the best outcomes.

\begin{figure}
    \centering
    \includegraphics[width=0.9\linewidth]{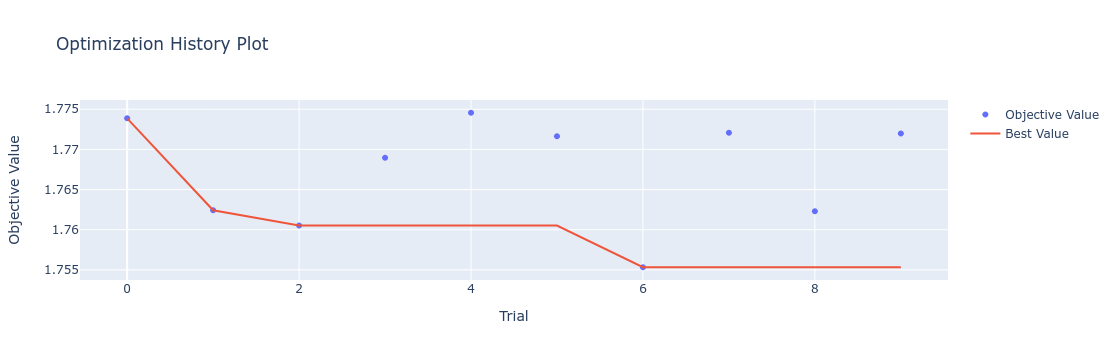}
    \caption{Objective function values through different trials}
    \label{fig:ObjectiveValues}
\end{figure}

Figure~\ref{fig:ObjectiveValues} illustrates the variations in the objective value function, specifically cross-entropy, across trials.
The discrepancy between the worst and best evaluation scores is not significant, with a difference of only 0.02 points.
This small difference suggests that the model converges to relatively optimal solutions under different hyperparameter configurations.
We attribute the adequate convergence of the models, in part, to the utilization of the Noam optimizer.
The Noam optimizer normalizes the learning rate in Adam, thereby facilitating improved convergence.
This optimizer incorporates the hidden dimension to normalize the learning rate and is commonly employed with an initial learning rate of 0, which adapts dynamically during training.
This phenomenon also explains why the random seed emerges as the most significant hyperparameter, as it is not accounted for in the optimizer.
However, as previously noted, the fluctuations in evaluation values are relatively minor.

\begin{figure}
    \renewcommand{\thesubfigure}{} 
    \centering
    \subfigure[Parallel coordinates]{\includegraphics[width=0.9\linewidth]{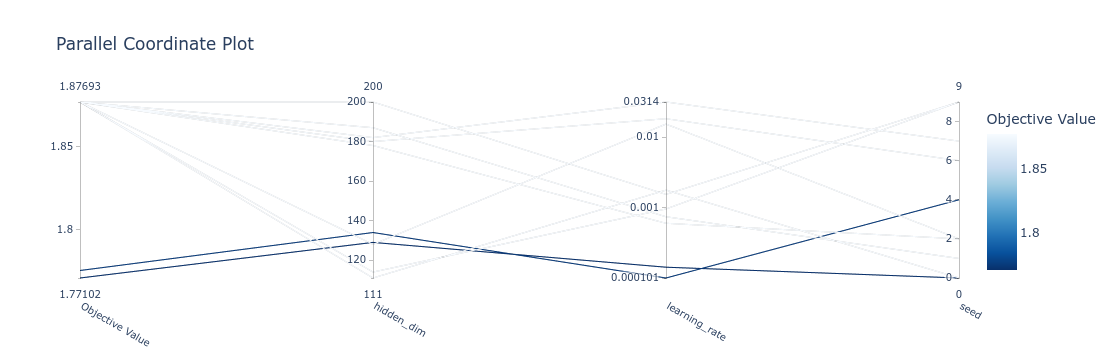}}
    \subfigure[Significance]{\includegraphics[width=0.85\linewidth]{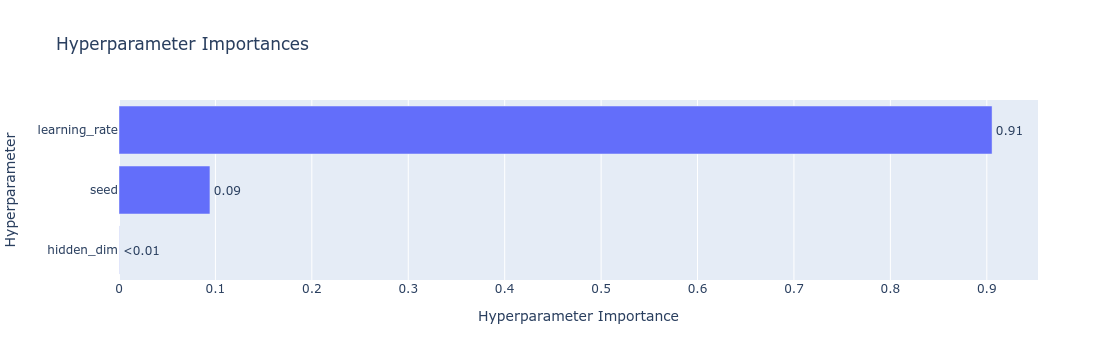}}
 \caption{Sensitivity analysis using Adam optimizer over 10 trials}
    \label{fig:Adam}
\end{figure}

To further investigate this, we conducted a secondary sensitivity analysis under identical conditions, except utilizing the Adam optimizer.
Figure~\ref{fig:Adam} presents the results of this analysis.
As evident from the figure, the variation between the objective function values is more pronounced, ranging from 1.77 to 1.87, resulting in a 0.1 difference.
Notably, the learning rate emerges as the most significant hyperparameter in this scenario.
This finding is pertinent, as it corroborates our hypothesis that the choice of optimizer substantially impacts the convergence of the model training process.
The prominence of the learning rate when using the Adam optimizer suggests that how the optimizer normalizes the learning rate during training profoundly influences model convergence.
This analysis motivates the use of the Noam optimizer in our experiments.

\section{EuLearn Datasheet}
\label{App:Datasheet}

Here we include a Datasheet describing our EuLearn dataset, following the guidelines of \cite{10.1145/3458723}.

\begin{enumerate}
    
\item \emph{ For what purpose was the dataset created? } To be able to train machine learning and deep learning systems to understand topological variation in a uniformly controlled way. 
\item \emph{ What do the instances that comprise the dataset represent (for example, documents, photos, people, countries)? }
The instances are embedded and possibly knotted $2$-dimensional surfaces in $3$-space. Each object has a mesh (.stl), a scalar field (.txt), and a smoothed mesh (.stl).

\item \emph{ How many instances are there in total (of each type, if appropriate)? }
We have a total of 3~300 instances, with 300 surfaces for each genus, from $0$ to $10$. 
The total number of files is 9~900, as each surface has 3 associated files.

\item \emph{  Does the dataset contain all possible instances or is it a sample (not necessarily random) of instances from a larger set? }
The dataset is a random sample, whose construction builds upon parameterizations of random knots.

\item \emph{ What data does each instance consist of? }
Three files, two in .stl format and another in .txt format.

\item \emph{ Is there a label or target associated with each instance? }
The target associated with each instance is its topological genus.
The name labeling indicates the surface genus, knot type, its six associated parameters (frequencies and phases), frequency, constant and amplitude for the sinusoidal variation, minimum and maximum radius, and number of voxels per axis.

\item \emph{ Are relationships between individual instances made explicit (for example, users’ movie ratings, social network links)? } 
Yes, multiple instances are generated by a single random knot type, thus these are related by belonging to the same isotopy type of embedding. 

\item \emph{ Are there recommended data splits (for example, training, development/validation, testing)? }
Due to local memory restrictions on our GPU units, for our reported experiments we adapted a data split method that saves training and testing subsets to our local hard disk for  batch processing. This method randomly selects part of the data to train and to evaluate a model, saving these sets in a local directory, so the model can call the necessary batches one at a time.
If memory availability is not a concern, you may use the entire dataset at once, using standard data splits.

\item \emph{ Are there any errors, sources of noise, or redundancies in the dataset?}
None that we are aware of. Please do not hesitate to contact us if any appear to be there.

\item \emph{ Is the dataset self-contained, or does it link to or otherwise rely on external resources (for example, websites, tweets, other datasets)? }
The EuLearn dataset is self-contained.

\item  \emph{ Does the dataset contain data that might be considered confidential
(for example, data that is protected by legal privilege or by doctor–patient
confidentiality, data that includes the content of individuals’ non-public communications)? }
None of the data is confidential.

\item \emph{ Does the dataset identify any subpopulations (for example, by age, gender)? }
There are no population identifications.

\item \emph{ Is it possible to identify individuals (that is, one or more natural persons), either directly or indirectly (that is, in combination with other data) from the dataset? }
This dataset does not include individual data.

\item \emph{ Does the dataset contain data that might be considered sensitive in any way (for example, data that reveals race or ethnic origins, sexual orientations, religious beliefs, political opinions or union memberships, or locations; financial or health data; biometric or genetic data; forms of government identification, such as social security numbers; criminal history)? }
No.

\item \emph{ How was the data associated with each instance acquired? Was the data directly observable (for example, raw text, movie ratings), reported by subjects (for example, survey responses), or indirectly inferred/ derived from other data (for example, part-of-speech tags, model-based guesses for age or language)? }
This is a synthetic mathematical dataset, the data associated with each instance was created specifically to produce a uniform genus distribution.

\item \emph{ What mechanisms or procedures were used to collect the data (for example, hardware apparatuses or sensors, manual human curation, software programs, software APIs)? }

The parameterizations needed to generate each of the instances were drawn from a pool limited to frequencies up to 11 and the phase was taken from the set $\{\pi/2, \pi/3, \pi/5, \pi/7\}$.
For this purpose, a code script was designed to compute the number of self-intersections for each triple of frequencies with one of those phases.
The triples follow $n_x < n_y < n_z$ and the phase was set in $x$, $y$, or $z$.
Once we knew the number of self-intersections, then we manually add the corresponding parameters to a list ordered by genus.
Finally, we ran the code to generate the corresponding objects: scalar field, sharp surface, and smooth surface.

The dataset development integrated several Python modules into the workflow. 
Parallel computations of the scalar field were performed via pyCUDA and numpy, and scipy's differential evolution method was used to identify knot self-intersections. 
The modules trimesh, numpy-stl, and the scikit-image implementation of the Marching Cubes algorithm assisted the mesh file generation.
We used the Blender software for mesh smoothing.

Our workflow's modular design carried out the computations and mesh development in multiple servers to reduce the dataset generation time. 
The dataset was primarily generated in an Intel i7 server with 64GB of RAM and RTX2070 GPU, and the self-intersection detection and training tasks were achieved using two servers, one server with an Xeon Gold processor with 64 GB of RAM and two Tesla V100 GPUs, and a second one equipped with a GeForce GTX 980 videocard and 32GB of RAM.

\item \emph{ If the dataset is a sample from a larger set, what was the sampling strategy (for example, deterministic, probabilistic with specific sampling probabilities)?}

 The sampling strategy was deterministic, since we manually selected the most complete subset of instances from a larger set.
For some genera, there were more than 15 parameterizations and so we selected those that completed the 20 frequencies for the sinusoidal variation (some did not because the genus was altered due to the contact of the surface with itself when the radius exceeded the reach), and if there were still more than 15, we manually selected the best-looking ones.

\item \emph{ Who was involved in the data collection process (for example, students, crowdworkers, contractors) and how were they compensated (for example, how much were crowdworkers paid)?}
Faculty, postdocs and graduate students, receiving a salary or a fellowship. %

\item \emph{ Over what time frame was the data collected? }
The research into the data design process started in fall 2021. 
The data generation procedure for the final version spanned two months, November-December 2024.

\item \emph{ Did you collect the data from the individuals in question directly, or obtain it via third parties or other sources (for example, websites)?}
No individuals were involved.
We relied on HPC supercomputing resources, obtained through a national grant application. %

\item \emph{ Were any ethical review processes conducted (for example, by an institutional review board)?}
None were needed, none were conducted.

\item \emph{ Has an analysis of the potential impact of the dataset and its use on data subjects (for example, a data protection impact analysis) been conducted? }
It has not, as this is a synthetic dataset and no data protection analysis is required.

\item \emph{ Was any preprocessing/cleaning/labeling of the data done (for example, discretization or bucketing, tokenization, part-of-speech tagging, SIFT feature extraction, removal of instances, processing of missing values)? }
Yes, a scalar field was created in order to construct the final mesh object for each instance, and a post-processing smoothing procedure was performed.

\item \emph{ Was the “raw” data saved in addition to the preprocessed/cleaned/ labeled data (for example, to support unanticipated future uses)? }
While we did not save the curve data that the database relies on as a 3D object, we do have a list of the parameters used to create the associated curves. 
The available data may be considered the raw data itself. 

\item \emph{ Is the software that was used to preprocess/clean/label the data available? }
Yes. Namely, an implementation of Marching Cubes, Blender for smoothing, and CUDA for computing the scalar fields. 

\item \emph{ Has the dataset been used for any tasks already? }
Not yet.

\item \emph{ Is there a repository that links to any or all papers or systems that use the dataset?  }
\blue{\url{https://huggingface.co/datasets/appliedgeometry/EuLearn}} %

\item \emph{ What (other) tasks could the dataset be used for? }
It could be of crucial assitance in model pretraining, so that a proposed architecture is guaranteed to have the ability to distinguish varying topological types in 3D data.
We mentioned specific engineering and scientific applications where these techniques are relevant in the main body. 

\item \emph{ Is there anything about the composition of the dataset or the way it was collected and preprocessed/cleaned/labeled that might impact future uses? }
The files are in .stl and .txt formats. As long as these remain in use, there should be no foreseeable impact on future users. 

\item \emph{ Are there tasks for which the dataset should not be used? }
Not that we are aware of. 

\item \emph{ Will the dataset be distributed to third parties outside of the entity (for example, company, institution, organization) on behalf of which the dataset was created? }
It will be made openly and publicly available.

\item \emph{ How will the dataset be distributed (for example, tarball on website, API, GitHub)? }
Through HuggingFace and in an institutional repository.

\item \emph{ When will the dataset be distributed? }
The dataset will be made available upon submission of the paper.%

\item \emph{ Will the dataset be distributed under a copyright or other intellectual property (IP) license, and/or under applicable terms of use (ToU)? }
With a MIT license. %

\item \emph{ Have any third parties imposed IP-based or other restrictions on the data associated with the instances? }
None.

\item \emph{ Do any export controls or other regulatory restrictions apply to the dataset or to individual instances? }
No.

\item \emph{ Who will be supporting/hosting/maintaining the dataset? }
The authors.

\item \emph{ How can the owner/curator/ manager of the dataset be contacted? }
Via email, found in the associated publication.

\item \emph{ Is there an erratum? }
Not yet.

\item \emph{ Will the dataset be updated (for example, to correct labeling errors, add new instances, delete instances)? }
Perhaps after the submission revision process.

\item \emph{ If others want to extend/augment/build on/contribute to the dataset, is there a mechanism for them to do so? }
This includes two possibilities. 
For those interested in augmenting the available instance in the genus range $\{0, \ldots , 10 \}$, they would have to construct curves that use parameters beyond our original parameter list. Then they could construct surfaces using a Marching Cubes method, with varying radial distances to the initial curve, again following a sinusoidal function.
For those interested in contributing surfaces with higher genus $g\geq 11$, they would have to replicate our entire workflow.

\end{enumerate}

\end{document}